%% file: teza.tex
\documentclass[12pt]{report}

\usepackage[croatian,english]{babel} 

\usepackage[utf8x]{inputenc}

\usepackage[a4paper,top=3cm,bottom=3cm,right=3cm,left=3cm]{geometry}

\usepackage[onehalfspacing]{setspace}

\usepackage{multicol}

\usepackage{afterpage}

\usepackage{caption}
\usepackage{fancyhdr}
\fancypagestyle{plain}{
	\fancyhf{}
	\fancyfoot[R]{\thepage}
	
}

\usepackage{amsmath}
\usepackage{amssymb}

\usepackage{import}

\usepackage[numbers,compress]{natbib}

\usepackage[T1]{fontenc}

\usepackage[linktocpage]{hyperref}
\hypersetup{colorlinks,allcolors=black} 

\usepackage{url}

\usepackage{graphicx}

\usepackage{verbatim} 
\usepackage{rotating} 

\usepackage[titletoc]{appendix}

\usepackage{eurosym}

\usepackage{subfig}

\usepackage{longtable}
\usepackage[shortlabels]{enumitem}

\usepackage[nottoc,notlot,notlof]{tocbibind}
\AtBeginDocument{\renewcommand{\bibname}{References}}

\newcommand{\markedchapter}[2]{\chapter[#2]{#2%
\chaptermark{#1}}
\chaptermark{#1}}

\newcommand{\markedsection}[2]{\section[#2]{#2%
\sectionmark{#1}}
\sectionmark{#1}}

\renewenvironment{abstract}{\rightskip1in\itshape}{}


\def\mn#1   {{\color{red} \scshape \textbf{ [MN: #1]} }}

\def\utw{\ensuremath{\smash{\rlap{\lower5pt\hbox{$\sim$}}}}}
\def\udtw{\ensuremath{\smash{\rlap{\lower6pt\hbox{$\approx$}}}}}

\let\OLDthebibliography\thebibliography
\renewcommand\thebibliography[1]{
  \OLDthebibliography{#1}
  \setlength{\parskip}{2pt}
  \setlength{\itemsep}{0pt plus 0.3ex}
  \setlength{\bibsep}{0pt plus 0.3ex}
}

\import{./}{journals.tex}

\makeatletter
\def\NAT@def@citea{\def\@citea{\NAT@separator}}
\makeatother

\usepackage{graphicx}

\usepackage{xcolor}

\usepackage{amsmath}
\usepackage{amsfonts}
\usepackage{amssymb}
\usepackage{bm}
\usepackage{bbm}
\usepackage{physics}
\usepackage{tikz}
\usepackage{scalerel}

\usepackage{times}
\usepackage{cleveref}

\usepackage{bibentry}
\nobibliography*
\input{mydefs.tex}

\begin{document}

\newgeometry{top=3cm,bottom=3cm,right=3cm,left=3cm}
\pagenumbering{gobble}

\import{./}{title.tex}

\newgeometry{top=2.5cm,bottom=2.5cm,right=2.5cm,left=2.5cm,includehead,includefoot} 

\chapter*{Supervisor information}

\textbf{Hrvoje Buljan} is a Professor at the Physics department at the Faculty of Science, University of Zagreb. His main research interests are in optics and photonics, graphene plasmonics and ultracold atomic gases, while his recent contributions include research on anyons. At the time of writing, he has authored/co-authored 80 publications, with 3931 citations (according to WoS). 
He has mentored five defended PhD theses, and currently mentors two. He leads/has lead ten research projects, including the 5MEUR project funded by the European Structural and Investment Funds 2017, the \textit{Scientific Center of Excellence for Quantum and Complex Systems and Representations of the Lie Algebra} 2015, the  \textit{Pseudomagnetic forces and fields for atoms and photons} funded by the Unity through Knowledge Fund 2013, and the \textit{Nonlinear phenomena and wave dynamics in photonic systems} funded by the Ministry of Science in Croatia 2007.
In 1995, he was awarded the Rector’s Award by the Rector of the University of Zagreb. He co-authored the most cited Phys. Rev. B paper published in 2009. He is the recipient of the Annual State Award for Science, awarded by the Croatian parliament in 2010, and the "Andrija Mohorovičić" award, awarded by the University of Zagreb in 2019.
\vspace{0.5cm}

\noindent\textbf{Dario Jukić} is an Assistant Professor of Physics at the Faculty of Civil Engineering, University of Zagreb. The main topics of his research include the quantum dynamics in many-body cold atomic gases (especially low-dimensional systems), light-matter interaction in such systems, and nonlinear phenomena in the fields of optics and photonics. His recent contributions include the study of anyons, and the interplay of nonlinearity and topology in photonics. At the time of writing, he has authored/co-authored 21 publications.
He is the recipient of the Award by the Faculty of Science, University of Zagreb, for exceptional success as a student in 2007. In 2009, he received annual award "Znanost" by the National Science Foundation for the best student
paper in natural science. He was awarded in 2019 and 2021 by the Faculty of Civil Engineering, University of Zagreb.

\chapter*{Acknowledgements}

This work was supported in part by the Croatian Science Foundation Grant No. IP-2016-06-5885 SynthMagIA; QuantiXLie Center of Excellence, a project cofinanced by the Croatian Government
and European Union through the European Regional Development Fund Competitiveness and Cohesion Operational Programme (Grant No. KK.01.1.1.01.0004); 
the National Key R\&D Program of China under Grant No. 2017YFA0303800; the National Natural Science Foundation (11922408, 91750204, 11674180); PCSIRT; the 111 Project (No. B07013) in China. We acknowledge support from the Croatian-Chinese Scientific \& Technological Cooperation.



\chapter*{Abstract}

\textbf{Keywords}: \textit{Berry phase, Berry curvature, topology, topological quantum matter, photonics, Zitterbewegung, symmetry broken honeycomb lattice, valley degree of freedom, SSH model, optical solitons, nonlinearity, anyons, fractional statistics, topological quantum computation, quantum Hall effect, synthetic anyons}
\vspace{.37cm}

Nontrivial topology in physical systems is the driving force behind many interesting phenomena. Notably, phases of matter must be classified in part by their topological properties. Phases with topological (nonlocal) order, such as the fractional quantum Hall effect (QHE), can support anyonic excitations obeying fractional statistics with potential application in topological quantum computing. States lacking intrinsic topological order can still belong to topological phases, provided certain symmetries are imposed. On their own, these symmetry-protected phases do not support anyons, but they can still have other interesting features, such as protected boundary states.
	In this thesis we present an original research into several topologically nontrivial systems. First, we present the experimental and theoretical results on light propagation in the valley modes of inversion-symmetry broken honeycomb lattices (HCL). We find that a rotating spiral pattern, leading to Zitterbewegung, arises in the intensity profile of the beam as a result of the nontrivial topology of the HCL valleys. Next, we present the numerical demonstration of dynamical topological phase transitions driven by nonlinearity of the photonic medium, which occur in soliton SSH lattices. The phase transitions, marked by the appearance of topological edge states in the band gap, occur due to the setup which enables continually changing relative values of the intracell and intercell soliton couplings. Finally, we propose a scheme for creating and manipulating synthetic anyons in a noninteracting system by perturbing it with specially tailored localized probes. The external probes are needed because noninteracting systems do not possess the kind of topological order required to support anyonic excitations. We start from a noninteracting 2D electron gas in a uniform magnetic field, which is in an integer QHE state, and introduce thin solenoids carrying a fractional magnetic flux. We present the solution for a suitable ground state, and demonstrate the fractional braiding statistics in the coordinates of the solenoids.


\pagenumbering{roman} 


\pagebreak
\begingroup
	\singlespacing
	\tableofcontents
	\listoffigures
\endgroup

\pagebreak
\pagenumbering{arabic}

\chapter{Introduction}
\input{0_intro/intro.tex}

\chapter{Theoretical background}
\label{ch:intro}
\input{1_introduction/introduction.tex}

\markedchapter{Topological Photonics}{Topological Photonics}\label{ch:topofoto}
\input{2_photo/photo.tex}

\markedchapter{Anyons}{Proposal for realization of Abelian anyons} \label{ch:prb}
\input{3_anyons/synth_anyons.tex}

\markedchapter{Conclusion}{Conclusion} \label{ch:concl}
\input{4_concl/concl.tex}

\newpage
\begingroup
\footnotesize
\singlespacing
\renewcommand{\bibname}{References}
\typeout{}

\bibliography{bibtex}
\endgroup

\chapter*{Curriculum vitae}
\addcontentsline{toc}{chapter}{Curriculum vitae}

Frane Lunić was born on September 25th, 1991 in Split (Croatia). In 2014, he obtained his Bachelor's degree in Physics, and in 2017, his Master's degree in Physics (major in Computational Physics) from the Faculty of Science, University of Split. Since 2018, he has been working as a research and teaching assistant in the group of Hrvoje Buljan at the Faculty of Science, University of Zagreb. He has co-authored four papers in journals cited in the Web of Science, being the first or co-first author on two of them. He has presented his work at 5 international scientific conferences and workshops.

\begingroup
\singlespacing
\import{Y_additional_materials/}{papers.tex}

\endgroup

\chapter*{Prošireni sažetak}
\addcontentsline{toc}{chapter}{Prošireni sažetak}
\markboth{Prošireni sažetak}{}
\import{./}{sazetak.tex}

\textbf{Ključne riječi}: \textit{Berryjeva faza, Berryjeva zakrivljenost, topologija, topološka kvantna materija, fotonika, Zitterbewegung, saćasta rešetka s narušenom simetrijom, dolinski stupanj slobode, SSH model, optički solitoni, nelinearnost, anyoni, frakcijska statistika, topološko kvantno računanje, kvantni Hallov efekt, sintetički anyoni}
\vspace{.37cm}

\end{document}

%% file: journals.tex
\def\arxivprefixesep{:}

\newcommand{\eprint}[2][]{%
{\tt\if!#1!#2\else#1\arxivprefixesep\ignorespaces#2\fi}%
}

%% file: mydefs.tex
\newcommand{\DOT}[1]{\overset{\bm .}{#1}}

\newcommand{\m}{{m}}

\newcommand{\Epsilon}{\mathcal E}

\newcommand{\mr}[1]{\mathrm{#1}}
\newcommand{\conj}[1]{\overline{#1}\,}
\newcommand{\Ne}{{N_e}}

\newcommand{\pdd}{\partial}
\newcommand{\vr}{{\bm{r}}}

\newcommand{\VP}{\bm A}
\newcommand{\B}{{\bm B}}
\newcommand{\gab}{\gamma_{AB}}

\newcommand{\A}{\bm{\mathcal{A}}}
\newcommand{\scalA}{\mathcal{A}}

\newcommand{\p}{\bm {p}}
\newcommand{\vk}{\bm{k}}
\newcommand{\q}{{\bm{k}}}
\newcommand{\sq}{k}

\newcommand{\oble}[1]{\left( #1 \right)}
\newcommand{\uglate}[1]{\left[ #1 \right]}
\newcommand{\Nabla}{\bm \nabla}

\newcommand{\C}{\mathcal C}

\newcommand{\uvec}[1]{\hat{\bm{#1}}}

\newcommand{\ointerval}[2]{\left\langle #1,#2 \right\rangle}
\newcommand{\cointerval}[2]{\left[ #1,#2 \right\rangle}

\newcommand{\cinterval}[2]{\left[#1,#2\right]}

\newcommand{\R}{{\bm{R}}}
\newcommand{\Rt}{\bm{R}(t)}
\newcommand{\nR}{\ket{n(\R)}}
\newcommand{\nRt}{\ket{n(\Rt)}}
\newcommand{\mR}{\ket{m(\R)}}

\newcommand{\naR}{\ket{n_a(\R)}}
\newcommand{\naRof}[1]{\ket{n_a(\R(#1))}}
\newcommand{\psint}{\ket{\psi_n(t)}}
\newcommand{\gnt}{\gamma_n(t)}
\newcommand{\An}{\A ^n}
\newcommand{\AnR}{\A ^n (\R)}
\newcommand{\curv}{\bm \Omega ^n}
\newcommand{\curvtensor}{\curv_{\mu \nu}}
\newcommand{\gradR}{\Nabla_{\!\!\R}\,}
\newcommand{\UnC}{U^n_\C}

\newcommand{\ketn}[1]{\ket{\psi_n(#1)}}

\newcommand{\psinq}{\psi_{n\q}}
\newcommand{\unq}{u_{n\q}}
\newcommand{\unqprim}{u_{n'\q'}}

\newcommand{\flqnt}{\Phi_0}
\newcommand{\sh}{\sigma_\mr H}
\newcommand{\Gauss}{\exp(-\frac{|z|^2}{4l_B^2})}
\newcommand{\Gaussj}{\exp(-\frac{|z_j|^2}{4l_B^2})}
\newcommand{\GaussMB}{\exp(-\sum_i^{N_e}\frac{|z_i|^2}{4l_B^2})}
\newcommand{\ceta}{{\conj\eta}}
\newcommand{\Aeta}{\mathcal{A}_{\eta}}
\newcommand{\Aceta}{\mathcal{A}_{\,\ceta}}
\newcommand{\pdeta}{\pdd_{\eta}}
\newcommand{\pdceta}{\pdd_{\,\ceta}}

\newcommand{\veta}{\bm\eta}

\newcommand{\psiup}{\psi_\frac{1}{2}}
\newcommand{\psidn}{\psi_{-\frac{1}{2}}}
\newcommand{\psink}{\psi_{n\vk}}
\newcommand{\betank}{\beta_{nk}}
\newcommand{\Nnk}{N_{nk}}
\newcommand{\gupN}{g_{\frac{1}{2},0}}
\newcommand{\gupV}{g_{\frac{1}{2},-1}}
\newcommand{\gdnV}{g_{-\frac{1}{2},+1}}
\newcommand{\gdnN}{g_{-\frac{1}{2},0}}

\DeclareMathOperator{\sgn}{sign}


%% file: title.tex


\begin{titlepage}
    \begin{center}
        \fontfamily{phv}\selectfont
   		\fontsize{16}{19}\selectfont
   		
        \includegraphics[bb=0 0 1200 1200, width=0.15\textwidth]{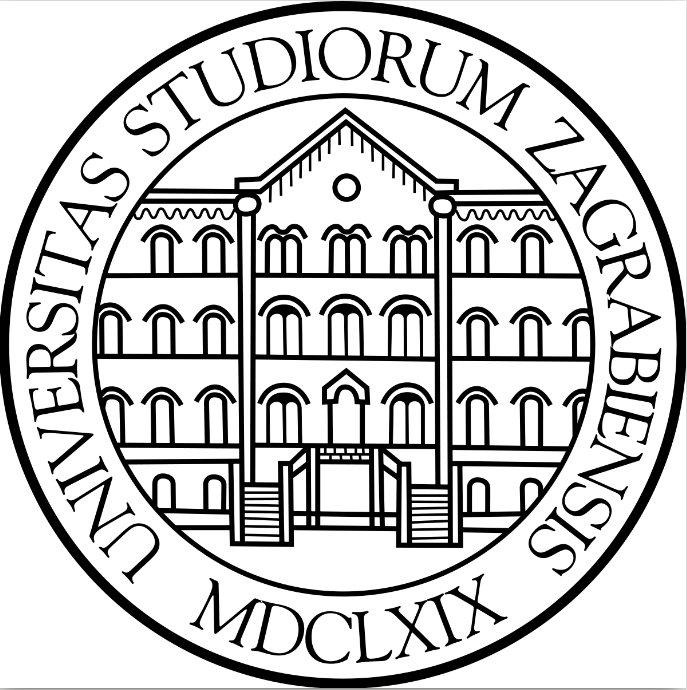}\\
        ~\\
        
        University of Zagreb\\ 
        Faculty of Science\\
        Department of Physics\\
        ~\\
        ~\\

        Frane Lunić \\
        ~\\
        ~\\
        \fontsize{22}{33}\selectfont
        \textbf{Anyons, Zitterbewegung and dynamical phase transitions in topologically nontrivial systems}\\
        \fontsize{16}{19}\selectfont
        ~\\
        ~\\
        DOCTORAL THESIS\\

        \vfill
        
         \fontsize{14}{21}\selectfont
         Zagreb, 2022
    
    \end{center}
\end{titlepage}


\begin{titlepage}
    \begin{center}
        \fontfamily{phv}\selectfont
   		\fontsize{16}{19}\selectfont
   		
        \includegraphics[bb=0 0 1200 1200, width=0.15\textwidth]{Y_additional_materials/unilogo.png}\\
        ~\\
        University of Zagreb\\ 
        Faculty of Science\\
        Department of Physics\\
        ~\\
        ~\\
        Frane Lunić \\
        ~\\
        ~\\
        \fontsize{22}{33}\selectfont
        \textbf{Anyons, Zitterbewegung and dynamical phase transitions in topologically nontrivial systems}\\
        
        \fontsize{16}{19}\selectfont
        ~\\
        ~\\       
        DOCTORAL THESIS\\
		~\\
		~\\
		Supervisor:\\
		prof.\ dr.\ sc.\ Hrvoje Buljan\\
		doc.\ dr.\ sc.\ Dario Jukić
		        
        \vfill
        
         \fontsize{14}{21}\selectfont
         Zagreb, 2022
    
    \end{center}
\end{titlepage}


\begin{titlepage}
    \begin{center}
		\fontfamily{phv}\selectfont
		\fontsize{16}{19}\selectfont
		 		
		\includegraphics[bb=0 0 1200 1200, width=0.15\textwidth]{Y_additional_materials/unilogo.png}\\
		~\\
		
		Sveučilište u Zagrebu\\
		Prirodoslovno-matematički fakultet\\
		Fizički odsjek\\
		
		~\\
		~\\
		
		Frane Lunić \\
		~\\
		~\\
		
		\fontsize{22}{33}\selectfont
		\textbf{Anyoni, Zitterbewegung i dinamički fazni prijelazi u topološki netrivijalnim sustavima}\\
		
		\fontsize{16}{19}\selectfont
		~\\
		~\\
		DOKTORSKI RAD\\
		~\\
		~\\
		Mentor:\\
		prof.\ dr.\ sc.\ Hrvoje Buljan\\
		doc.\ dr.\ sc.\ Dario Jukić

		\vfill
		
		 \fontsize{14}{21}\selectfont
		 Zagreb, 2022

    \end{center}
\end{titlepage}

%% file: 0_intro/intro.tex
	In contemporary condensed matter physics and related areas, the role of topology has been receiving an increasing amount of attention. There are various ways in which topological features of a system can impact its properties, some of which we will encounter in this thesis. Topology is usually described as a branch of mathematics related to geometry which studies those properties of geometric objects which are unchanged under continuous deformations. For example, even though a sphere and (the surface of) a cube are clearly different geometric shapes, it is the case that every point on a sphere can be mapped to a unique point on a cube according to a continuous function, and back according to the continuous inverse. The function satisfying the stated properties is called a homeomorphism. On the other hand, a torus, i.e. the surface of a donut with a hole in the middle, can never be homeomorphically mapped to a sphere. Here, the number of holes is a topological property which cannot be changed by a continuous deformation. Another interesting example in topology is a Möbius strip, created by twisting and then buckling a belt. Due to the half-twist, the topology of a Möbius strip is different from a regularly buckled belt. The consequence is that its surface is nonorientable, i.e. it does not have the inner and outer surfaces, as both sides of the strip can be reached by moving along its length. No continuous transformation (i.e. without unbuckling) may change this fact.
	
	When a topological property of a system is behind a physical phenomenon, this can make the phenomenon more robust, as it becomes insensitive to continuous transformations which preserve the topological property. This may protect the phenomenon against disorder, or preserve it under different parameters of the system. In a sense, topology is ubiquitous in physical science. For instance, topological defects can be present in crystalline, and other media, and they can be important in various fields, from electronics and chemistry all the way to cosmology. It is therefore not surprising (with hindsight) that quantum states can be classified according to their topological properties, which differentiate between topological phases. When nontrivial topological phases are possible, we are dealing with topological quantum matter, which is among the topics explored in this thesis. A ubiquitous topologically protected property found in topological matter are the boundary modes between different topological phases, which can be chiral and protected against backscattering. It is not difficult to imagine that such modes could find technological applications. Another notable example which we will touch on is the topological protection some systems possess against information "leakage" into the environment, which has been the bane of quantum computing.
	
	The main purpose of this thesis is to compile the results of the published original research the author has contributed to. Each of the papers explores a different topic, all of which, in their own way, relate to topology. Two of the papers are in the field of photonics, first of which also presents experimental results. In this paper, the propagation of light through photonic symmetry-broken honeycomb lattices under specific circumstances is explored. The effect that was found is a consequence of a momentum-space topological singularity (a kind of vortex). In the second photonics paper, a nonlinear medium is used which allows for soliton beams, and mediates their interaction. The configuration of beams is arranged so as to lead to an emergent 1D system which can assume two topological phases at different points during the propagation of the beams. The final included paper presents solutions to a model for a noninteracting system which features synthetic anyons. Anyons obey fractional statistics, different to the (integral) bosonic and fermionic statistics. Their existence is allowed under the constraints of a two-dimensional space, and they are thought to occur as excitations in topologically nontrivial materials. Anyons are crucial to the subject of topological quantum computing.
	
	The thesis is arranged as follows. In \cref{ch:intro}, we provide a theoretical introduction to some of the topics which are relevant to the later presented research. Where it seems warranted, we expand our discussion beyond what is strictly necessary for a basic understanding of our results, in order to provide a fuller overview. In \cref{sec:geomph}, we introduce the geometric (Berry) phase and the related concepts, which are invaluable for understanding the subjects of this thesis, as they  can correspond to the mathematical quantities which determine the topology (such as curvature), or reveal the fractional statistics. In \cref{sec:TQM}, we introduce the topological quantum matter, and discuss some specific pertaining topics, including the quantum Hall effect, anyons, and the symmetry-protected topological states. In \cref{ch:topofoto}, we present the results of the aforementioned photonics papers. In \cref{sec:zb}, we give the results on single-valley light propagation through the inversion-symmetry-broken honeycomb photonic lattices. This includes the experimental results, as well as theoretical analysis from the "low-energy" point of view. The results feature a rotating spiral diffraction pattern, and the Zitterbewegung effect. In \cref{sec:ssh}, we present the numerical results on propagation of soliton beams arranged in a bipartite (SSH) lattice pattern through a nonlinear medium which facilitates soliton interaction. The results suggest the occurrence of dynamically emerging topological phase transitions. In \cref{ch:prb}, we present the solutions of a model for synthetic anyons in a noninteracting system. A derivation of the statistics is outlined, and the interpretation of the synthetic anyons is discussed. Finally, we summarize and conclude in \cref{ch:concl}.

%% file: 1_introduction/introduction.tex

In this chapter, we introduce the basic concepts, and discuss some of the relevant theoretical background. In \cref{sec:geomph}, we introduce the geometric phase. In \cref{sec:geomph:bph}, we discuss the simple case of the Abelian Berry phase, relevant to the nondegenerate systems. In \ref{sec:geomph:NAGPh}, we generalize to the degenerate case, which includes the possibility of non-Abelian geometric phases (more precisely, Berry holonomies). Section \ref{sec:TQM} deals with topological quantum matter. The topics covered individually in this section are the integer quantum Hall effect in \cref{sec:TQM:IQHE}, the fractional quantum Hall effect in \cref{sec:TQM:FQHE}, anyons in \cref{sec:TQM:anyons}, symmetry-protected topological phases in \cref{sec:TQM:SPT}, and finally, some notable experimental platforms for realization of topological phases of matter in \cref{sec:TQM:platforms}.

\section{Geometric phase}\label{sec:geomph}

Geometric phases (Berry phases) are the phases determined by the geometry of the paths traced in the parameter space as the system Hamiltonian is varied. This phenomenon was discovered by Pancharatnam in polarization optics in 1956 \cite{pancharatnam}, and the Aharonov-Bohm effect \cite{AB}, a special case of quantum geometric phase, was discovered in 1959. However, the true importance and ubiquity of geometric phases in quantum mechanics was only appreciated following the seminal work of Sir Michael Berry in 1984 \cite{berry1984}. Roughly along the lines of Berry's original paper, we will review the geometric phase in the general case for nondegenerate energy levels under assumptions of adiabatic and cyclical evolution. We will refer to this phase as the Berry phase. We will then touch on the more general non-Abelian geometric phases that arise when no-degeneracy requirement is relaxed. We will discuss the role these phases play in the physics of electronic bands. Adiabatic and cyclical evolution is assumed throughout this presentation, even though the existence of geometric phases is not predicated on these assumptions \cite{aharonov&anandan,samuel1988}.

	\subsection{Berry phase}\label{sec:geomph:bph}
	Let the Hamiltonian of a system depend on a number of parameters $R_i$, written compactly as the vector $\R=(R_1,R_2,\dots)$ in the parameter space; $H=H(\R)$. As the parameters are slowly tweaked between times $0$ and $T$, the system adiabatically traces a path through the parameter space. The cyclicity assumption is expressed as $\R(T)=\R(0)\equiv\bm R_0$.
	
	At any point during the evolution, we may choose an orthonormal "natural" basis $\mR$ such that
	\begin{equation}\label{eq:berry_eigeneq}
		H(\R)\mR = E_m(\R)\mR.
	\end{equation}	 
If the system is initially prepared in a specific nondegenerate basis state $\ket{\psi_n(0)}=\ket{n(\bm R_0)}$, the adiabatic theorem guarantees that the state it assumes at time $t$, i.e. $\ket{\psi_n(t)}$, only differs from $\nRt$ by a phase factor, provided no degeneracy is encountered along the way. To find the phase we search for a solution to the time-dependent Schr\"odinger equation
	\begin{equation}\label{eq:berry_tdse}
		i\hbar\dv{t}\psint = H(\Rt)\psint
	\end{equation}
	in the form
	\begin{equation}\label{eq:berry_psint}
		\psint = e^{i\gnt} \exp[-\frac{i}{\hbar} \int\limits_0^t \dd t' E_n(\R(t')) ] \nRt ,
	\end{equation}
	where the second exponential is the dynamical phase factor, and the first is associated with the nonvanishing time derivative $\dv{t}\nRt=-i\DOT{\gamma}_n(t)\nRt$. Integrating, we find that
	\begin{equation}
		\gnt = \int\limits_0^t \An \cdot\DOT\R \ \dd t',
	\end{equation}
	having defined the Berry connection:
	\begin{equation}\label{eq:berry_connection}
		\An(\R) \equiv i \braket{n(\R)}{\gradR n(\R)}.
	\end{equation}
	We use the notation $\ket{\gradR n(\R)} \equiv \gradR \nR$ to stress that $\gradR$ is the gradient operator in the parameter space, instead of the Hilbert space. Upon traversing the path corresponding to a closed curve $\C$ and returning to the initial position,  we are left with the Berry phase
	\begin{equation}\label{eq:berry_phase_contour}
		\gamma_n \equiv \gamma_n(T) = \oint_\C \dd\R \cdot \An(\R) .
	\end{equation}
	
	Since the Berry phase corresponds to a closed path, it is gauge invariant, in contrast to the phase $\gnt$ which corresponds to an open path and can be removed by a gauge transformation. The connection $\An$ is gauge-dependent, and transforms according to $\AnR \rightarrow {\AnR + \gradR \lambda (\R)}$ where $\lambda(\R)$ is a scalar function corresponding to some gauge transformation. This motivates the analogy between the Berry connection and the magnetic vector potential. It makes sense to define the gauge invariant quantity
	\begin{equation}\label{eq:berry_curvtensor}
		\curvtensor \equiv \partial_{R_\mu} \scalA^n_\nu - \partial_{R_\nu} \scalA^n_\mu,
	\end{equation}
	called the Berry curvature tensor. Then by the Stokes theorem \cite{stanescu}
	\begin{equation}\label{eq:berry_phase_flux_tensor}
		\gamma_n = \int_{\mathcal S(\C)} \dd R_\mu \wedge \dd R_\nu \, \frac{1}{2} \curvtensor (\R),
	\end{equation}
	where $\mathcal S(\C)$ is the surface bounded by the path $\C$, and $\wedge$ represents the wedge product, which is used to obtain the area elements (see pages 47 and 58 in \cite{stanescu}). It is often the case that the relevant parameter space is 3-dimensional (e.g. when the parameters varied are the components of the real magnetic field). In this case, the gauge invariant Berry curvature vector, analogous to the magnetic field, can be defined as 
	\begin{equation}\label{eq:berry_curv}
		\curv \equiv \gradR \cross \An, 
	\end{equation}
	and related to the curvature tensor by $\curvtensor  = \epsilon_{\mu\nu\lambda} \Omega^n_\lambda$. Now, the Berry phase is 
	\begin{equation}\label{eq:berry_phase_flux}
		\gamma_n = \int_{\mathcal S(\C)} \dd \bm S \cdot \curv,
	\end{equation}
	making obvious the analogy between the Berry phase and the magnetic flux through $\mathcal S(\C)$.
	
	An interesting behaviour can occur when $\C$ lies close to a point of degeneracy $\R^*$. Consider a 2-level system with conical dependence of energy on $\tilde\R=\R-\R^*$ around a degeneracy point in a 3D parameter space. Close to $\R^*$, the Hamiltonian is of the form $H(\R) \propto \tilde\R\cdot\bm\sigma$. The Berry curvature can be shown to be a monopole field $\bm\Omega^{\pm} \propto \pm\frac{\tilde\R}{2R^3}$ \cite{berry1984}. This phenomenon occurs in, for example, Weyl semimetals, where the parameter space corresponds to the Brillouin zone which features pairs of oppositely charged degeneracies (Weyl points).
	
	As Berry has shown, it is possible to reinterpret the Aharonov-Bohm phase in terms of the geometric phase. The Aharonov-Bohm effect \cite{AB} is an entirely quantum phenomenon wherein charged particles feel the effect of electromagnetic potentials, even in regions with no finite fields. This effect manifests as a phase shift, and can be demonstrated by splitting a coherent beam of electrons, having them pass by a magnetic flux confined inside a solenoid on opposite sides, and then rejoining them to produce interference. The connection to the Berry phase can be made evident by considering the following setup. Let the charged particle be confined in a box centred on $\R$. Now the box is slowly transported on the contour $\C$ around an infinite solenoid carrying the flux $\Phi_\B$, but never penetrating it. The wave function of the particle in the vector potential $\VP(\vr)$ is given by
	\begin{equation}\label{eq:berry_boxwf}
		n(\vr; \R) = e^{\frac{iq}{\hbar} \int_\R^\vr \VP(\vr') \cdot \dd \vr'} \psi_n^0(\vr - \R), 
	\end{equation}
	where $\psi_n^0(\bm \chi)$ is the state in absence of the magnetic flux. The Aharonov-Bohm result tells us the phase shift upon completing the circuit is
	\begin{equation}\label{eq:berry_AB}
		\gab = \frac{q}{\hbar} \oint_\C \VP \cdot \dd \R = \frac{q\Phi_\B}{\hbar}.
	\end{equation}
	 However, this process involves a loop in the space of parameter $\R$. Using \eqref{eq:berry_boxwf} and \eqref{eq:berry_connection}, we find the Berry connection
	 \begin{equation}
	 	\AnR = \frac{q}{\hbar} \VP(\R),
	 \end{equation}
	 and using to \eqref{eq:berry_phase_contour}, we see that $\gamma_n = \gab$. Evidently, the Aharonov-Bohm phase is a special case of the geometric Berry phase.
	
\subsection{Non-Abelian geometric phases}\label{sec:geomph:NAGPh}
	Cyclic transport as discussed above can lead to non-Abelian geometric phases when the system is prepared in a degenerate state \cite{wz1984,stanescu}. It is assumed this degeneracy holds for all $\R$. Once again, we choose an orthonormal basis $\naR$ such that $H(\R) \naR = E_n(\R) \naR$, for $a=1,2,\dots,g_n$, where $g_n$ is the degeneracy of the $n$-th energy level. Having chosen an initial state within the $n$-th level subspace 
	\begin{equation}\label{eq:berry_nonabel_initwf}
		\ketn 0 = \sum_a C_a^n (0) \naRof 0,
	\end{equation}	 
we let $\R$ adiabatically trace a closed path $\C$ between times $0$ and $T$. The result is given by
	\begin{equation}\label{eq:berry_nonabel_finalwf}
		\ketn T = U_\mathrm{dyn} \UnC \ketn 0.
	\end{equation}
	Here, $U_\mathrm{dyn}$ is the familiar dynamical phase, while $\UnC\in U(g_N)$ is a unitary matrix, the so-called Berry holonomy which may always be expressed in terms of a Hermitian matrix $\Gamma_n$ defined as $e^{i\Gamma_n} \equiv \UnC$. An expression analogous to \eqref{eq:berry_phase_contour} may be given for the Berry holonomy:
	\begin{equation}\label{eq:berry_nonabel_holonomy}
		\UnC = \mathcal P  \exp( \oint_\C \dd \R \cdot \AnR ),
	\end{equation}
	where $\mathcal P$ is the path-ordering operator, and the  Berry connection is now a vector whose components are Hermitian matrices with elements
	\begin{equation}\label{eq:berry_nonabel_connect}
		{\scalA^n_\mu}_{(ab)}(\R) = i \braket*{n_a(\R)}{\pdd_{R_\mu} n_b(\R)}.
	\end{equation}
	The possibility of non-Abelian nature of these matrices, i.e. $[\scalA^n_\mu(\R), \scalA^n_\mu(\R')] \neq 0$ for $\R\neq\R'$, makes the path ordering necessary. 
	 The effect of this unitary transformation is to rotate the original vector within the degenerate subspace. Hence, final superposition is not the same as the initial ($\abs{C^n_a(T)}\neq \abs{C^n_a(0)}$). This is strikingly different from the nondegenerate case where the final state was parallel to the initial state, differing only by a phase.
	 
	Finally, let us review the behaviour under gauge transformation. The connections will transform as $\scalA_\mu^n \rightarrow U^{-1} \scalA_\mu^n U + iU^{-1}\pdd_{R_\mu} U$. In contrast to the gauge invariant Abelian Berry holonomy, the non-Abelian holonomy transforms as $\UnC \rightarrow U^{-1}(\R_0) \UnC U(\R_0)$ \cite{stanescu}. Hence, a gauge invariant quantity must be found in order to measure the effect of holonomy. Among such quantities are the eigenvalues of $\UnC$, and the so-called Wilson loop $W=\Tr[\UnC]$.

\subsection{Role in Bloch bands}\label{sec:geomph:Bloch}
	The Bloch's theorem guarantees that each eigenstate of a spatially periodic single-particle Hamiltonian is the product of a plane wave and a cell-periodic function \cite{A&M}
	\begin{equation}\label{eq:berry_blochthm}
		\psinq (\vr) = e^{i\q \cdot \vr} \unq (\vr),
	\end{equation}
	with $\unq(\vr +\sum_i m_i\bm a_i) = \unq(\vr)$, where $\bm a_i$ are the lattice vectors, $m_i\in \mathbb Z$, and $i$ is the coordinate index. The functions $\unq(\vr)$ are the eigenstates of the Bloch Hamiltonian $H(\q)=e^{-i\q\cdot\vr} H e^{i\q\cdot\vr}$. Since this Hamiltonian is parametric, a cyclic change in $\q$ may give rise to a geometric phase, or more generally, a holonomy. In the Abelian case, the Berry phase is obtained according to \eqref{eq:berry_phase_contour}, with the Berry connection defined as:  
	\begin{equation}\label{eq:berry_band_connecton}
		\An(\q) = i \braket{\unq}{\Nabla_{\q}\unq}.
	\end{equation}
	In a 3-dimensional lattice, the Berry curvature is naturally defined as $\curv=\Nabla_{\q} \times \An(\q)$. This amounts to a magnetic field acting in momentum space. Additionally, an explicit time dependence of the Hamiltonian may be thought of as introducing a geometric scalar potential in addition to the time-dependent vector potential $\An(\q, t)$ \cite{stanescu}.
	
	By working with the Bloch Hamiltonian, the lattice Brillouin zone becomes the parameter space, instead of it being imposed externally. However, it is the effect of an external perturbation that drives the cyclic evolution in the Brillouin zone. For example, upon imposing a homogeneous time-dependent vector potential $A(t)$, a 1-dimensional periodic system still admits a solution essentially of the type \eqref{eq:berry_blochthm}, except $u_{n\sq}(x) \rightarrow u_{n\sq(t)}(x)$, where $\sq(t)=\sq-\frac{e}{\hbar}A(t)$\footnote{Note that in this gauge the plane wave factor remains time-independent.}. This vector potential can be realized by simply applying a homogeneous electric field. The only way to obtain a Berry phase in a 1D lattice is to have the perturbation drive $\sq(t)$ across the Brillouin zone. A Berry phase, called the Zak phase, then appears due to its periodicity \cite{zak}
	\begin{equation}\label{eq:Zak}
		\gamma_\mathrm{Zak} = \int\limits_{\sq_0}^{\sq_0+\frac{2\pi}{a}} \dd \sq \, \mathcal{A}^n(\sq).
	\end{equation}	 
	This phase is due to the circle-like topology of the periodic Brillouin zone, since no curvature can be enclosed by a contour in 1D space. It assumes any value in general, but in inversion-symmetric systems, it is constrained to either $0$ or $\pi$ \cite{zak}.

\section{Topological quantum matter}\label{sec:TQM}
	By the end of the 20th century, it became clear that topology has a role to play in classification of the phases of matter. The classical approach, based on the Landau symmetry-breaking theory \cite{landau}, ascribes transitions between phases to changes in symmetry of the system. The solution of a Hamiltonian which obeys certain symmetries does not necessarily obey all of the same symmetries. Instead, it will inherit a subset of its symmetries. The solutions obeying different symmetries are distinct phases, and cannot be connected without encountering a phase transition. A reduction in symmetry is associated with the appearance of some kind of ordering of the state, as indicated by a nonzero value of some local order parameter, such as magnetization in the case of ferromagnetic phase transition. However, it is now known that classification based on symmetry is not complete, since some zero-temperature phases of matter that share the same symmetries are nevertheless separated by a phase transition \cite{stanescu}. 
These phases are characterized by discrete topological invariants (sometimes requiring certain protective symmetries) that cannot be changed without going through the so-called topological phase transitions. The topological states that we will consider are gapped, and the topological phase transitions involve closing and reopening the gap. At the point when two bands are merged, the topological invariants associated with each band are not well defined. This facilitates the discontinuous change in the value of the invariants during the phase transition. A variety of interesting phenomena, like topological degeneracy of the ground state (i.e. a degeneracy robust against local perturbations that do not affect the topology of the state), exotic excitations, and robust, perfectly conducting edge states may arise due to the topological nature of these states \cite{stanescu}. Currently, the most attractive potential application for the former two is in quantum computing \cite{nayak}, while the latter may have applications in fields such as electronics and photonics.

	Historically, the need to go beyond local order has lead to the notion of topological order \cite{wen1990}. 
All (intrinsic) topological orders are a consequence of long-range entanglement (LRE). LRE states are the many-body states that cannot be transformed into product states by any local unitary transformation (arising from a gap-preserving adiabatic deformation of the Hamiltonian). Different patterns of LRE which cannot be connected by a local unitary transformation correspond to different topological orders \cite{chen2010}. The topologically ordered states may have features such as topologically protected gapless boundary modes, topologically degenerate ground states and exotic excitations obeying fractional Abelian or non-Abelian statistics \cite{stanescu} (see \cref{sec:TQM:anyons}). These nontrivial excitations are a feature of strongly-interacting states, such as the fractional quantum Hall effect (see \cref{sec:TQM:FQHE}) and quantum spin liquids \cite{kalmeyer1987,savary2016} and are sometimes considered as a defining criterion for topological order. However, the weakly-interacting integer quantum Hall effect (\cref{sec:TQM:IQHE}) states lack the nontrivial excitations, but do possess LRE according to the above definition \cite{stanescu}.

	Not all topological phases possess LRE. These short-range entangled (SRE) states lack the full topological protection of the topologically ordered phases\footnote{By topological order, we mean the intrinsic topological order. The SRE states are sometimes said to possess a different kind of topological order.}, implying they can be connected without a phase transition by adiabatic deformations, provided these deformations break certain symmetries of the Hamiltonian. As long as the required symmetries are preserved, these symmetry-protected topological (SPT; see \cref{sec:TQM:SPT}) phases  remain separate. SPT phases lack the topological degeneracy and nontrivial excitations, but  they possess the topologically protected boundary states 
\cite{stanescu}. 

	These boundary states are a consequence of the bulk-boundary correspondence which relates the topological properties of the system's bulk to the topological properties of its boundary. Due to their topological nature, they are protected against perturbations that preserve the gap and, in the case of SPT phases, the required symmetries. For example, in 2D, if the boundary is gapless, strips of states connecting the valence to the conduction band will be seen in the dispersion relation plot, while in 1D, one or more states associated with each edge will appear at zero energy. These states are the interface phenomena localized to the region where two different topological phases meet. This motivates an intuitive understanding of the gapless boundary states. Essentially, the gap is closed at the interface, thus enabling the discontinuous change of the topological invariants.
 
In this section, we take a closer look at the integer and the fractional QHE, as well as fractional statistics, i.e anyons. We also briefly discuss the SPT phases, and introduce the Su-Schrieffer-Heeger model. We conclude by mentioning some notable experimental platforms that can be used for realization of topological phases of matter.

\subsection{Integer quantum Hall effect}\label{sec:TQM:IQHE}
\newcommand{\psia}{\psi_\alpha}
\newcommand{\psib}{\psi_\beta}
\newcommand{\Ea}{E_\alpha}
\newcommand{\Eb}{E_\beta}
\newcommand{\pdphi}[1]{\pdd_{\Phi_{#1}}}
\newcommand{\pdth}[1]{\pdd_{\theta_{#1}}}

\newcommand{\torus}{\bm{\mathrm T}^2}
\newcommand{\fluxtorus}{\bm{\mathrm T}^2_\Phi}

\begin{figure}[htb]\centering
		\includegraphics[width=.85\textwidth]{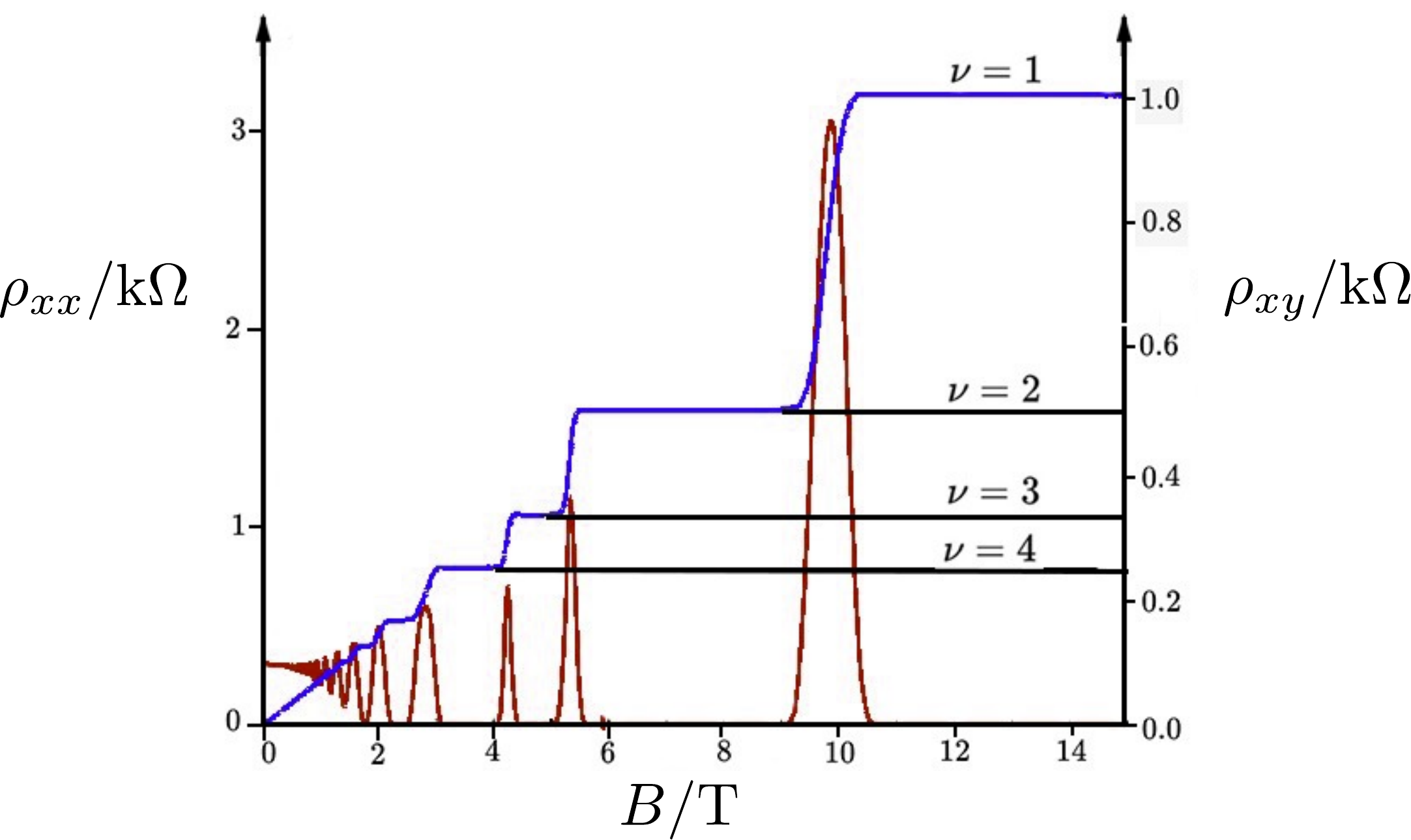}
		\caption[Integer quantum Hall effect]{Schematic representation of Hall (blue) and longitudinal (red) resistivity dependence on magnetic field for a 2DEG. Created by Alba Cazorla (modified). Licensed under \href{https://creativecommons.org/licenses/by-sa/4.0/deed.en}{CC BY-SA 4.0}. \href{https://upload.wikimedia.org/wikipedia/commons/3/38/Rhoxy.jpg}{Link to original.}}\label{IQHE:fig:iqhe}
	\end{figure}

	The quantum Hall effect (QHE), first measured in 1980 by Klitzing, Dorda and Pepper \cite{kdp}, refers to the quantization of Hall conductivity in two-dimensional electronic systems (or analogous atomic and other systems) with broken time-reversal symmetry (TRS). When a voltage is applied across some direction, a Hall current density $j_\mr H = \sigma_\mr H E$ will appear in the orthogonal direction. The Hall (transverse) conductivity is quantized in units of $q^2/h$ \cite{kdp}
	\begin{equation}\label{IQHE:eq:hall_cond}
		\sigma_\mr H = \frac{q^2}{2\pi\hbar} \nu,
	\end{equation}
	where $q$ is the carrier charge and $\nu$ is an integer in case of the integer QHE (IQHE). As we shall see, $\nu$ is a topological invariant characteristic of the IQHE state in question. Typically, the TRS is broken by a magnetic field, and $\sh$ is a function of the field characterized by wide plateaus corresponding to the quantized values, connected by transition regions \cite{tong} (see \cref{IQHE:fig:iqhe}). With respect to the longitudinal conductivity, the system is normally an insulator, but becomes conductive in the transition regions.
	
	\subsubsection{Continuum noninteracting 2D electron gas model}
\begin{figure}[htb]\centering
		\includegraphics[width=.95\textwidth]{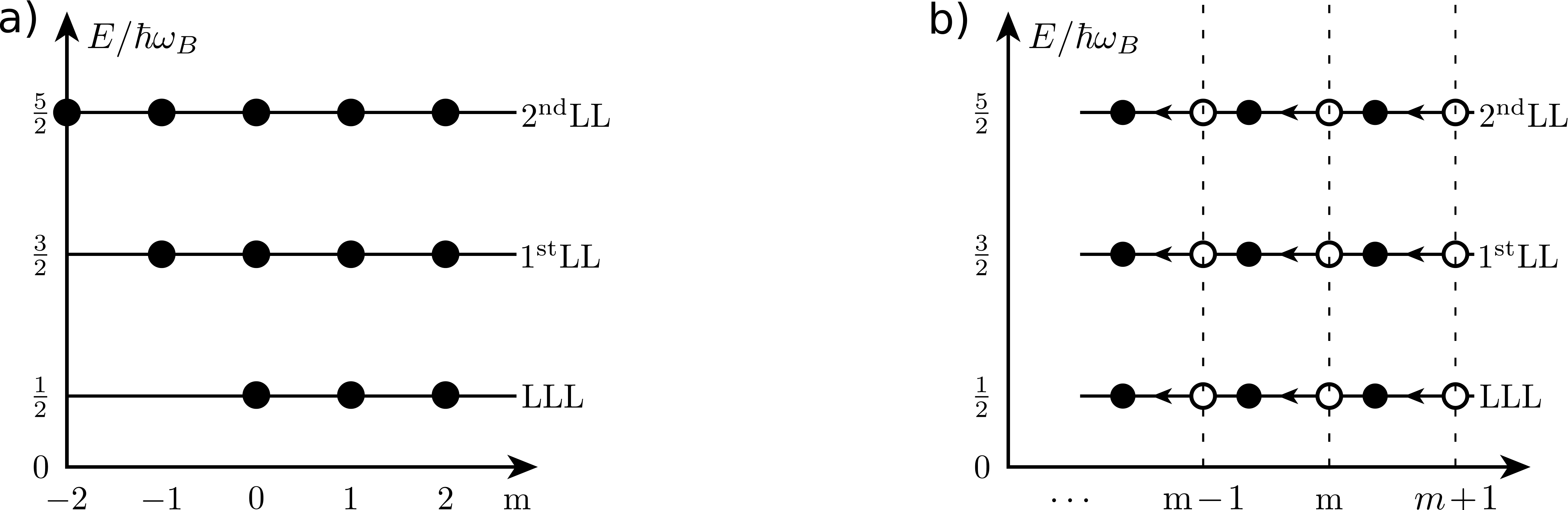}
		\caption[Schematic representation of Landau levels and spectral flow.]{Schematic representation of Landau levels and spectral flow. Solid circles represent the shared basis of the symmetric gauge single-particle Hamiltonian and the canonical angular momentum operator. (a) Unperturbed spectrum (infinite system). (b) Spectrum of a system on a Corbino ring after the flux through the hole has been adiabatically increased from $0$ to $\alpha=\Phi/\flqnt <1$. Empty circles represent the initital eigenstates $m$ for $\alpha=0$ and the arrows indicate the direction of the spectral flow.}
		\label{IQHE:fig:LL}
	\end{figure}

	An archetypal system featuring the IQHE is a continuous noninteracting 2D electron gas (2DEG) under a uniform magnetic field in the perpendicular direction ($z$). A moderate amount of disorder is required to obtain the extended conductivity plateaus. The Hamiltonian of the system is given by:
	\begin{equation}\label{IQHE:eq:cont_H}
		H	= \sum\limits_{j=1}^{N_e} \uglate{ \frac{1}{2m} \left( 
						\bm{p_j} - q\bm{A}(\bm{r_j}) \right)^2 + V(\vr_j) },
	\end{equation}	where $N_e$ is the number of electrons, $V(\vr)$ is a scalar potential, and $\VP(\vr)$ is the vector potential due to the uniform field $\B=(0,0,B)$ in the symmetric gauge:
	\begin{equation}
		\bm{A}(\bm r) = \frac{1}{2} \bm{B} \times \bm r.
	\end{equation}
	We will first consider the simple case $V(\vr)=0$, and will later discuss the effect of a random potential. The spectrum of the single-particle Hamiltonian (inside square brackets in \eqref{IQHE:eq:cont_H}) is composed of flat, massively degenerate Landau levels at energies $E_n = \hbar\omega_B (n+1/2)$, where $n=0,1,\dots$, and ${\omega_B=|q|B/m}$ is the cyclotron frequency (see \cref{IQHE:fig:LL}(a)). If we assume $B>0$ and $q<0$, the lowest Landau level (LLL) single-particle states are given by $\psi_\mr{LLL}(z_j,\conj{z_j}) = f(\conj{z_j}) \Gaussj$, where $\conj {z_j}$ is the complex conjugate of $z_j=x_j+iy_j$, $l_B=\sqrt{\hbar/B|q|}$ is the magnetic length, and $f(\conj z)$ is any antiholomorphic function \cite{tong,prb}. Given the azimuthal symmetry, it makes sense to work in the basis labeled by the angular momentum quantum number:
	\begin{equation}\label{IQHE:eq:LLL_wf}
		\psi_{\mr{LLL},m} (z,\conj z) \equiv \psi_m = \conj z^\m \Gauss, \qquad m=0,1,2,\dots
	\end{equation}
	The normalization factor is implied, but will usually be omitted.
	The degeneracy of each Landau level per unit area equals the number of flux quanta per unit area ($B/\Phi_0$) \cite{tong}. The $\psi_m$ states are peaked around the radius $r_m=\sqrt{2m} l_B$ which encloses a region containing $m$ flux quanta.
	The many-body ground state is obtained by constructing a Slater determinant from all filled single-particle states $\psi_{n,m}$. 
	
	We restrict now to the case of $n$ completely filled Landau levels, i.e. the Fermi energy lies between the $E_n$ and $E_{n+1}$. The longitudinal conductivity vanishes since the system is now a band insulator.  To explore the Hall conductivity, we introduce a scalar potential restricting the electron gas to a ring between the radii $r_1<r_2$ (the Corbino ring). Then an additional magnetic flux $\Phi$ can be threaded through an infinite solenoid placed at the origin without changing the magnetic field inside the ring. This setup was explored in \cite{halperin}. We express the flux in units of the magnetic flux $\flqnt=\frac{2\pi\hbar}{|q|}$ as $\alpha=\Phi/\flqnt$. Suppressing the Gaussian factor, we write the LLL states with $r_1<r_m<r_2$ \footnote{More precisely, we require that \(r_m-r_1\gg l_b\), and \(r_2-r_m\gg l_B\)} for $\alpha=0$ in the symmetric gauge as 
	\begin{equation}\label{IQHE:eq:LLL_corbino}
		\phi^0_m=\conj{z}^m=e^{-im\varphi}r^m.
	\end{equation}
	The vector potential due to the flux $\alpha$ is $\VP_\alpha=\frac{\Phi}{2\pi r}\hat{\varphi}$, and is locally pure gauge. Therefore, turning it on would only shift the phase of any localized state by $e^{i\alpha\varphi}$, which can be removed via a gauge transformation. However, it is clear from \eqref{IQHE:eq:LLL_corbino} that the LLL states are extended around the ring. The topology of extended states does not in general allow for this kind of gauge transformation since it would disrupt phase coherence around the ring, except for ${\alpha\in \mathbb{Z}}$. Since the flux is then not pure gauge, it may affect the spectrum, but we expect the spectrum to be intact for integer values of $\alpha$. Let us consider an adiabatic evolution of $\alpha$ from $0$ to $1$ over time $T\gg \omega_B^{-1}$. Due to the azimuthal symmetry, the state $\phi^\alpha_m$ remains an $m$-eigenstate of the canonical angular momentum operator $L_z^\alpha=\abs{\vr \times (\p^0+q\VP_\alpha)}=L_z^0+\sgn(q)\hbar\alpha$ during the evolution. In the end we have
	\begin{equation}
		L_z^1 \phi_m^1 = \oble{L_z^0 - \hbar} \phi_m^1= -\hbar m \phi_m^1
	\end{equation}
	for electrons, which implies that $L_z^0 \phi_{m}^1 = - \hbar (m-1) \phi_{m}^1$. Therefore, due to the discrete nature of gauge invariance of the extended states, changing the flux from $0$ to $1$ amounts to mapping the states according to $m\rightarrow m-1$. During the evolution, the states gradually shift inward from radius $r_m$ to $r_{m-1}$, as can be seen by inspecting the wave function\footnote{See Appendix A in \cite{prb} for derivation.}
	\begin{equation}\label{IQHE:eq:LLL_corbino_alpha}
		\phi_m^\alpha = \abs{z}^{-\alpha} \conj z^m,
	\end{equation}
	while remaining in the LLL with unperturbed energy.
	This phenomenon, schematically depicted in \cref{IQHE:fig:LL}(b), is called spectral flow, and it can be intuitively understood as the ring-like states contracting in order to keep the enclosed number of flux quanta constant and therefore avoid reacting to the induced emf due to Faraday's law $\mathcal E=-\flqnt/T$ \footnote{No bulk azimuthal current is produced. This does not apply to the edges.}. On the other hand $\Epsilon$ does produce a Hall current in the radial direction due to spectral flow. From what we have seen, we conclude that the LLL contributes the current of $i_r=-q/T$ since the net worth of one charge carrier is adiabatically pumped from the outer to the inner edge over time $T$. Similarly, every one of the $n$ filled Landau levels contributes a single charge $q$ to the current, leading to a net conductivity $\sigma_H = \frac{i_r}{\Epsilon}=-\frac{q}{\Phi_0}n=\frac{q^2}{2\pi\hbar}n$, which is consistent with \eqref{IQHE:eq:hall_cond}, if $\nu$ corresponds to the filling factor. 
	
	A ring system bounded in one direction as discussed above may provide an intuitive understanding of the connection between the Landau levels and the plateau values of the Hall conductivity. However, the pumping process described cannot go on indefinitely. Even though the model ignores the build-up of Hall potential opposing the current, the process is stopped when all of the electrons have been pumped to the inner edge. Furthermore, the upward curving of the Landau levels near the edges leads to the appearance of net azimuthal currents \cite{tong,halperin}. This may in fact be understood as as a Hall current due to the potential difference between the states on the inner and outer edge, caused by the pumping. These boundary effects obfuscate the profound role of topology in the IQHE. For this reason, we will next consider a system defined on a space with a toroidal topology.
	
\begin{figure}[htb]\centering
		\includegraphics[width=.4\textwidth]{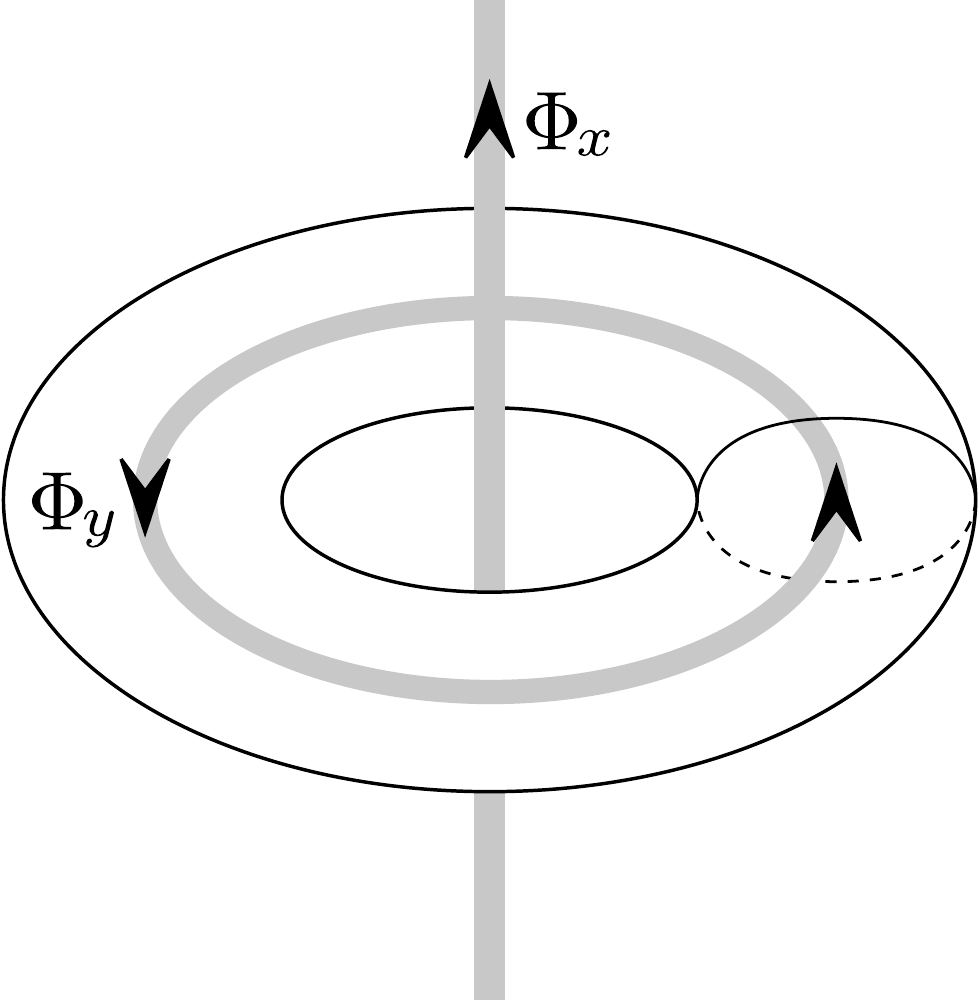}
		\caption{A torus with threaded magnetic fluxes $\Phi_x$ and $\Phi_y$.}\label{IQHE:fig:torus}
	\end{figure}

	A rectangle with periodic boundary conditions in both directions is topologically equivalent (homeomorphic) to a torus, since there are two distinct possible classes of nontrivial loops that cannot be contracted into a point without breaking them open (see \cite{stanescu} for a more rigorous  treatment). This is akin to loops around the two holes of a spatial torus that can be created by welding together the two ends of a metal pipe. We will work in a rectangular coordinate system $(x,y)$ that may be visualised as curving around the spatial torus, and we will call these nontrivial loops the $x-$cycle and the $y-$cycle. A presence of a hole allows us to thread a flux, as in the previous example. In this case, we have two possible fluxes $\Phi_x$ and $\Phi_y$. One may visualise this by imagining the pipe was welded around an infinite solenoid carrying $\Phi_x$, and another toroidal solenoid is carrying the flux $\Phi_y$ on the inside of the pipe, as shown in \cref{IQHE:fig:torus}. Note that the visualisations involving the spatial torus are crude, since the periodic rectangle we are considering does not curve in space, and is not possible to construct in practice. It is, nevertheless, a conceptually sound simplification.
	
	The vector potential due to the uniform magnetic field and the fluxes $\Phi_x$ and $\Phi_y$ is
	\begin{equation}\label{IQHE:eq:torus_VP}
		\VP = \VP_B + \frac{\Phi_x}{L_x}\hat{\bm e_x} + \frac{\Phi_y}{L_y}\hat{\bm e_y},
	\end{equation}
	where $L_x$ and $L_y$ are the rectangle dimensions. We will consider the induced field  a weak perturbation, so it suffices to consider the first order in perturbation theory. We may then use the linear response (Kubo) formula for the Hall conductivity \cite{stanescu,tknn}
	\begin{equation}\label{IQHE:eq:Kubo}	
		\sigma_{xy} = \frac{q^2 \hbar}{iL_xL_Y} \sum\limits_{\alpha}^{(E_\alpha<E_F)} \sum\limits_{\beta}^{(\beta\neq\alpha)}
			\frac{
	\mel{\psia}{v_x}{\psib} \mel{\psib}{v_y}{\psia} - \mel{\psia}{v_y}{\psib} \mel{\psia}{v_x}{\psib}
				}{(\Ea-\Eb)^2},
	\end{equation}
	where $\alpha=(n,m)$ and $\beta=(n',m')$, $\psia$ and $\psib$ are the unperturbed single-particle Hamiltonian eigenstates in the $n$-th and $n'$-th Landau level\footnote{In this form, the formula assumes nondegeneracy of single-particle states within each Landau level. This will be borne out in the realistic case due to level broadening. One may approach this more rigorously by working with the many-body ground state instead.}, and $v_i$ are the components of the velocity operator ${\bm v=(-i\hbar\Nabla-q\VP)/m}$. 
	One may verify that (for single-particle Hamiltonian $\mathcal H$)
	\begin{equation}\label{IQHE:eq:velocity_magic}
		\mel{\psib}{v_j}{\psia} =  \mel{\psib}{-\frac{L_j}{q}\pdv{\mathcal H}{\Phi_j}}{\psia}
								= -\frac{L_j}{q} (\Ea-\Eb) \braket*{\psib}{\pdphi j \psia}.
	\end{equation}
	Upon substituting into the Kubo formula, and some manipulation, we arrive at
	\begin{equation} 
		\sigma_{xy} = i\frac{q^2}{\hbar} \sum\limits_\alpha^{(\Ea<E_F)} \uglate{ \pdth y \braket{\psia}{\pdth x \psia} - \pdth x \braket{\psia}{\pdth y \psia} },
	\end{equation}
	where $\theta_i=2\pi\Phi_i/\flqnt$. Since fluxes that differ by an integer are related by a gauge transformation, the parameter space is essentially a torus $\fluxtorus$. The angular variables $\theta_i$ were introduced in recognition of this fact since they parametrize a torus in a natural way. After absorbing the part of summation going over $m$, the expression in the brackets becomes proportional to the Berry curvature $\curv=\Nabla_\theta\times\An$, where ${\An=i\sum_m \braket{\psia}{\Nabla_\theta\psia}}$, and so we have
	\begin{equation}
		\sigma_{xy} = - \frac{q^2}{\hbar} \hat{\bm e}_3 \cdot\! \sum\limits_n^{(E_n<E_F)} \curv,
	\end{equation}
	where $\hat{\bm e}_3$ is the unit vector perpendicular to the 2D plane. The Hall conductivity is related to the flow of the states on the torus, and we do not expect this to depend on the exact values $\theta_i$\footnote{This is obvious for our example, since there is no curvature and nothing special about any point on the rectangle.}. Therefore, we may average it over the parameter torus
	\begin{equation}\label{IQHE:eq:conductivity_Chern1}
		\sigma_{xy} = -\frac{q^2}{\hbar} \sum\limits_n^{(E_n<E_F)} \int_{\fluxtorus} \frac{\dd^2\theta}{(2\pi)^2} \, \hat{\bm e}_3\cdot\curv = \frac{q^2}{2\pi\hbar} \sum\limits_n^{(E_n<E_F)} C^n,
	\end{equation}
	where we have defined the first Chern number (TKNN\footnote{Thouless-Kohmoto-Nightingale-den Nijs \cite{tknn}} invariant) of the $n$-th band as
	\begin{equation}\label{IQHE:eq:Chern}
		C^n \equiv C_1^n = -\frac{1}{2\pi} \int_{\fluxtorus} \dd^2\theta \,\hat{\bm e}_3\cdot\curv.
	\end{equation}
	
\begin{figure}[htb]\centering
	\includegraphics[width=.4\textwidth]{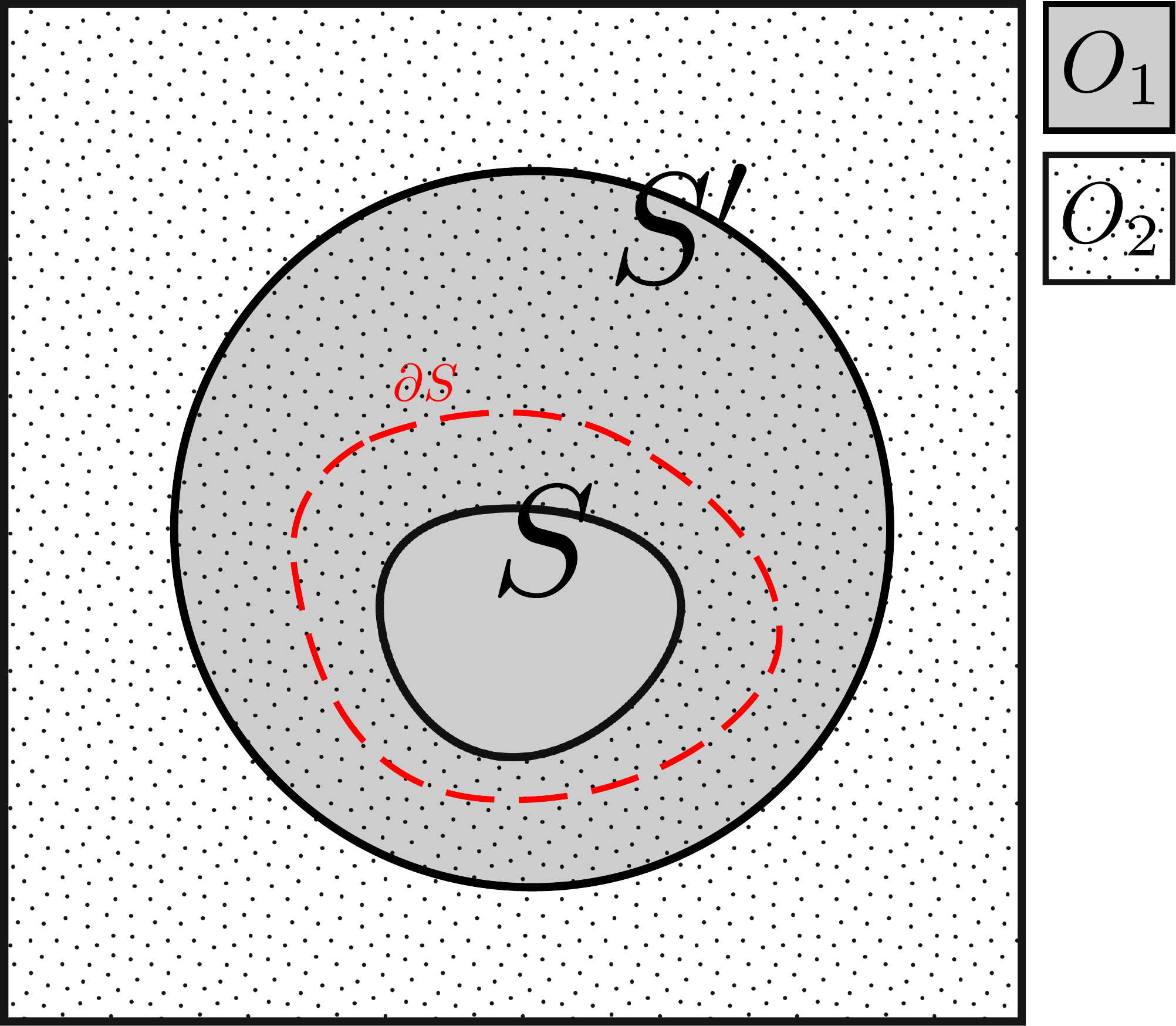}	
	\caption[Toroidal parameter space divided into the "interior" and the "exterior" regions]{The toroidal parameter space divided into the "interior" region $S$ and the "exterior" region $S'$, which lie (respectively) inside the $O_1$, and the $O_2$ subsets of the parameter space.} 
	\label{IQHE:fig:covering}
	\end{figure}

	Provided the Berry connection $\An$ can be defined smoothly over the entire $\fluxtorus$, we may apply the Stokes theorem to obtain $\C^n = -\frac{1}{2\pi}\oint_{\pdd\fluxtorus} \dd\bm l_\theta \cdot \An\equiv0$, since the torus has no boundary. It is, however, not generally the case that $\An$ is defined smoothly on the whole parameter space, and a nonzero Chern number indicates a breakdown of the Stokes theorem due to nontrivial topology of the parameter space. We will state without proof that $C^n \in \mathbb Z$. However, this is easy to show if we assume the parameter space can be covered by two open subsets $O_1$ and $O_2$ that allow for smooth definition of $\An_1$ and $\An_2$. In the overlap region of $O_1$ and $O_2$, the two definitions of the Berry connection are related by a gauge transformation $\An_2=\An_1-\Nabla\chi$. We imagine a closed contour $\pdd S$ lying entirely in the overlap region, and dividing the parameter space into two disjunct regions, the "interior" $S$, and the exterior $S'$ (see \cref{IQHE:fig:covering}). Now we may write the Chern number as the sum of the Berry curvature flux through the regions $S$ and $S'$, and apply the Stokes theorem to both integrals:
	\begin{align}\label{IQHE:eq:CinZ_proof}
		C^n &= -\frac{1}{2\pi} \uglate{ \int_S \dd^2\theta \,\curv + \int_{S'} \dd^2\theta\,\curv } \nonumber \\
			&= -\frac{1}{2\pi} \uglate{ \oint_{\pdd S} \dd \bm l_\theta \cdot \An_1
			- \oint_{\pdd S} \dd \bm l_\theta \cdot \An_2 } = \frac{1}{2\pi}(\gamma_2-\gamma_1).
	\end{align}	
	As we can see, the Chern number is now the difference between the Berry phases along the same path, calculated by using gauge shifted Berry connections. Since the Berry phase is gauge invariant, the difference must be a multiple of $2\pi$, and hence $C^n\in\mathbb Z$.	
	Specifically, for each Landau level, $\C^n=\pm1$ \cite{ozawa}. We may now conclude that when the Fermi energy lies in the $N$-th band gap, the Hall conductivity for a Landau level system defined on a torus is $\sigma_{xy}=\frac{q^2}{2\pi\hbar}\nu$, where $\nu=\sum_n^N C^n=N$ is the total IQHE state Chern number, and equal up to a sign to the filling factor. 
	
	The discrete nature of the Chern number points to its topological nature, since it cannot be gradually changed by deforming the quantum states. Indeed, it is a topological invariant characterizing an IQHE state, and it can only change at the point of a phase transition when the gap separating the valence and the conduction band closes and reopens. On the other hand, closing and reopening of the lower gaps may change the Chern number of the specific bands in question, but their sum is unaffected \cite{avron1983}. 
	
	\subsubsection{IQHE in lattice models}
	
	Thus far, we have been considering the IQHE for a continuum model of a noninteracting 2DEG in a magnetic field. Let us now turn our attention to lattice models. We again assume an insulating state with $E_F$ in a gap. As we have seen in \cref{sec:geomph:Bloch}, spatial periodicity in a Hamiltonian allows us to view the problem in terms of the Bloch Hamiltonian
	\begin{equation}\label{IQHE:eq:BlochH}
		H_{\q} = \frac{1}{2m} \oble{-i\hbar\Nabla + \hbar \q - q\VP(\vr)}^2 + U(\vr),
	\end{equation}
	 parametrically dependent on the crystal momentum which assumes values from the Brillouin zone, $\q\in BZ$. Due to its periodicity, the Brillouin zone itself is homeomorphic to a $\torus$ torus. In this case, the parameter space is intrinsic to the system, as opposed to the externally imposed perturbations discussed for a continuum system confined to a torus, but the consequences of its topology are similar. We may once again use the Kubo formula \eqref{IQHE:eq:Kubo} to find the Hall conductivity, replacing the states $\ket{\psia}$ by the eigenstates $\ket\unq$ of the Bloch Hamiltonian $H_\q$, and the summation indices $\alpha=(n,m)$ by $(n,\q)$. In this case, we may replace $v_j$ by $\frac{1}{\hbar}\pdv{H}{\sq_j}$, which does not change the off-diagonal matrix elements:
	\begin{equation}
		\mel{\unq}{v_j}{\unqprim}	= \frac{1}{\hbar}\mel{\unq}{\pdv{H_\q}{\sq_j}}{\unqprim} 
									= (E_{n'\q'} - E_{n\q}) \braket{\unq}{\pdd_{k_j}\unqprim},
	\end{equation}
	and the summation over $\q$ by an integral, $\sum_\q \rightarrow \frac{L_xL_y}{(2\pi)^2} \int_{BZ}\dd^2 k$. Once again, we arrive at the Hall conductivity of the form 
	\begin{equation}\label{IQHE:eq:conductivity_Chern2}
		\sigma_{xy} = \frac{q^2}{2\pi\hbar} C,
	\end{equation}
	with the Chern number given by
	\begin{equation}
		C= \sum\limits_n^{(E_n<E_F)} C^n = \sum\limits_n^{(E_n<E_F)} (-)\frac{1}{2\pi} \int_{BZ} \dd^2\sq \, \uvec{e}_3 \cdot \curv.
	\end{equation}
	 The same remarks on the integral values of the Chern number, and its topological nature apply here.

	\begin{figure}[htb]\centering
		\includegraphics[width=.5\textwidth , trim={0 0 5cm 0}, clip]{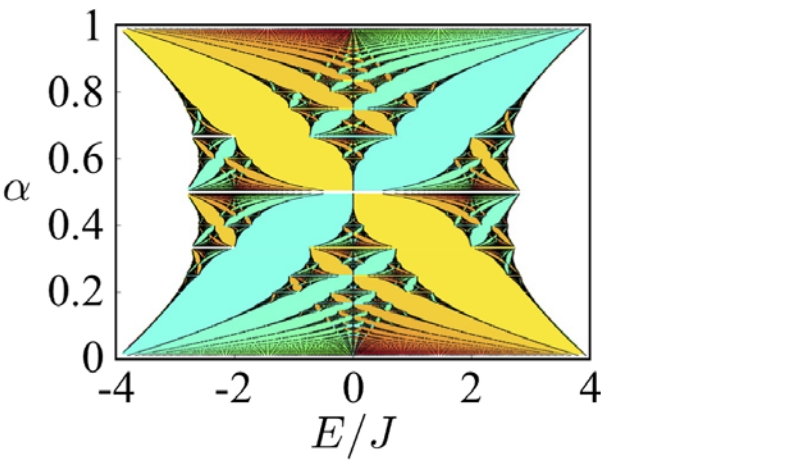}
		\caption[Spectrum of the Harper-Hofstadter model (the Hofstadter butterfly)]{The spectrum of the Harper-Hofstadter model (the Hofstadter butterfly) for flux per plaquette $\alpha\in\cinterval{0}{1}$ in units of flux quantum. The colors indicate the Chern number of the gap, i.e. the sum of the Chern numbers of all levels below the gap. Warm (cold) colors indicate a negative (positive) Chern number. Adopted from \cite{ozawa}.}			
		\label{IQHE:fig:reptil}
	\end{figure} 
	
	The specific value of the Chern number, and hence the topological phase of the IQHE system, will depend on the specifics of the system in question. The simple Landau level picture of the continuum problem above does not carry over to the lattice version of the same problem, as can be seen, for example, in the case of the square lattice (the Harper-Hofstadter model). Fig. \ref{IQHE:fig:reptil} shows the spectrum of Harper-Hofstadter model \cite{hofstadter} in relation to the magnetic flux per plaquette ($\alpha$) in units of $\flqnt$; the famous Hofstadter butterfly. 
	For rational values of ${\alpha=\frac{p}{q}}$, $p$ and $q$ being coprime integers, the topologically trivial lowest Bloch band decomposes into $q$ subbands with nontrivial Chern numbers, while for irrational values of $\alpha$, the spectrum fractalizes \cite{tong,ozawa}. Decomposition is the result of a vector potential breaking the discrete translation symmetry of the lattice, as evident from the Landau gauge tight-binding (TBA) Hamiltonian \cite{ozawa}:
	\begin{equation}
		\hat{H}_{HH} = -J \sum\limits_{x,y}(\hat a^\dagger_{x+a,y}\hat a_{x,y} + e^{i2\pi\alpha x/a} \hat a^\dagger_{x,y+a}\hat a_{x,y} + \mathrm{H.c.}),
	\end{equation}	  	
	where $a$ is the lattice constant, $\hat a_{x,y}$ is the annihilation operator for a particle at site $(x,y)$, and $J$ is the hopping amplitude to the neighbouring site, modified in the $y$ direction by the symmetry-breaking Peierls phase factor $e^{i2\pi\alpha x/a}$ . For rational fluxes, translation symmetry still applies in the Landau gauge, but for the so-called magnetic unit cell of size $(qa,a)$. In this case, we may define the magnetic Brillouin zone that is $q$ times smaller than the original BZ. This leads to the appearance of $q$ subbands.

\begin{figure}[htb]\centering
		\includegraphics[width=.95\textwidth]{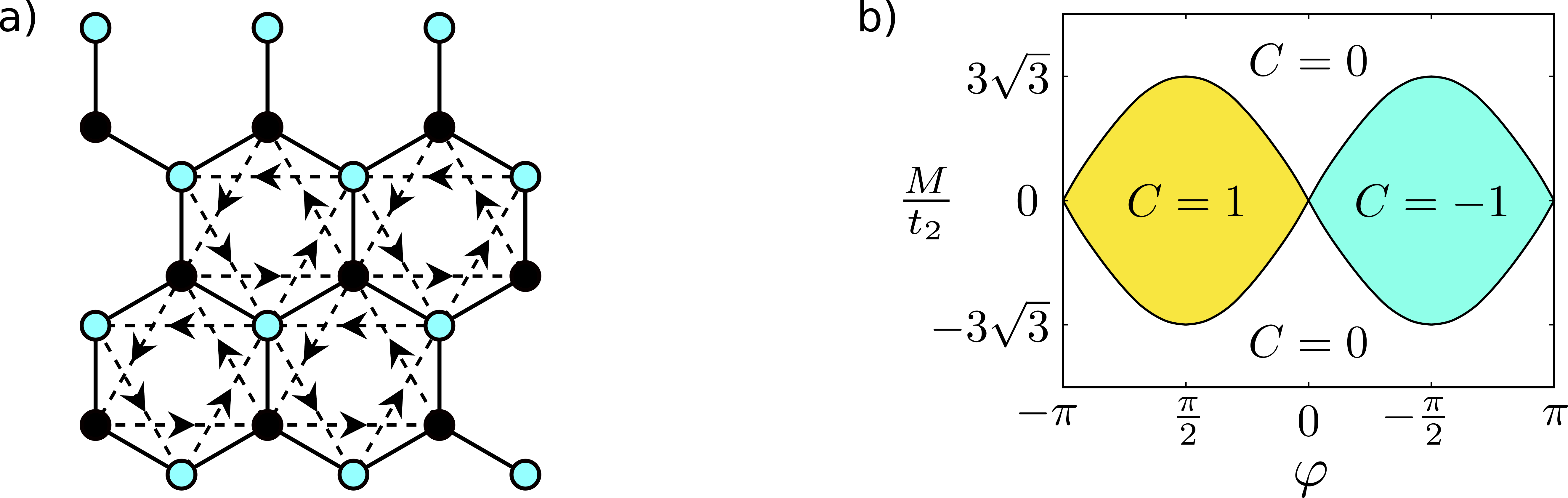}
		\caption[Haldane model]{(a) The honeycomb lattice. Next-nearest neighbour hopping along the arrows produces the Peierls phase $+\varphi$, while hopping against the arrows produces $-\varphi$. (b) The phase diagram (sketch) of the Haldane model. $C$ denotes the Chern number, $M$ the mass parameter, and $t_2$ the next-nearest neighbour hopping amplitude.}
		\label{IQHE:fig:HaldaneModel} 
	\end{figure}

	As mentioned at the beginning of this section, time-reversal symmetry breaking is a necessary condition for the QHE. The reason is that TRS causes localization in 2D, and has to be broken to allow for the existence of current-carrying extended states \cite{halperin}. A net magnetic flux through each plaquette will achieve TRS breaking, since hopping in opposite directions produces opposite Peierls phases, but it is not necessary for quantized Hall conductivity, as demonstrated by Haldane \cite{haldane1988}. The TBA Haldane model Hamiltonian is given by \cite{BH2013}
	\begin{equation}\label{IQHE:eq:Haldane}
		\hat H = t_1\sum\limits_{\ev{i,j}} \hat a_i^\dagger \hat a_j + t_2\sum\limits_{\ev{\ev{i,j}}} e^{-i\nu_{i,j}\varphi} \,\hat a_i^\dagger \hat a_j + M \sum\limits_i \epsilon_i \hat a_i^\dagger \hat a_i,
	\end{equation}
	where $\hat a_i$ is the annihilation operator for a particle at site $i$; $\ev{i,j}$ and $\ev{\ev{i,j}}$ denote respectively the summation over the nearest and next-nearest neighbours for all sites of the honeycomb lattice (HCL), while $t_1$ and $t_2$ are respectively the nearest and next-nearest neighbour hopping amplitudes; $\nu_{i,j}=-\nu_{j,i}=\pm 1$ when $i$ and $j$ are next-nearest neighbours (see \cref{IQHE:fig:HaldaneModel} (a)), $\varphi$ is a constant (see \cref{IQHE:fig:HaldaneModel} (a)), $M$ is the mass parameter, and $\epsilon_i=\pm 1$, depending on the sublattice. The HCL has two sites per cell, so the lowest Bloch band splits into two. 
	
	The first term in the Hamiltonian accounts for hopping between nearest neighbours with amplitude $t_1$. On its own, this term is a simple model for graphene, and it features the familiar gapless Dirac points in the six corners of the BZ, two of which are nonidentical. The dispersion is conical in their vicinity, resembling that of massless relativistic electrons obeying the 2D Dirac Hamiltonian $H_D = \hbar v_D (q_x\sigma_x+q_y\sigma_y)$ in $\q$-space, where $\bm q$ is the momentum measured from a Dirac point, $\sigma_i$ are the Pauli matrices, and $v_D\equiv 3t_1/2$ is the Dirac velocity. 
	
	The Dirac points are known to be protected by the combined TR and inversion (IS) symmetries  \cite{BH2013}. The last two terms break one of those symmetries each, which has the effect of adding a term proportional to $\sigma_z$ to the $\q$-space Hamiltonian. The third term is simply an on-site energy, breaking the IS by alternating the sign between the two sublattices. This term opens a gap by adding a mass ($M$) term to the Dirac Hamiltonian, but does not induce nontrivial band topology. The second term breaks the TRS (when $\varphi\neq0,\pi$) due to the opposite Peierls phases for next-nearest neighbour hopping in opposite directions. This allows the two bands to be nontrivial in certain regions of the phase diagram (\cref{IQHE:fig:HaldaneModel} (b)), with the Chern numbers $C^\pm = \pm 1$. However, the Peierls phases arranged in this way do not produce a net flux through a plaquette, and preserve the original translation symmetry, unlike the previous example of the Harper-Hofstadter model.
	
	\subsubsection{Role of disorder}
\begin{figure}[htb]\centering
		\includegraphics[width=.4\textwidth]{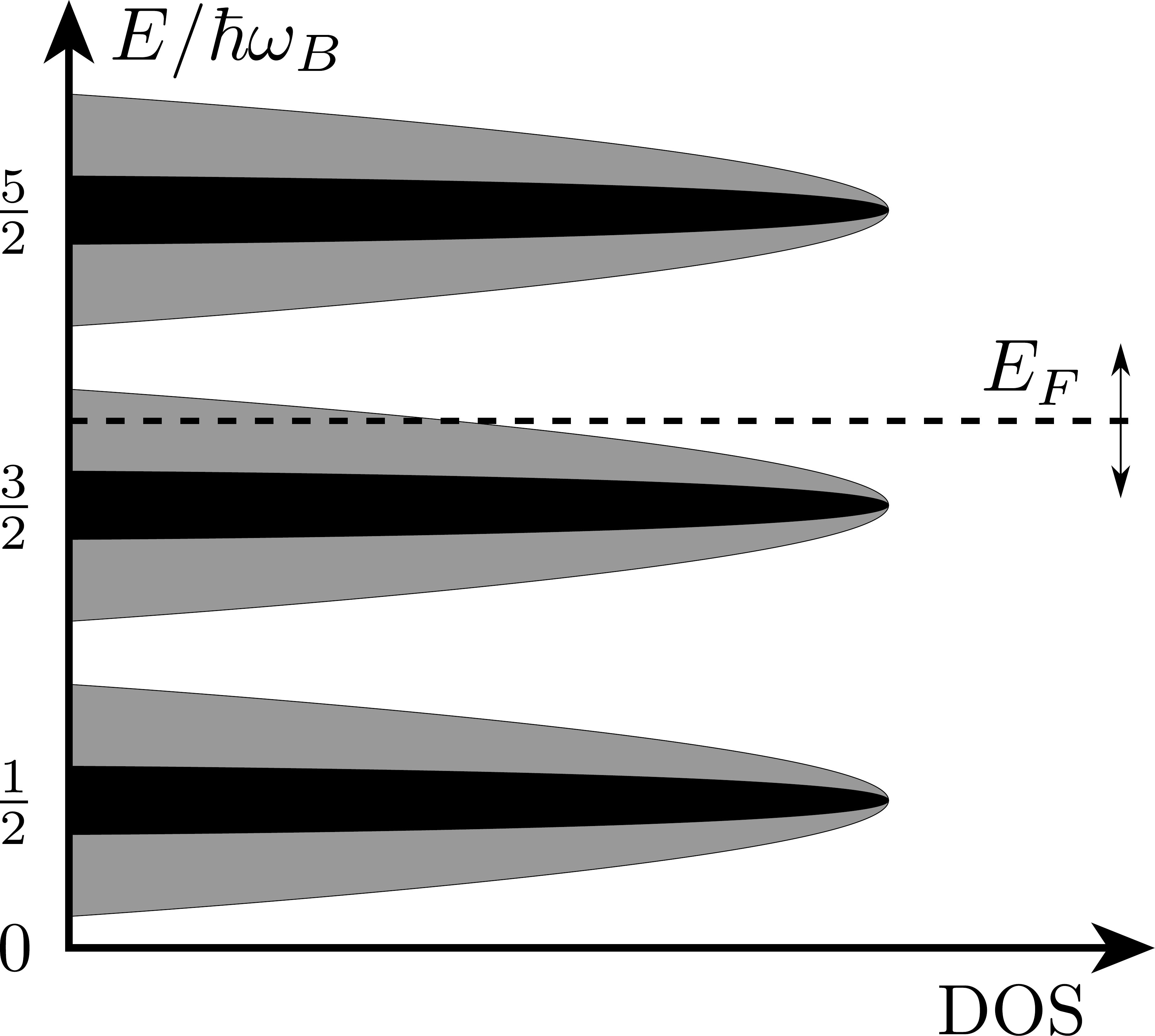}
		\caption[Density of states for broadened Landau levels]{A density of states (sketch) characteristic of Landau levels broadened by disorder. The regions in black represent the bands of extended states, while the broader regions in grey represent the whole bands, including localized states. The Fermi energy $E_F$ shifts vertically as the magnetic field is varied.}
		\label{IQHE:fig:DOS}
	\end{figure}

	So far, we have explained the Hall conductivity values for the case of completely filled bands or Landau levels ($E_F$ lying in a gap), in terms of the Chern number. These values correspond to the resistivity plateaus shown in \cref{IQHE:fig:iqhe}. However, in order to explain the width of the plateaus, we must consider partially filled bands. For a 2DEG with electron density $\rho_0$ in a magnetic field, the filling factor is controlled by the flux density, i.e. the magnetic field. The filing factor is an integer $\nu$ when the field is given by $\rho_0\flqnt /\nu$. Between these values, we expect the resistivity to increase smoothly, as opposed to stepwise. This picture is changed in presence of disorder. Taking the random potential into account in the Hamiltonian \eqref{IQHE:eq:cont_H}, the Landau levels will broaden, due to the degeneracy being lifted. We assume the disorder is not strong enough to completely disrupt the Landau level picture. The randomly distributed potential hills and valleys may bind states, and localize them to the region of increased or decreased potential, leading to an increase or reduction in their energy. The remaining extended states will be perturbed, but their energies will tend to shift less \cite{tong}, so they will tend to occupy the central regions of their respective bands in the density of states (\cref{IQHE:fig:DOS}). Let us recall the Corbino ring geometry example from earlier. Threading a flux through the ring does not affect the localized states beyond a gauge transformation, but the extended states still have to flow, due to the discrete nature of gauge invariance. When flux is increased to $\flqnt$, the spectrum maps back onto itself, and each extended state takes the place of the next available state. Therefore, provided all extended states are filled, a single electron is mapped from one edge to another, regardless of the number of available extended states. As argued in \cite{halperin}, some extended states will survive, as long as the disorder is not strong enough to disrupt the Landau level picture. 
	
	To explain the plateaus, we now simply have to consider what happens as the magnetic field is adiabatically varied, starting from a particular strength, such that $\nu$ bands are completely filled. As we gradually increase the field, the band centers in the density of states shift upwards. Since the energy axis in \cref{IQHE:fig:DOS} is scaled by $\hbar\omega_B$, the levels appear stationary, and the Fermi energy shifts downwards instead. As the Fermi energy crosses the localized states on the top of the highest occupied band, these states become unoccupied since the system remains in the ground state. However, the longitudinal and transversal conductivities are unaffected, since they are only dependent on the extended states. As the Fermi energy crosses the extended portion of the band, the system becomes a conductor, and the Hall conductivity transitions between the plateaus. This corresponds to a particular transition region in \cref{IQHE:fig:iqhe}. The new plateau value of conductivity is then retained until the Fermi energy reaches the next band of extended states. An important point on which this explanation relies is the topological protection of the band Chern number against disorder that is not strong enough to merge the bands of extended states.

\let\psia\undefined
\let\psib\undefined
\let\Ea\undefined
\let\Eb\undefined
\let\pdphi\undefined
\let\pdth\undefined
\subsection{Fractional quantum Hall effect}\label{sec:TQM:FQHE}

\begin{figure}[htb]\centering
		\includegraphics[width=.95\textwidth]{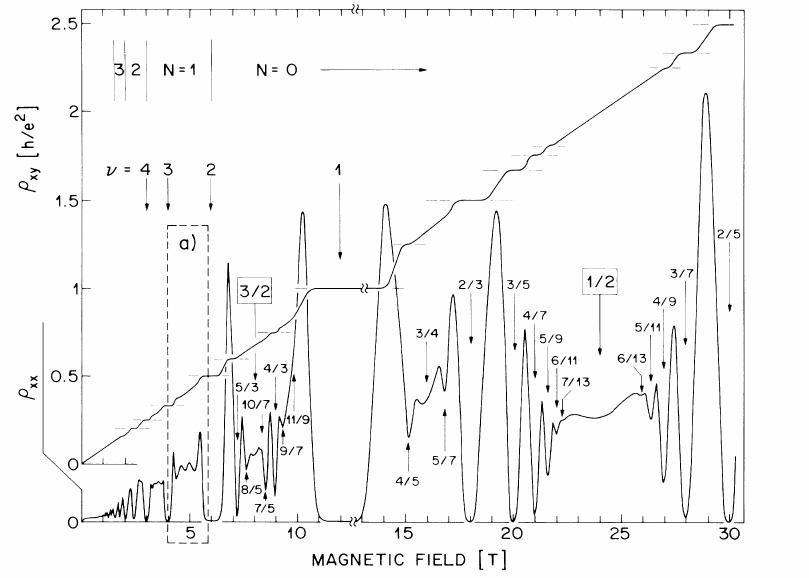}
		\caption[Fractional quantum Hall effect]{Overview of Hall and longitudinal resistivity over a range of magnetic field values measured in a $\mathrm{GaAs/AlGaAs}$ heterostructure. Plateaus at fractional fillings are visible in the Hall resistivity. From \cite{willett1987}.}
		\label{FQHE:fig:fqhe}
	\end{figure}

	Not long after the discovery of the IQHE, Tsui, Stormer, and Gossard discovered a conductivity plateau at a fraction ($\nu=1/3$) of $q^2/h$ \cite{TSG}. Other fractional values followed, some of them depicted in \cref{FQHE:fig:fqhe}. This phenomenon is called the fractional quantum Hall effect (FQHE) and it is a rich source of emergent and often unintuitive physics. The key to fractionalizaion of Hall conductivity is that the Coulomb interaction be prominent enough to break the relatively simple IQHE picture. This requires that the Coulomb energy scale should be large in comparison to the scale of the disorder potential, but it should be small when compared to the cyclotron energy $\hbar\omega_B$ \cite{tong}, so that the Landau level structure is perturbed, but not destroyed.

	\subsubsection{FQHE at fillings $\nu=1/m$}
	We will again consider a 2DEG in a uniform magnetic field, but this time we include the interactions between electrons. Perhaps the simplest system of this kind is given by the Hamiltonian
	\begin{equation}\label{FQHE:eq:H}
		H = \sum\limits_j^{N_e} \uglate{ \frac{1}{2m} (\p_j-q\VP(\vr_j))^2 + V(\vr_j)}
				+ \frac{q^2}{4\pi\epsilon_0} \sum\limits_{j>k}^{N_e} \frac{1}{|\vr_j-\vr_k|}, 
	\end{equation}
	where $V$ is a neutralizing potential due to a positively charged uniform background (for electrons). In practice, finding an exact ground state is impossible for large $N_e$, but the topologically protected features of the FQHE states ought to be captured by any wave function that is adiabatically deformable into the correct ground state. Such wave functions have been proposed for various FQHE states, most notably the Laughlin states \cite{laughlin1983} which model the $\nu=1/m$ FQHE states (fermionic for odd $m$):
	\begin{equation}\label{FQHE:eq:LaughlinState}
		\psi_m(\{z_k\}) = \prod\limits_{i<j}^{N_e} \conj{z_i-z_j}^\m \GaussMB,
	\end{equation}
	where we take the $\{z_k\}$ dependence to include both $z_k$ and $\conj z_k$ for each $k$.
	These states have high overlaps with the true ground states of the Hamiltonian \eqref{FQHE:eq:H} for small numbers of particles, but are not expected to be a good approximation for large systems \cite{tong}. Instead, they are useful entirely because they are characterized by the same kind of topological order as the true ground state. Note that for the completely filled LLL ($m=1$), the interactions become unimportant in the context of QHE, which is reflected in the fact that the Laughlin state corresponds to the many-body ground state for the noninteracting system, i.e. the Slater determinant comprised of the states \eqref{IQHE:eq:LLL_wf}.
	
	It can be argued by employing the plasma analogy (explained below) that the preferred number density of electrons in a Laughlin ground state is given by $\rho_0=\frac{1}{2\pi l_B^2 m}$ \cite{tong,laughlin-lec}. In fact, it behaves like an incompressible liquid, meaning it admits no gapless excitations. This indicates that the QHE at a fractional filling $1/m$ should be explained as a consequence of the interactions opening a gap in the LLL at this filling. We may repeat the Corbino ring argument here. The spectrum is sensitive to the value of the flux threaded through the ring, but it must be indistinguishable for different integer values of flux. Threading a negative flux $\alpha=-1$ will cause the states to flow outwards, since the number of flux quanta in the area is reduced. We assume an infinitely thin flux-carrying solenoid, and the argument is then unchanged if we consider the system on a disk instead of the ring geometry. In this case, electron density must be depleted in a region around the position of the solenoid due to the outward flow. Since the new state is again an eigenstate of the Hamiltonian, we will consider this depletion a quasihole. Since only $1/m$ of the LLL is filled, we may anticipate that only $q/m$ charge is transferred radially. This may be confirmed by examining the probability distribution resulting from the wave function describing a quasihole at position $\eta$ \cite{laughlin1983,tong}
	\begin{equation}\label{FQHE:eq:quasihole}
		\psi_m^\eta( \{z_l\} ) = \prod_k^{N_e} \conj{z_k-\eta} \prod\limits_{i<j}^{N_e} \conj{z_i-z_j}^\m \GaussMB
	\end{equation}
	through the lens of plasma analogy \cite{laughlin1983}. To do this we write this distribution as a Boltzmann distribution $|\psi_m^\eta(\{z_k\})|^2=\exp[-\beta U(\{z_l\}) ]$, with the inverse "temperature" $\beta=2/m$ and
	\begin{equation}\label{FQHE:eq:plasma_potential}
		U(\{z_l\})= -m^2\sum_{i<j}^\Ne \log(\frac{|z_i-z_j|}{l_B}) + \frac{m}{4l_B^2}\sum_{i}^\Ne |z_i|^2
					- m \sum_i^\Ne \log(\frac{|z_i-\eta|}{l_B} ),
	\end{equation}
	where we have restored the scale $l_B$ in the logarithms from the omitted normalization factor. This potential function corresponds to a 2D plasma with a charged impurity at $\eta$. The first term describes the Coulomb repulsion between particles of "charge" $m$ (corresponding to electrons) and the second term the interaction of each "charge" $m$ with a neutralizing background "charge" density of $-1/2\pi l_B^2$. The first two terms are also present for the ground state, whereas the third term is specific to a state with a hole at $\eta$, and it describes the interaction of "charges" $m$ with an impurity of "charge" $1$. At a distance much greater than $l_B$, the impurity "charge" will be screened by a redistribution of the plasma particles. Since the impurity "charge" to be neutralized is $m$ times less than the particle "charge", it follows that the actual charge of the quasihole is $-q/m$, and therefore, that the charge $q/m$ has been radially transferred away from the hole. Alternatively, a positive flux would cause an inward transfer of the same charge. The accumulation of additional charge $q/m$ around $\eta$ indicates the presence of a quasiparticle. 
	
	Combined with the effect of disorder, the fractional charge of the Laughlin quasiparticles and quasiholes explains the plateaus of conductivity for $1/m$ states. However, an apparent contradiction may arise with the result \eqref{IQHE:eq:conductivity_Chern1} that implies Hall conductivity quantization in terms of the (integral) Chern number for a system defined on a torus. To resolve this, we note that the Kubo formula in the form \eqref{IQHE:eq:Kubo} only applies when the ground state is nondegenerate. However, the ground state on a Riemann surface of genus $g$ is not unique, but is characterized by a topological degeneracy $m^g$ \cite{WenNiu}. This gives an $m$-fold degeneracy in case of a torus ($g=1$). This degeneracy is a topological invariant characteristic of the topological order inherent to the FQHE system on a torus.
	
\begin{figure}[htb]\centering
		\includegraphics[width=.95\textwidth]{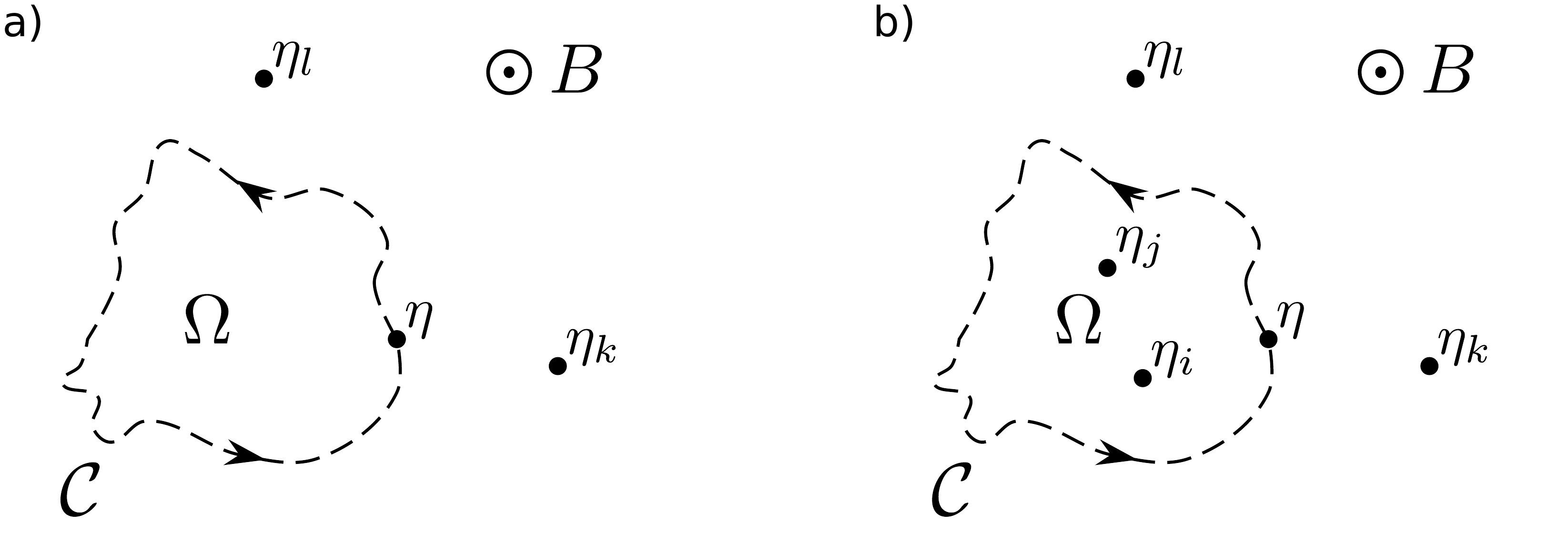}
		\caption[Quasihole path for calculation of statistical phase]{A possible path $\mathcal C$ traversed by the quasihole $\eta$, enclosing the region $\Omega$. The direction of the magnetic field is indicated in the top-right corners. a) When no quasiholes are enclosed, integration in \eqref{FQHE:eq:bph_rho} leads to the Aharonov-Bohm phase. b) The path encloses $N_\Omega=2$ quasiholes, leading to the total Berry phase \eqref{FQHE:eq:bph_FINAL}, which is the sum of the Aharonov-Bohm phase, and the statistical phase. }\label{FQHE:fig:bph}
	\end{figure}

	\label{QPSTATS1}	
	In addition to fractional charge, another remarkable property of the Laughlin quasiparticles is their fractional (anyonic) exchange statistics \cite{halperin1984,arovas}. We will derive the statistical parameter for $1/m$ Laughlin quasiholes  in the following, and will further discuss anyons in the next section. Similar to Arovas {\it et al.} \cite{arovas}, we consider a quasihole at $\eta$, and let $\eta$ adiabatically traverse a closed path $\mathcal C$, enclosing the region $\Omega$ (see \cref{FQHE:fig:bph}).
	 The resulting Berry phase is composed of the Aharonov-Bohm contribution due to the charge of the quasihole, and the statistical phase which occurs if the path encloses other identical quasiholes\footnote{We argue in the next section that taking a path around another particle can be regarded as equivalent to two subsequent exchanges.}. The wave function for a state containing $N$ quasiholes is given by:
	\begin{align} \label{FQHE:eq:eta_state}
		\psi( \{z_l\}; \{ \eta_\alpha \} ) 
			&=  \prod_{\beta=1}^N \prod_k^{N_e} \conj{z_k-\eta_\beta} \prod\limits_{i<j}^{N_e} \conj{z_i-z_j}^\m \GaussMB \nonumber \\
			&= \prod_k^{N_e} \conj{z_k-\eta} \: \psi( \{z_i\}; \{ \eta_{\alpha>1}\} ),
	\end{align}
	where in the second line we single out the quasihole that is to be taken around a loop. Since we are using complex notation for the coordinates, the Berry phase is given by
	\begin{equation}\label{FQHE:eq:bph_cplx}
		\gamma = i \oint_\mathcal C (\Aeta \dd\eta + \Aceta \dd\ceta ),
	\end{equation}
	where $\Aeta$ and $\Aceta$ are the holomorphic and antiholomorphic Berry connections. The functional form of the normalization factor is needed to derive $\Aeta$ and $\Aceta$, so we introduce the explicitly normalized state $\Psi = \sqrt{Z}^{\,-1} \psi$ and write
	\begin{align}\label{FQHE:eq:connections_cplx}
		\Aeta &= i \braket{\Psi}{\pdeta\Psi} = \frac{i}{Z}\braket{\psi}{\pdeta\psi} - \frac{i}{2} \frac{1}{Z}\pdeta Z 		\nonumber \\
		\Aceta &= i \braket{\Psi}{\pdceta\Psi} = \frac{i}{Z}\braket{\psi}{\pdceta\psi} - \frac{i}{2} \frac{1}{Z}\pdceta Z,
	\end{align}
	where $\pdeta=\frac{1}{2}(\pdd_x-i\pdd_y)$ and $\pdceta=\frac{1}{2}(\pdd_x+i\pdd_y)$ are the Wirtinger derivatives. From the Cauchy-Riemann equations, it follows that $\pdceta f=0$ for any holomorphic function, and equivalently $\pdeta f=0$ for antiholomorphic functions. Since $\psi$ is antiholomorphic in $\eta$, we have  $\braket{\psi}{\pdeta\psi}=0$, \\$\pdceta Z=\pdceta\braket{\psi}=\braket{\psi}{\pdceta\psi}$, and $\pdeta Z=\braket{\pdeta\psi}{\psi}=\braket{\psi}{\pdceta\psi}^*$. Finally, it is easily shown that $\braket{\psi}{\pdceta\psi}=-\ev{\sum_j^{N_e} \conj{z_j-\eta}^{-1}}{\psi}$. By \eqref{FQHE:eq:connections_cplx} and \eqref{FQHE:eq:bph_cplx}, the Berry phase now assumes the form
	\begin{equation}\label{FQHE:eq:bph_ev}
		\gamma = \frac{i}{2} \oint_\mathcal C \dd\eta \ev{\sum_j^{N_e}\frac{1}{z_j-\eta}}{\Psi} - \frac{i}{2} \oint_\mathcal C \dd\ceta \ev{\sum_j^{N_e}\frac{1}{\conj{z_j-\eta}}}{\Psi}
	\end{equation}
	Using the definition of the single-particle density $\rho(z;\{\eta_i\})=\ev{\sum_j^{N_e} \delta (z_j-z)}{\Psi}$, we may rewrite  in the form
	\begin{equation}\label{FQHE:eq:bph_rho}
		\gamma = \frac{i}{2} \int \dd^2 r 	\uglate{ \oint_\mathcal C \dd\eta \,\frac{\rho(z;\{\eta_i\})}{z-\eta}
										-	\oint_\mathcal C \dd\ceta \,\frac{\rho(z;\{\eta_i\})}{\conj{z-\eta}} } 
	\end{equation}
	Far away from the quasihole, the density is uniform, so we may write $\rho(z)=\rho_0+\delta\rho(\{z-\eta_i\})$. In the thermodynamic limit, the plasma analogy leads to $\delta\rho(\{z-\eta_i\})=-\frac{1}{m}\sum_k^{N} \delta^2(z-\eta_k)$ (see \cite{marija2018} and references therein). This result ignores the short-distance behaviour and only accounts for charge expulsion, but it may be used to approximate the Berry phase when the surface enclosed by the traversed contour is large compared to $l_B^2$. The uniform density $\rho_0$, being both holomorphic and antiholomorphic, allows us to use the residue theorem to solve the contour integral $\oint_\mathcal C\dd\eta\frac{\rho_0}{z-\eta}=-2\pi i \rho_0 \,\Theta_{z\in\Omega}\,$ and its complex conjugate, where the step function $\Theta_{z\in\Omega}\,$ will act to constrain the surface integration in \eqref{FQHE:eq:bph_rho} to region $\Omega$. We first derive the Aharonov-Bohm contribution, which corresponds to the Berry phase for $\rho(z\in\Omega)=\rho_0$, i.e. no quasiholes in region $\Omega$, as in \cref{FQHE:fig:bph} a). We arrive at
	\begin{equation}\label{FQHE:eq:ABph}
		\gamma_0 = 2\pi \ev n_\Omega = 2\pi \frac{\Phi_\Omega}{\flqnt}\frac{1}{m},
	\end{equation}
	where $\ev n_\Omega=\rho_0 A_\Omega$ is the mean number of electrons in the enclosed region $\Omega$. The result corresponds to the Aharonov-Bohm phase \eqref{eq:berry_AB} for a particle of charge $q*=q/m$, thus confirming the fractional charge of the quasiholes.

	Let us now imagine that the path $\mathcal C$ encloses $N_\Omega$ quasiholes as in \cref{FQHE:fig:bph} b). The  Aharonov-Bohm phase remains unchanged, and the $\delta\rho$ term in \eqref{FQHE:eq:bph_rho} is easily evaluated by first performing the surface, followed by the contour integral
	\begin{align}\label{FQHE:eq:bph_FINAL}
		\gamma &= \gamma_0 + 
			\frac{i}{2} \uglate{
			\oint_\mathcal C\dd\eta \int\dd^2r \frac{\frac{-1}{m}\sum_{k=2}^N \delta^2(z-\eta_k)}{z-\eta} -
			\oint_\mathcal C\dd\ceta \int\dd^2r \frac{\frac{-1}{m}\sum_{k=2}^N \delta^2(z-\eta_k)}{\conj{z-\eta}}
			} 																		\nonumber \\
		&= \gamma_0 - 
			\frac{i}{2m} \sum_{k=2}^N 
			\uglate{ \oint_\mathcal C\frac{\dd\eta}{\eta_k-\eta} - \oint_\mathcal C\frac{\dd\ceta}{\conj{\eta_k-\eta}} } 
			= 2\pi \frac{\Phi_\Omega}{\flqnt}\frac{1}{m} -
			\frac{i}{m} \sum_{k=2}^N	(-)2\pi i \, \Theta_{\eta_k\in\Omega}		\nonumber \\
		&= 2\pi ( \frac{\Phi_\Omega}{\flqnt} - N_\Omega )  \frac{1}{m}				\nonumber \\
		&= 2\pi \ev n_\Omega
	\end{align}
	Once again, the total phase is proportional to the mean number of electrons enclosed, but this number is now reduced by $1/m$ for each enclosed quasihole. The statistical phase is simply the difference between $\gamma$ and $\gamma_0$, which for a single enclosed quasihole amounts to
	\begin{equation}\label{FQHE:eq:statph}
		\gamma_S = \gamma_1 - \gamma_0 = -2\pi\frac{1}{m}
	\end{equation}
	The sign of the phase here depends on whether the contour $\mathcal C$ is traversed in clockwise or anticlockwise sense, which is a feature of anyonic statistics, along with the fractional phase (in units of $2\pi$) upon double exchange.
	\label{QPSTATS2}
	
	\subsubsection{FQHE at other filling fractions}
	FQHE at filling fractions $\nu$ other than $1/m$ may be explained by hierarchical construction \cite{haldane1983,halperin1984}, starting from Laughlin states. For small deviations of the magnetic field from the $\rho_0\Phi_0m$ value, the change in filling fraction may be viewed as a net excess in the number of quasiparticle excitations, defined as the difference between the number of quasiparticles and quasiholes, $N_\mathrm{ex}=N_\mathrm{qp}-N_\mathrm{qh}$. Since quasiparticles are (fractionally) charged, they feel the background magnetic field, and may condense into a Laughlin-like state. The allowed shapes of the wave function are imposed by statistics. We have seen that factors of the form $\conj{z_i-z_j}^\alpha$ with $\alpha=m=2p-1$ appear for each particle pair in case of fermions. More generally, $\alpha=2p+\theta$, where $\theta$ is the statistical parameter. In case of bosons, we would have $\theta=0$, resulting in Laughlin states with even $m$. For anyonic quasiparticles(holes) discussed above, $\theta=\chi/m$, with $\chi=+1$ for quasiparticles and $\chi=-1$ for quasiholes, and the magnetic length is $\sqrt m l_B$, so the appropriate wave function is 
	\begin{equation}\label{FQHE:eq:Lauglin_QPs}
		\psi(\{\eta_k\}) = \prod_{i<j}^{N} \conj{\eta_i-\eta_j}^{2p+\chi\frac{1}{m}} 
		\exp(-\sum_i^N \frac{|\eta_i|^2}{4ml_B^2} ),
		\qquad p = 1,2,3,\dots
	\end{equation}
	where, in order to accommodate for the possibility of positively and negatively charged quasiparticles, we define $\eta_k=\eta_x+i\chi\eta_y$. Given the starting Laughlin state, the restriction $p\in\mathbb N$ lets us predict a series of allowed filling fractions. It can be shown using the plasma analogy that the quasiparticle excitation density in the state \eqref{FQHE:eq:Lauglin_QPs} is $\rho_\mathrm{ex}=\chi\uglate{2\pi (2p-\chi/m)ml_B^2}^{-1}$. Since each excitation corresponds to an excess or shortage of $1/m$ electrons, their combined effect is a contribution to electron density of $\Delta\rho_\mathrm{el}=\rho_\mathrm{ex}/m$ above the density at filling factor $1/m$. Knowing that the electron density of the filled Landau level is $\rho_{\nu=1}=1/2\pi l_B^2$, we obtain the quasiparticle contribution to the filling factor as the ratio $\Delta \nu = \Delta \rho_\mathrm{el} / \rho_{\nu=1}$, and the total filling factor as
	\begin{equation}
		\nu = \frac{1}{m}+\Delta\nu =\frac{1}{m}+ \frac{\chi}{2pm^2 - \chi m} = \frac{1}{m-\frac{\chi}{2p}}
	\end{equation}
	The new "child" state at filling $\nu$ can have its own characteristic excitations which may also condense into a QHE state, giving rise to a hierarchy of QHE states at filling fractions
	\begin{equation}\label{FQHE:eq:hierarchy}
		\nu = \dfrac{1}{m-\dfrac{\chi_1}{2p_1-\dfrac{\chi_2}{2p_2-\dots}}}
	\end{equation}
	
	The hierarchies starting with $1/m$ for odd $m$ contain every rational number with an odd denominator \cite{haldanePandG}. It depends on the details of the system whether a QHE state at a certain filling is realized. In practice, the QHE states that are higher in the hierarchy (closer to the $1/m$ state) tend to be more stable, and similarly those with smaller $m$ and $p_1$, since interactions are more prominent at larger densities \cite{haldane1983}. Thus, the hierarchical construction explains the existence of FQHE at various fillings. An alternative way to explain the FQHE in terms of the IQHE of composite fermions has been proposed by Jain \cite{jain1989}. A composite fermion refers to a bound state of an electron and an even number of flux quanta, which feels a lower effective magnetic field, such that the filling factor is an integer. 
	
	We have ignored the role of spin in our discussion, which implies the assumption of complete spin polarization. This is sometimes justified due to the Zeeman effect, but other times it will be necessary to work with unpolarized or partially polarized states. It is also important to note that the FQHE states obtained in the way we described are Abelian, as the exchanges of their quasiparticle excitations merely produce a phase, as opposed to a noncommutative unitary transformation. However, other kinds of FQHE states are possible, sometimes at even-denominator fillings, and some some possibly non-Abelian \cite{nayak}. A notable example is the $\nu=5/2$ state, which was the first even-denominator FQHE state observed \cite{willett1987}, and is thought to be non-Abelian, specifically a spin-polarized Moore-Read Pfaffian state \cite{nayak}.

\subsection{Anyons}\label{sec:TQM:anyons}

	It is known that identical particles in quantum mechanics are indistinguishable, which implies that an exchange of two particles cannot affect the physical state in any measurable way. However, this requirement does not prevent a wave function from acquiring a phase factor. There is a common naive argument that since two subsequent exchanges of a pair of particles bring the system back to the initial configuration, the physical exchange must be analogous to a double permutation of particle labels. Since a double permutation cannot affect the wave function, it follows that the effect of a single permutation is multiplication by a square root of $1$, i.e $\pm1$. Therefore, the wave functions should either be symmetric or antisymmetric under an exchange of identical particles. The former case corresponds to bosons, while the latter corresponds to fermions.
	
	The effect of the requirements of (anti)symmetry on quantum statistics was already recognized in the 1920s, and for the better part of the 20th century, it was thought that Bose and Fermi statistics were the only two possibilities. This is unsurprising, given that the conclusion is true in 3D, and consequently, most easily accessible particles obey one or the other. It turns out, however, that the conclusion does not hold in general. It is now understood that the reason it holds in 3D is of topological nature, and that different possibilities arise in 2D. The significance of the difference between 3D and 2D space was recognized in 1977 by Leinaas and Myrheim \cite{LeinaasMyrheim}, and the name anyon for a particle obeying exotic statistics was coined by Frank Wilczek in 1982 when he showed that objects composed of a charge bound to a magnetic flux may behave unlike bosons or fermions \cite{wilczek1982}. 
	
	\begin{figure}[h]\centering
		\includegraphics[width=.95\textwidth]{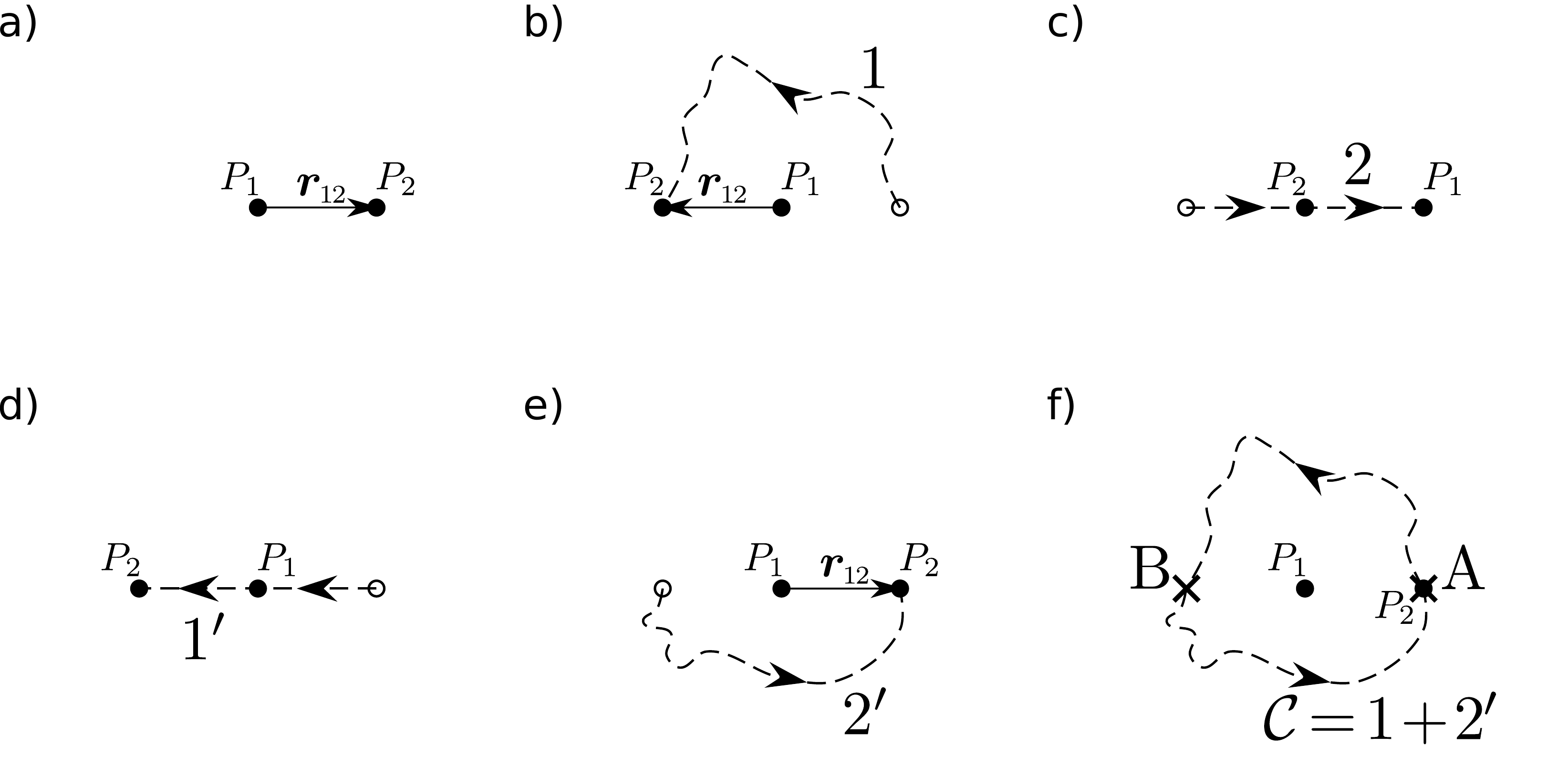}
		\caption[Anyon exchange process]{a)-e) Stepwise depiction of a process that achieves two subsequent exchanges of particles $P_1$ and $P_2$ in the anticlockwise sense. The dashed lines represent the paths taken during the current step. Solid (empty) circles represent the particle positions after (before) each step is performed, and $\vr_{12}$ is the relative coordinate vector.
			a) The starting position. b) $\vr_{12}$ is inverted by taking $P_2$ along path $1$. c) Rigid translation (path $2$) of both points so they match each others starting positions, completing single exchange. d) Rigid translation in the opposite direction (path $1'$). e) $\vr_{12}$ is inverted again by taking $P_2$ along path $2'$.
			f) Complete path of double exchange $\mathcal C$, excluding rigid translations. }
		\label{Anyons:fig:exchange}
	\end{figure}
	\subsubsection{Topology and exchange statistics}
	
	We now describe how topology affects the exchange phases (or transformations) in 2D and 3D. We consider a pair of hardcore\footnote{Hardcore repulsion implies a diverging repulsive potential for particle configurations with two or more particles in the same place, such that these configurations are effectively excluded from the configuration space (see eq. \eqref{Anyons:eq:confspace}).} particles lying in some plane (xy), and we ignore all physics except for the exchange statistics. Out of an infinite number of ways by which a single exchange may be achieved, we choose a two-step process depicted in \cref{Anyons:fig:exchange} a)-c). In step $1$, the relative coordinate is inverted ($\vr_{12}\rightarrow -\vr_{12}$) by taking particle $P_2$ in the anticlockwise sense (relative to particle $P_1$) along a path lying in the xy-plane, and terminating on the other side of $P_1$. Then in step $2$, both particles are rigidly translated so that the new position of each corresponds to the starting position of the other. A second exchange (\cref{Anyons:fig:exchange} d)-e)) may then be achieved by first undoing the rigid translation (step $1'$), and then completing the anticlockwise path of particle $P_2$, until it is back in its starting position (step $2'$). Clearly, we may get rid of both rigid translations when considering a double exchange (\cref{Anyons:fig:exchange} f)), and conclude that taking one of the particles along an anticlockwise path ($\mathcal C=1+2'$) around the other should be considered an equivalent process. If we assume that the exchange statistics is a topological effect, the final result should only depend on the winding of the path of one particle around the other.
	The path taken was mostly chosen arbitrarily in the xy-plane. The nonarbitrary aspects are the encirclement (anticlockwise winding) and the two fixed points marked in \ref{Anyons:fig:exchange} f). The point A is fixed by the initial and final conditions, but the point B, marking a single exchange, may be relaxed without affecting the result. 
	
	\begin{figure}[htb]\centering
		\includegraphics[width=.95\textwidth]{{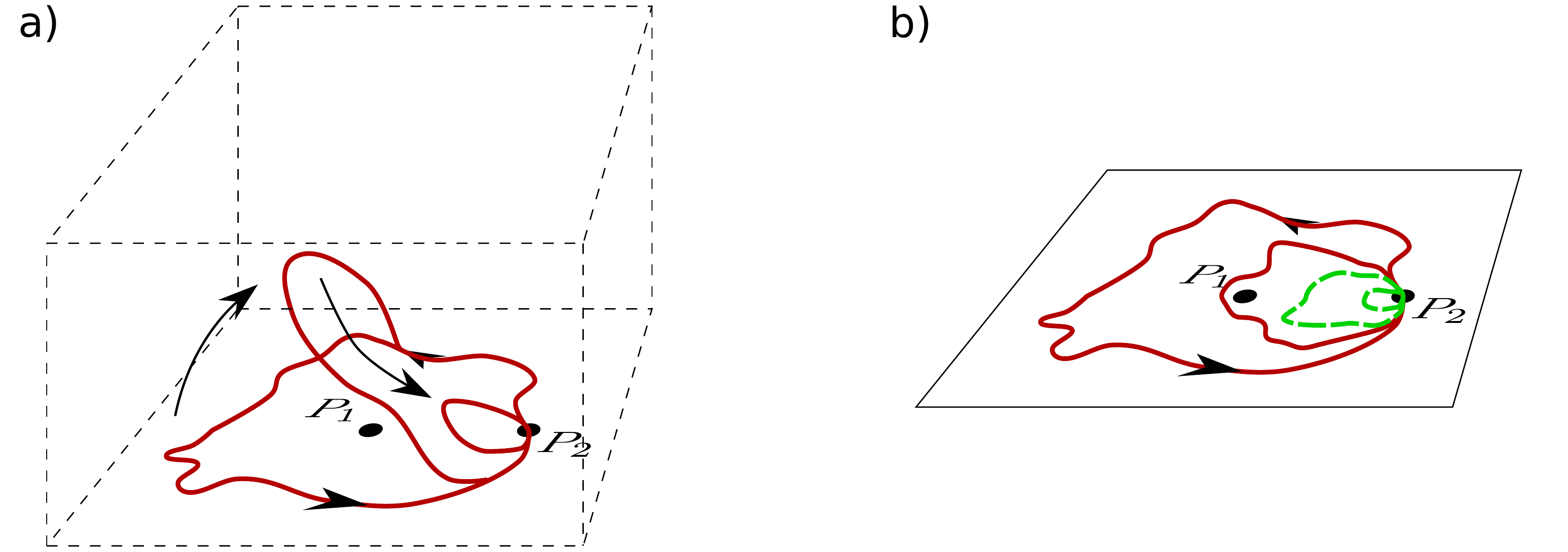}}
		\caption[Difference in topology between 3D and 2D space with an excluded point]{Difference in topology between 3D and 2D space with an excluded point ($P_1$). a) All loops starting at $P_2$ may be continuously contracted into $P_2$ in 3D. b) In 2D, the red loops wind around $P_1$ once, and, unlike the green loops, cannot be continuously deformed into $P_2$. }
		\label{Anyons:fig:exch_homotopy}
	\end{figure}
	
	Now, the difference between different dimensional spaces becomes clear. If the dimension is greater than $2$, the path can be deformed outside of the xy-plane, as shown in \cref{Anyons:fig:exch_homotopy} a). Therefore, it belongs to the class of loops that may be continuously contracted into a vanishing loop. Since a vanishingly small loop cannot wind around $P_1$, the conclusion is that winding is not well defined in 3D, and the double exchange must be equivalent to the identity operation (multiplication by 1). Only the $2\pi n$ phase shifts satisfy this requirement. On the other hand, as shown in \cref{Anyons:fig:exch_homotopy} b), the option of deforming $\mathcal C$ outside of the xy-plane is not available in 2D. Therefore, a continuous contraction into a point is not possible, since the loop would have to cross $P_1$, and such paths are forbidden due to hardcore repulsion. In topological terms (see \cite{stanescu}), loops winding around $P_1$ a different number of times belong to different homotopy classes. This is a consequence of the exclusion of $P_1$, making the space infinitely connected. The group of loop homotopy classes of a topological space is called the fundamental group. In 3D case,  the space is simply connected, and so this group is trivial, consisting of only the identity element (contractible loops). In 2D case, every integer has a corresponding homotopy class. The fundamental group is then isomorphic to the the $\mathbb Z$ group, i.e. integers under addition. The winding of a path is therefore well defined by its homotopy class, and the final result need not be trivial for nonzero winding numbers. Note that the particles $P_1$ and $P_2$ were not assumed to be identical. Mutual statistics may be defined for distinguishable particles in 2D when considering double exchanges \cite{nayak}.

	We may already draw some preliminary conclusions about single exchanges of indistinguishable particles. When dealing with one-component wave functions, it seems that a $\pm1$ factor, i.e. bosons or fermions, is the only possibility in 3D, since a double exchange entails winding which is not well defined. We run into no such constraint in 2D. However, to get a proper mathematical picture of single exchanges, we must look into the properties of the configuration space of the system. As before, all configurations with two particles in the same point must be excluded due to the hardcore condition. But furthermore, due to indistinguishability, we must identify all configurations that can be changed into one another by simply permuting particle labels. This can be easily visualised in the 2D, two-particle case, since the configuration space can be thought of as a product of the center-of-mass coordinate space and the relative coordinate space. The center-of-mass coordinate is unaffected, but the relative coordinate $\vr_{12}$ is identified with its inverse, $-\vr_{12}$. Visually, this means that a half of the plane is removed, e.g. bottom half, and the positive x-axis is folded onto the negative x-axis, creating a cone without the tip (as the origin is unphysical). In this topology, a single exchange is a closed loop around the cone, and is clearly noncontractible due to the missing tip. The fundamental group of this space (and by extension the full configuration space) is once again $\mathbb Z$, its elements distinguished by their winding number, but we must keep in mind that a single winding now corresponds to a single, instead of a double exchange. It is more difficult, but possible (see \cite{LeinaasMyrheim,lerda}), to visualise that the relative coordinate space in 3D is doubly connected, i.e. permitting contractible and a single kind of noncontractible loops. The latter corresponds to single, and the former to  double exchanges (equivalent to identity). The fundamental group is $\mathbb Z_2$, the group of integers $\{0,1\}$ under addition modulo 2, or equivalently, the permutation (symmetric) group of two elements $S_2$. 
	
\begin{figure}[htb]\centering
		\includegraphics[width=.95\textwidth]{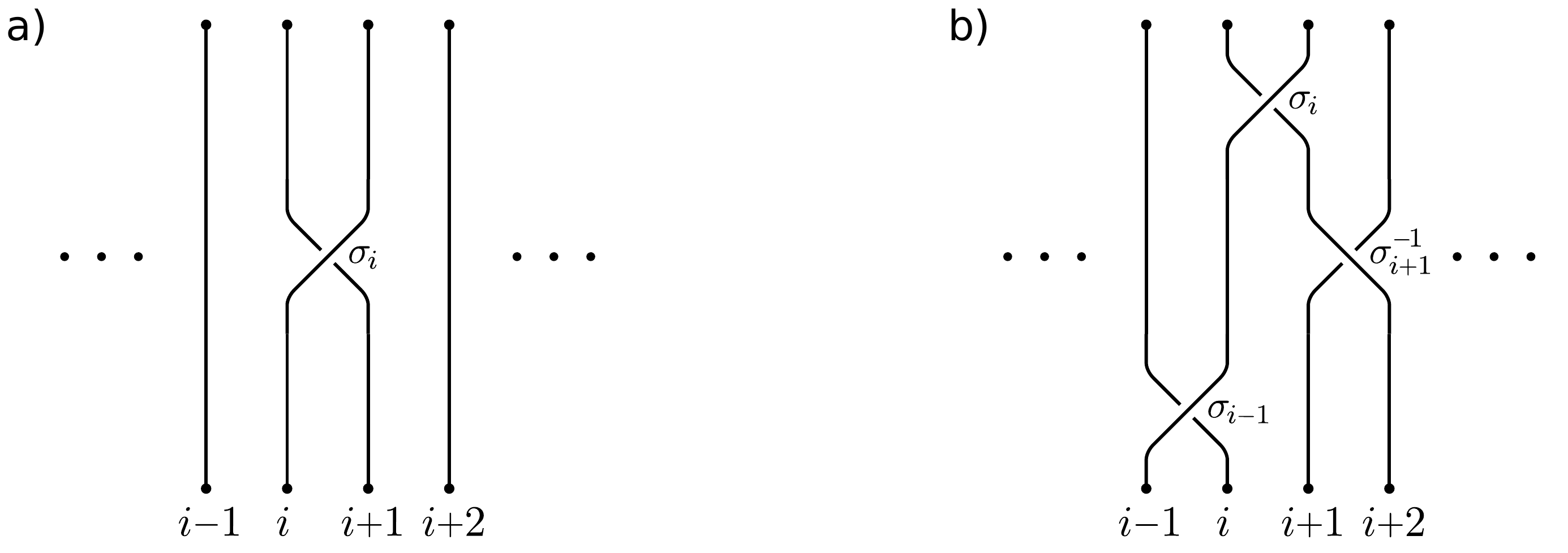}
		\caption[Schematic representation of braids]{A schematic representation of: a) The braid constructed by a single application of the $\sigma_i$ operator. b) Braid $\alpha=\sigma_i\sigma_{i+1}^{-1}\sigma_{i-1}$, constructed by a sequential application (product) of the braid group generators and their inverses. The strands may be interpreted as $(2+1)$D world lines, in which case the horizontal axis represents the 2D space, and the vertical axis represents time.}
		\label{Anyons:fig:braids}
	\end{figure}

	In the general case of $N$ particles in a $d$-dimensional space, the configuration space is denoted \cite{wu1984}
	\begin{equation}\label{Anyons:eq:confspace}
		M^d_N = \frac{(\mathbb R^d)^N-D}{S_N} ,
	\end{equation}
	Here, subtracting the generalized diagonal $D$, i.e. the set of configurations with $\vr_i=\vr_j$ for any $i\neq j$, imposes the hardcore condition, and "dividing" by the permutation group $S_N$ imposes the indistunguishability of identical particles by identifying configurations distinguished only by a particle permutation. The fundamental groups are known for this class of spaces (see \cite{lerda} and references therein) ,
	\begin{equation}\label{Anyons:eq:fundamental_group}
		\pi_1(M^d_N) =
		\begin{cases}
			 S_N & \textrm{for } d \geq 3,\\
			 B_N & \textrm{for } d = 2.
		\end{cases}
	\end{equation}
	where $B_N$ is a braid group of $N$ strands. Its elements are called braids and they can be obtained from the generating set $\sigma_1,\dots,\sigma_{N-1}$ satisfying
 	\begin{alignat}{3}\label{Anyons:eq:braid_rules}
		&\qquad &	\sigma_i \sigma_j &= \sigma_j \sigma_i	 &\qquad & \textrm{for }	|i-j|\geq 2 ,
		\nonumber\\
		&&	\sigma_i \sigma_{i+1} \sigma_i &= \sigma_{i+1} \sigma_i \sigma_{i+1}	&& \textrm{for } 1\leq i\leq N-2 .
 	\end{alignat}
 	A braid can be thought of as a specific way of connecting a sequence of $N$ fixed starting points to a sequence of $N$ fixed ending points by $N$ strands. Two braids are considered distinct only if they cannot be deformed into each other without any strands crossing, or moving any of the fixed points.
	The generator $\sigma_i$ can be thought of as the operation that winds (braids) the $i$-th and the $(i\!+\!1)$-th strand around each other in the positive sense so that their ends are exchanged, as shown in \cref{Anyons:fig:braids} a), and its inverse $\sigma_i^{-1}$ represents winding in the negative sense. Any braid $\alpha$ can be constructed by a product, i.e. sequential application, of the generators $\sigma_i$ and their inverses (see \cref{Anyons:fig:braids} b) for an example). 
	This is a very appealing picture, since each strand can be thought of as a world line of a particle in $(2+1)$ dimensions, representative of the equivalence class of those world lines it can be deformed into.  
	
	The infinite braid group $B_N$ is in some sense similar to the finite permutation group $S_N$ in that both describe exchanges of the elements of some set (particles). However, braids can keep track of the topology of the paths taken in $(2+1)$D, which may be nontrivial even if the final particle configuration matches the initial. On the other hand, in $(3+1)$D, an additional degree of freedom is available, and all world lines connecting matching configurations can be unwound. In this case, the only topological difference is between the $N!$ different configurations, which is encoded by the permutation group. This intuitively explains the result \eqref{Anyons:eq:fundamental_group}. Note that in the special case of $2$ particles, the braid group $B_2$ is isomorphic to the $\mathbb Z$ group, while the permutation group $S_2$ is isomorphic to the cyclic group $\mathbb Z_2$, as a consequence of the fact that any permutation is its own inverse. 

	This understanding of the topology of the configuration space can be applied to determine the outcome of particle exchanges. In the path integral framework, an amplitude for the system to transition from configuration $q$ at time $t$ to configuration $q'$ at $t'$ is obtained by summing over all paths with weights equal in magnitude, but their phases determined by the action. Since different path homotopy classes in a multiply connected space $M_N^d$ are not continuously connected, integration over paths must be done separately for each homotopy class $\alpha$ to obtain partial amplitudes
	\begin{equation}\label{Anyons:eq:partial_ampl}
		K_\alpha (q',t';q,t) = \int_{q(\tau)\in \alpha} \mathcal D q(\tau) \exp(\frac{i}{\hbar}\int_t^{t'} \dd s L[q(s),\dot{q}(s)]),
	\end{equation}
 	$L(q,\dot q)$ being the Lagrangian, and $\int_{q(\tau)\in \alpha} \mathcal D q(\tau)$ denoting integration over all paths $q(\tau)$ satisfying $q(t)=q$ and $q(t')=q'$. To obtain the total amplitude, the partial amplitudes are summed with weights \cite{laidlaw1971}
 	\begin{equation}\label{Anyons:eq:total_ampl}
 		K(q,t;q',t')=\sum_{\alpha\in \pi_1(M^d_N)} \rho(\alpha) K_\alpha (q,t;q',t').
	\end{equation}
	The total amplitude is the propagator that determines the wave function at $t'$, given the initial condition at $t$ according to $\psi(q',t') = \int_{M^d_N} \dd q K(q',t';q,t)\psi(q,t)$. Clearly, the weights $\rho(\alpha)$ are expressed in the final result, and they cause it to be dependent on the topology of the contributing paths. To find the possible values for $\rho(\alpha)$, we require that the transition amplitudes combine according to $K(q'',t'';q,t)=\int_{M^d_N}\dd q' K(q'',t'';q',t') K(q',t';q,t)$ (see \cite{lerda,laidlaw1971}), which leads to
	\begin{equation}\label{Anyons:eq:representation}
		\rho(\alpha)\rho(\beta)=\rho(\alpha\beta),\quad \forall \alpha,\beta\in\pi_1(M^d_N).
	\end{equation}
	This is a defining relation for a representation of the fundamental group $\pi_1(M^d_N)$. Assuming scalar quantum mechanics, $\rho$ can only be a one-dimensional representation. It can also be shown \cite{laidlaw1971} that the representation is unitary (i.e. a phase factor). The only unitary scalar representations of the permutation group $S_N$ are the trivial ($\rho(\alpha)=1$), and the sign ($\rho(\alpha)=\pm 1$, depending on permutation parity) representations. Therefore, in 3D, exchange of identical particles is governed by one of the two possible representations, resulting respectively in bosonic or fermionic statistics. 
	On the other hand, the braid group $B_N$ has an infinite number of possible unitary scalar representations. These can be parametrized by the statistical parameter $\theta\in\cointerval{0}{2}$, and it can be shown by employing \eqref{Anyons:eq:representation} and the first relation of \eqref{Anyons:eq:braid_rules} that \cite{wu1984}
	\begin{equation}\label{Anyons:eq:abel}
		\rho_\theta(\sigma_i)=e^{i\theta\pi}
	\end{equation}		
	 for any member of the generating set of $B_N$. The group representation satisfies ${\rho(\sigma_i)\rho(\sigma_i^{-1})=1}$, so ${\rho(\sigma_i^{-1})=e^{-i\theta\pi}}$.  We conclude that in 2D scalar quantum mechanics, a physical exchange is not equivalent to the unphysical permutation of particle labels. Instead it produces a phase shift dependent on the sense of the exchange path, with a spectrum of possible values. Bosonic and fermionic statistics correspond respectively to the special cases $\theta=0$ and $\theta=1$. 
	
	Let us briefly note that the $\pi$ in the exponent in \eqref{Anyons:eq:abel} comes from the change in the relative angle between particles $i$ and $i+1$ upon a single exchange. The representation for an arbitrary braid $\alpha$ involves summing over the changes of the relative angles between all particles, and may be expressed as \cite{wu1984,lerda}
	\begin{equation}
		\rho_\theta(\alpha)=\exp[i\theta\sum_{i<j}\int_t^{t'} d\tau \dv{\tau}\varphi_{ij}^{(\alpha)}(\tau)],
	\end{equation}
	where braiding operations are performed between time $t$ and $t'$. This allows us to incorporate the weights $\rho(\alpha)$ of \eqref{Anyons:eq:total_ampl} into the action in \eqref{Anyons:eq:partial_ampl}, by defining an effective Lagrangian ${L'=L+\hbar\theta \sum_{i<j} \dv{\tau}\varphi_{ij}^{(\alpha)}(\tau)}$. This suggests an alternative interpretation of fractional statistics as a form of topological interaction. The Lagrangian $L'$ may, for example, be that of the Chern-Simons theory, which is a topological quantum field theory effectively describing the low-energy properties of the Laughlin FQHE states discussed in the previous section \cite{stanescu,nayak}. 
	
	Since the representation \eqref{Anyons:eq:abel} is scalar, it must be commutative. Hence, it describes the so-called Abelian fractional (anyonic) statistics. However, braid group representations that govern particle exchanges are not always scalar. In case  wave functions are multiplets, higher dimensional unitary representations may be needed. If there is a set of $g$ degenerate states describing $N$ particles at fixed positions, the effect of braiding particles may be described by a $g\cross g$ unitary matrix \cite{nayak}
	\begin{equation}\label{Anyons:eq:braiding}
		\psi_{a}\rightarrow [\rho(\alpha)]_{ab} \psi_b,
	\end{equation}
	where $\{\psi_a|a=1,\dots,g\}$ is an orthonormal basis set of the degenerate subspace. Besides phase shifts, these unitary transformations may rotate the state $\psi_a$ within the degenerate subspace. The commutator $[\rho(\sigma_i),\rho(\sigma_j)]$ does not vanish in general, so these representations may describe non-Abelian statistics. Of course, permutation groups also have higher-dimensional representations which give rise to parastatistics, but this is not of fundamental importance as it can be viewed as ordinary statistics for particles with an additional quantum number \cite{nayak}.
	
	\subsubsection{Anyon fusion}	
	Two or more anyons can be fused, i.e. brought together, and considered a single particle with combined quantum numbers \cite{nayak}. This includes combined statistics (topological quantum number), which manifests if such fused anyons are braided. This means that systems allowing for anyons normally allow for more than one species, and to be able to characterize an anyonic system, a list of possible species is needed. Furthermore, an outcome of fusion is not always unique. Similar to addition of angular momenta, different fusion channels may be possible. These are governed by fusion rules denoted as
	\begin{equation}\label{Anyons:eq:fusion_rules}
		\phi_a\cross\phi_b = \sum_c N_{ab}^c\phi_c,
	\end{equation}
	meaning that when an anyon of species $a$ is fused with an anyon of species $b$, an anyon of each allowed species $c$ may be obtained in $N_{ab}^c$ distinct ways. When $a$ and $b$ are Abelian, there is only one possible outcome $c$ with $N_{ab}^c=1$, and $N_{ab}^{c'}=0$ when $c'\neq c$. For example if two anyons obeying statistics $\theta$ are fused, the outcome is a $4\theta$ anyon. This can be explained by noting that in braiding one fused anyon with another, each of its two constituent $\theta$ anyons is braided with the other two $\theta$ anyons, thus accumulating the $2\theta$ phase twice. \label{page:fusion}
	
	When viewed in isolation, a system of $N$ anyons fuses, i.e. observes a combined statistics. It may be prepared in some linear combination of states with different topological quantum numbers allowed by fusion rules. For example, since vacuum has the statistics of a topologically trivial (bosonic) particle, a system with a definite, trivial quantum number may be prepared by creating $N/2$ pairs from vacuum. To understand the fusion of $N$ anyons, we start by dividing the system into arbitrary pairs, which fuse into intermediate channels, and repeat the process until the final result is reached. This can be visualized as a fusion tree. For example, let us consider a system of three anyons of species $a$, $b$, and $c$. We may first fuse $a$ and $b$ into an intermediate channel $i$, and then fuse $i$ with the remaining anyon $c$ into the channel $d$:
	\begin{equation}\label{Anyons:eq:fusion_tree}
	\begin{aligned}
		\begin{tikzpicture}[x=1cm,y=0.5cm]
			\tikzset{interm/.style={circle}, inner sep=2pt}
			\draw (0,3)--(0.5,2); \draw (1,3)--(0.5,2);
				\draw (0,3) node[anchor=south]{$a$};
				\draw (1,3) node[anchor=south]{$b$};
				\draw (2,3) node[anchor=south]{$c$};
			\draw (0.5,2) --node[interm, anchor= north east]{$i$} (1,1); \draw (2,3)--(1,1);
			\draw (1,1)--(1.5,0);
				\draw (1.5,0) node[anchor=north]{$d$};
		\end{tikzpicture}
		\end{aligned}.
	\end{equation}
	Let us now assume that $a$, $b$, and $c$, as well as the outcome $d$ are known in advance. This means the total fusion space is restricted to a particular subspace, namely the fusion space with total quantum number $d$. In this case, $i$ may be any species consistent with the fusion rules, i.e. for which $N_{ab}^i\neq 0$ and $N_{ic}^d\neq 0$. The number of such species determines the dimension of the fusion space. The set of states \eqref{Anyons:eq:fusion_tree} for every allowed $i$ forms the basis. Clearly, there is more than one way to divide the system into pairs. For example, $b$ may initially fuse with $c$ to give the intermediate anyon $j$, which then fuses with $a$. Since this is not a physical change, the fusion space must be the same, and states with different $j$ comprise another basis. The basis change is parametrized by the so-called $F$ matrix
	\begin{equation}
	\begin{aligned}
		\begin{tikzpicture}[x=1cm,y=0.5cm]
			\tikzset{interm/.style={circle}, inner sep=2pt}
			\draw (0,3)--(0.5,2); \draw (1,3)--(0.5,2);
				\draw (0,3) node[anchor=south]{$a$};
				\draw (1,3) node[anchor=south]{$b$};
				\draw (2,3) node[anchor=south]{$c$};
			\draw (0.5,2) --node[interm, anchor=north east]{$i$} (1,1); \draw (2,3)--(1,1);
			\draw (1,1)--(1.5,0);
				\draw (1.5,0) node[interm, anchor=north]{$d$};
		\end{tikzpicture}
	\end{aligned}
		= \sum_j [F_{abc}^d]_{ij}
	\begin{aligned}
		\begin{tikzpicture}[x=1cm,y=0.5cm]
			\tikzset{interm/.style={circle}, inner sep=2pt}
			\draw (2,3)--(1.5,2); \draw (1,3)--(1.5,2);
				\draw (0,3) node[anchor=south]{$a$};
				\draw (1,3) node[anchor=south]{$b$};
				\draw (2,3) node[anchor=south]{$c$};
			\draw (0,3) -- (1,1); \draw (1.5,2) --node[interm, anchor=south east]{$j$} (1,1);
			\draw (1,1)--(1.5,0);
				\draw (1.5,0) node[anchor=north]{$d$};
		\end{tikzpicture}
	\end{aligned},
	\end{equation}
	where the indices within the brackets enumerate the initial and final anyons, while the ones outside enumerate the intermediate anyons. Basis transforamtions for larger fusion trees may still be represented by $F$ matrices with the same number of indices by treating subtrees independently. 
	Assuming degenerate fusion space and recalling \eqref{Anyons:eq:braiding}, we conclude that braiding may induce rotations within the fusion space. To describe braiding in this fusion framework, we require another ingredient, namely the $R$ matrices, which describe the effect of exchanging (in the positive sense) two particles in a fixed fusion channel. When all $N_{ab}^c\leq 1$, the $R$ matrices reduce to phases $R_{ab}^c$. This must be true, since the topological quantum number of the pair (i.e. their fusion channel) is fixed as long as the two can be viewed as an isolated system. Therefore, braiding a pair of anyons in a fixed fusion channel does not change the state in the fusion space. However, braiding one of the two with a third anyon may induce rotations, as we will demonstrate in the following. Together with the list of particle species and the fusion rules, the $F$ and $R$ matrices specify the braiding statistics of a given anyon model. We note that the $R$ and $F$ matrices are constrained by consistency relations, namely the pentagon and hexagon equations (see \cite{tong}).
	
	We will now describe a simple braiding operation which rotates a state in the four-anyon fusion space on a very simple example called the Ising anyon model (see e.g. \cite{stanescu}). This model allowes three species of anyons, namely, the trivial (vaccum) anyon $\bm 1$, the $\sigma$ anyon, and the $\psi$ anyon (fermion). The fusion rules are
	\begin{equation}\label{Anyons:eq:ising_rules}
	\begin{aligned}
		\sigma\cross\sigma& = \bm 1 + \psi, \quad \sigma\cross\psi=\sigma, \quad \psi\cross\psi=\bm 1, \\
		\bm 1 \cross x &= x, \quad \text{for } x=\bm 1, \sigma, \psi .
	\end{aligned}
	\end{equation}
	We consider two pairs of $\sigma$ anyons, $1$,$2$, and $3$,$4$, created from the vacuum. By \eqref{Anyons:eq:ising_rules}, two $\sigma$ anyons may fuse in the $\bm 1$ or the $\psi$ channel, but since they are created from vacuum, we know that $1$ and $2$ must fuse into $\bm 1$. Furthermore, the total fusion channel of the four-anyon system is also trivial, and by \eqref{Anyons:eq:ising_rules}, this means that $3$ and $4$ must always fuse in the same channel as $1$ and $2$. To change the fusion channel of $1$ and $2$, we must braid two particles from different pairs, e.g. $2$ and $3$, twice\footnote{Braiding $2$ and $3$ once is an exchange and would result in fusing $1$ with $3$ instead of $2$..} before measuring the quantum number of the pair. To determine the outcome, we must write the matrix $\rho(\sigma_{23})$ in terms of the $F$ and $R$ matrices, which can be obtained from the consistency relations. 
	We first rearrange the fusion tree by grouping particle $3$ with the fusion product of $1$ and $2$ instead of particle $4$:
	\begin{equation}\label{Anyon:eq:braid_step1}
		\begin{aligned}
		\begin{tikzpicture}[x=1cm,y=0.5cm]
			\tikzset{interm/.style={circle}, inner sep=2pt}
			\draw (0,3)node[interm, anchor=north ]{$\sigma$}--(0.5,2); 
			\draw (1,3)node[interm, anchor=north ]{$\sigma$}--(0.5,2);
			\draw (2,3)node[interm, anchor=north ]{$\sigma$}--(2.5,2); 
			\draw (3,3)node[interm, anchor=north ]{$\sigma$}--(2.5,2);
				\draw (0,3.5) node[anchor=south]{$1$};
				\draw (1,3.5) node[anchor=south]{$2$};
				\draw (2,3.5) node[anchor=south]{$3$};
				\draw (3,3.5) node[anchor=south]{$4$};
			\draw (0.5,2)-- node[interm, anchor=north east]{$\bm 1$} (1.5,0); 
			\draw (2.5,2)-- node[interm, anchor=north west]{$\bm 1$}(1.5,0);
			\draw (1.5,0)--(2,-1) node[interm, anchor=north east]{$\bm 1$};
		\end{tikzpicture}
		\end{aligned}
		\xrightarrow{\makebox[2cm]{\scaleto{F_{\bm1\sigma\sigma}^\sigma}{15pt}\hspace{0.2cm}}}	
		\begin{aligned}
		\begin{tikzpicture}[x=1cm,y=0.5cm]
			\tikzset{interm/.style={circle}, inner sep=2pt}
			
			\draw (0,3)node[interm, anchor=north ]{$\sigma$}--(0.5,2);
			\draw (1,3)node[interm, anchor=north ]{$\sigma$}--(0.5,2);
			\draw (2,3)--node[interm, anchor=north west]{$\sigma$}(1,1); 
				\draw (0,3.5) node[anchor=south]{$1$};
				\draw (1,3.5) node[anchor=south]{$2$};
				\draw (2,3.5) node[anchor=south]{$3$};
				
			\draw (0.5,2)-- node[interm, anchor=north east]{$\bm 1$} (1,1);
			\draw (1,1)-- node[interm, anchor=north east]{$\sigma$} (1.5,0); 
			\draw[dashed] (3,3)--node[interm, anchor=north west ]{$\sigma$}(1.5,0);
				\draw (3,3.5) node[anchor=south]{$4$};
			\draw[dashed] (1.5,0)--(2,-1) node[interm, anchor=north east]{$\bm 1$};
		\end{tikzpicture}
		\end{aligned}.
	\end{equation}
	This is a trivial transformation, as there is only one possible fusion tree on the right. We will omit the dashed branches in the following, and only consider the fusion of three $\sigma$ anyons with the total quantum number $\sigma$, since all nontrivial effects of the braiding operation in question are manifested in this subtree. We further rearrange so that $2$ fuses with $3$ instead of $1$:
	\begin{equation}\label{Anyon:eq:braid_step2}
		\begin{aligned}
		\begin{tikzpicture}[x=1cm,y=0.5cm]
			\tikzset{interm/.style={circle}, inner sep=2pt}
			
			\draw (0,3)node[interm, anchor=north ]{$\sigma$}--(0.5,2);
			\draw (1,3)node[interm, anchor=north ]{$\sigma$}--(0.5,2);
			\draw (2,3)--node[interm, anchor=north west]{$\sigma$}(1,1); 
				\draw (0,3.5) node[anchor=south]{$1$};
				\draw (1,3.5) node[anchor=south]{$2$};
				\draw (2,3.5) node[anchor=south]{$3$};			
			\draw (0.5,2)-- node[interm, anchor=north east]{$\bm 1$} (1,1);
			\draw (1,1)-- node[interm, anchor=north east]{$\sigma$} (1.5,0); 
		\end{tikzpicture}
		\end{aligned}
		\xrightarrow{\makebox[2cm]{\scaleto{F_{\sigma\sigma\sigma}^\sigma}{15pt}\hspace{0.2cm}}}
		\begin{aligned}
		\begin{tikzpicture}[x=1cm,y=0.5cm]
			\tikzset{interm/.style={circle}, inner sep=2pt}
			
			\draw (0,3)--node[interm, anchor=north east]{$\sigma$}(1,1);
			\draw (1,3)node[interm, anchor=north ]{$\sigma$}--(1.5,2);
			\draw (2,3)node[interm, anchor=north ]{$\sigma$}--(1,1); 
				\draw (0,3.5) node[anchor=south]{$1$};
				\draw (1,3.5) node[anchor=south]{$2$};
				\draw (2,3.5) node[anchor=south]{$3$};		
			\draw (1.5,2) --node[interm, anchor=north west]{$i$} (1,1);
			\draw (1,1)-- node[interm, anchor=north east]{$\sigma$} (1.5,0); 
		\end{tikzpicture}
		\end{aligned}.
	\end{equation}		
	The intermediate anyon $i$ may be either $\bm 1$ or $\psi$. We are now in position to exchange the particles $2$ and $3$ twice, which involves the phases $[R_{\sigma\sigma}^i]^2$. To complete the braid we move in the direction opposite of \eqref{Anyon:eq:braid_step2} by applying $[F_{\sigma\sigma\sigma}^\sigma]^{-1}$.
	To determine the braid matrix, we only need one fusion and one rotation matrix
	\begin{align}
		F_{\sigma\sigma\sigma}^\sigma = 	\frac{1}{\sqrt 2}	
		\begin{pmatrix}
			1 & 1 \\
			1 & -1
		\end{pmatrix},
		&&
		R_{\sigma\sigma} = 
		\begin{pmatrix}
			R_{\sigma\sigma}^{\bm 1} & 0 \\
			0 & R_{\sigma\sigma}^\psi
		\end{pmatrix},
	\end{align}
	with the phase shifts $R_{\sigma\sigma}^{\bm 1} = e^{-\pi i/8}$, $R_{\sigma\sigma}^\psi = e^{3\pi i/8}$ \cite{stanescu}.
	 This is done by taking the product of all transformations
	\begin{equation}
		\rho(\sigma_{23}) = [F_{\sigma\sigma\sigma}^\sigma]^{-1} [R_{\sigma\sigma}]^2 F_{\sigma\sigma\sigma}^\sigma = 
		e^{-i\pi/4}
		\begin{pmatrix}
			0 & 1\\
			1 & 0
		\end{pmatrix}.
	\end{equation}
	The anti-diagonal structure of the matrix indicates that braiding $2$ and $3$ results in changing the fusion channel of $1$ and $2$ from $\bm 1$ to $\psi$. It is then straightforward to reattach the 3-anyon tree to the larger 4-anyon tree, and transform back to the original particle grouping $(1,2)$ and $(3,4)$. Graphically, the braid is represented as
	\begin{equation}
		\begin{aligned}
		\begin{tikzpicture}[x=1cm,y=0.5cm]
			\tikzset{interm/.style={circle}, inner sep=2pt}
			\draw (0,3)node[interm, anchor=north ]{$\sigma$}--(0.5,2); 
			\draw (1,3)node[interm, anchor=north ]{$\sigma$}--(0.5,2);
			\draw (2,3)node[interm, anchor=north ]{$\sigma$}--(2.5,2); 
			\draw (3,3)node[interm, anchor=north ]{$\sigma$}--(2.5,2);
				\draw (0,3.5) node[anchor=south]{$1$};
				\draw (1,3.5) node[anchor=south]{$2$};
				\draw (2,3.5) node[anchor=south]{$3$};
				\draw (3,3.5) node[anchor=south]{$4$};
			\draw (0.5,2)-- node[interm, anchor=north east]{$\bm 1$} (1.5,0); 
			\draw (2.5,2)-- node[interm, anchor=north west]{$\bm 1$}(1.5,0);
			\draw (1.5,0)--(2,-1) node[interm, anchor=north east]{$\bm 1$};
		\end{tikzpicture}
		\end{aligned}
		\xrightarrow{\makebox[2cm]{\scaleto{\rho(\sigma_{23})}{15pt}\hspace{0.2cm}}}
		\begin{aligned}
		\begin{tikzpicture}[x=1cm,y=0.5cm]
			\tikzset{interm/.style={circle}, inner sep=2pt}
			\draw (0,3)node[interm, anchor=north ]{$\sigma$}--(0.5,2); 
			\draw (1,3)node[interm, anchor=north ]{$\sigma$}--(0.5,2);
			\draw (2,3)node[interm, anchor=north ]{$\sigma$}--(2.5,2); 
			\draw (3,3)node[interm, anchor=north ]{$\sigma$}--(2.5,2);
				\draw (0,3.5) node[anchor=south]{$1$};
				\draw (1,3.5) node[anchor=south]{$2$};
				\draw (2,3.5) node[anchor=south]{$3$};
				\draw (3,3.5) node[anchor=south]{$4$};
			\draw (0.5,2)-- node[interm, anchor=north east]{$\psi$} (1.5,0); 
			\draw (2.5,2)-- node[interm, anchor=north west]{$\psi$}(1.5,0);
			\draw (1.5,0)--(2,-1) node[interm, anchor=north east]{$\bm 1$};
		\end{tikzpicture}
		\end{aligned}.
	\end{equation}
	%
	
	\subsubsection{Topological quantum computation}
		The most significant reason for studying anyons is the potential application of non-Abelian anyons in realization of fault tolerant quantum computers. The idea behind quantum computation is to use the resources of the Hilbert space to solve computational problems by manipulating a set of qubits  according to appropriate quantum algorithms. A qubit is a two-state quantum systems serving as the unit of quantum information. Unlike classical bits, which can only assume states $0$ and $1$, qubits can assume any superposition state $a\ket0+b\ket1$. Different qubits can be entangled, and the state of an $N$-qubit system then belongs to a $2^N$-dimensional Hilbert space. Simulations of quantum systems whose complexity scales exponentially with system size, an intuitively appealing use for such a machine, was indeed among the earliest envisaged uses \cite{feynman1982}. On the other hand, probably the most famous quantum algorithm is the Shor's algorithm \cite{shor1994} for integer factorization. The time of execution scales favourably with the number of digits when compared to the fastest known classical algorithm. It is conjectured that for a certain class of problems, quantum algorithms exist that outperform any possible classical algorithm.
	
	In essence \cite{nayak}, quantum computation involves initializing a system in some known state $\ket{\psi_0}$ (input), and unitarily evolving the state according to a Hamiltonian $H(t)$, encoding the algorithm. The final state $U(t)\ket{\psi_0}$ is the output. While the state of a classical computer traverses only one trajectory during computation, the quantum computer traverses all the available states, since the qubits go through superpositions of the definite (classical) states between the final measurement and initialization. This is commonly likened to parallel computing, but there is a distinction, since there is only one final state, determined by the coherent sum of all trajectories.
	The main difficulty with quantum computation is the tendency of pure superposition states of many-qubit systems to become entangled with the environment. This causes decoherence through which quantum information is lost to the environment. This makes it difficult to achieve fault tolerance, i.e. the possibility of error correction at a rate faster then their occurrence. 
	
	One of the schemes for achieving fault tolerance is through topological quantum computation, employing non-Abelian anyons. As we have seen, multi-anyon system states belong to a fusion space and may be rotated within that space by certain braiding operations. An appropriate multi-anyon system may assume the role of a qubit in a quantum computer. The anyons may be created at the beginning, and fused at the end of a computation, thus forming links whose topology determines the output. The resistance to errors due to interaction with the environment stems from the fact that information is stored nonlocally in the links. Therefore, local perturbations have vanishing matrix elements within the fusion space and so the state of a qubit cannot be affected by local interactions with the environment \cite{nayak}. The only way to perform a computation is to affect the topology by braiding the anyons, and the only way errors can occur is via unwanted braiding with unaccounted anyons, either  thermally excited or those trapped by disorder. It is hoped that these sources of error are not detrimental, since unwanted excitations may be reduced by working at energies lower than the excitation gap and with weakly disordered systems.
	
	As an example, the $(1,2)$ pair of the four-anyon system discussed above may be regarded as a simple qubit. Indeed, while discussing the $\sigma_{23}$ braid, we have essentially described a NOT gate, changing the fusion channel from $\bm 1$ to $\psi$, which may now respectively correspond to the $\ket 0$ and $\ket 1$ qubit states. The anyons of the $5/2$ Moore-Read FQHE state are very similar to the Ising anyons discussed above. Should the physical $5/2$ state be confirmed to be the Moore-Read state, it may be a candidate for quantum computation. However, this kind of anyonic system cannot perform all possible unitary transformations, meaning that it must be supplemented with extra nontopological gates to be capable of universal quantum computation \cite{nayak}. On the other hand, some other anyon models, such as the Fibonacci model, containing two anyon species and obeying a single nontrivial fusion rule $\tau\cross\tau=\bm 1 + \tau$, are capable of universal topological quantum computation \cite{nayak}.


\subsection{Symmetry-protected topological states}\label{sec:TQM:SPT}
	As mentioned in the introduction to this section, gapped short-range entangled systems lacking topological order may also exist in topologically nontrivial phases. This requires that the Hamiltonian belongs to a certain symmetry class. These symmetry-protected topological phases then correspond to the regions in the phase diagram spanned by those parameters of the Hamiltonian that do not affect the symmetry class. 
	If the required symmetries are broken, the different topological phases merge, as they can be connected to the topologically trivial phase by adiabatic deformations without closing the gap (i.e. going through a topological phase transition). The trivial phase is topologically identical to the atomic insulator, as well as the vacuum. 
	%
	Unlike the topologically ordered phases, which cannot be connected to the vacuum by any gap-preserving deformation, the SPT phases do not support anyonic excitations in the bulk, but they still exhibit a variety of nontrivial boundary physics. 
	
	SPT phases may be interacting or noninteracting. Adding interactions to the noninteracting SPT states can lead both to merging of distinct noninteracting topological phases and to emergence of new phases. Topologically protected $(d\!-\!1)$D boundary states are present in all cases, and are always found in the band gap in noninteracting SPT phases. On the other hand, they need not be gapless in the interacting case, provided the boundary either spontaneously breaks a symmetry that enables the existence of the SPT phase in question (in 2D or higher dimensions), or that it carries an intrinsic topological order (in 3D or higher dimensions) \cite{stanescu}. 
	
	An example of a noninteracting SPT phase is the quantum spin Hall effect (QSHE). The QSHE is the spin current analogue of the charge current QHE. In 2005, a topologically nontrivial 2D QSHE system was proposed by Kane and Mele \cite{kane2005a,kane2005b}. It is a spin-$1/2$ system on a honeycomb lattice with a spin-orbit interaction. The system is similar to the Haldane model, but the alternating magnetic field (with a net-zero flux) felt by the spin up and spin down components is opposite. Additionally, processes that couple the two components are allowed. In the special case of uncoupled spins ($S_z$ conservation), this amounts to two copies of the Haldane model with broken time-reversal symmetry leading to the opposite Chern numbers $C_{\uparrow\downarrow}$. Even though the total system preserves TRS, and hence the total Chern number is ${C_\uparrow+C_\downarrow=0}$, each spin component has $|C_{\uparrow\downarrow}|$ topologically protected gapless chiral edge modes. The modes carry no net charge current, since they come in oppositely propagating pairs, but instead carry a robust spin current. Such a pair makes up a so-called helical edge mode, and the existence of one such topologically protected gapless mode is a defining feature of the QSH phase. The QSH phase can survive even if $S_z$ is not conserved, provided the gap remains open and TRS is not broken. The gapless edge mode is protected by TRS, since the Kramers' theorem guarantees the existence of a distinct state with momentum $-\vk$ and energy $E$ for each state with momentum $\vk$ and energy $E$ in any TR-symmetric fermionic system. This means that the edge states are doubly degenerate for TR-invariant momenta\footnote{A TR-invariant momentum $k_{TR}$ is separated by a reciprocal lattice vector from its time-reversed counterpart $-k_{TR}$.}, implying the gap must close at these momenta.
	
	The QSHE is a topological insulator (TI) characterized by a $\mathbb Z_2$ topological invariant \cite{kane2005b}, taking up values $0$ and $1$ in the trivial and the QSH phase, respectively. TIs are the noninteracting insulating phases topologically distinct from the trivial insulator \cite{stanescu}. In addition to the $\mathbb Z_2$ TIs, there are also $\mathbb Z$ TIs, which are characterized by a topological invariant taking any integer value. Technically, the first Chern number is a $\mathbb Z$ invariant, and the IQHE phases can be classified along with noninteracting SPT phases, although they are not protected by any symmetry. Higher-order IQH effects can exit in even dimensions higher than $2$, and are characterized by $\mathbb Z$ invariants, namely the higher Chern numbers \cite{ozawa}. Other non-IQHE $\mathbb Z$ and $\mathbb Z_2$ TIs, as well as topological superconductors are possible for various dimensions and combinations of time-reversal, particle-hole, and chiral (sublattice) symmetries. Classification of TIs and topological SCs is reviewed in chapter 5 of ref. \cite{stanescu}.
	
	In \cref{sec:ssh}, we will encounter an example of a 1D $\mathbb Z$ model with topological phases protected by chiral symmetry, namely the Su-Schrieffer-Heeger (SSH) model \cite{barisic1,barisic2,barisic3,ssh1979}. 
	For noninteracting tight-binding models, chiral symmetry means that the lattice can be divided into two sublattices, and hopping occurs only between them. Then the momentum space Hamiltonian of a periodic chiral-symmetric system with $n$ sites per sublattice can be written as \cite{ozawa}
	\begin{equation}\label{SPT:eq:genericH}
		H_k =
		\begin{pmatrix}
			\bm 0	& Q(k)^\dagger \\
			Q(k) 	& \bm 0
		\end{pmatrix},
	\end{equation}
	where $\bm 0$ and $Q(k)$ are $n\cross n$ matrices, and $Q(k)=Q(k+2\pi/a)$, $a$ being the lattice spacing. When there is a gap at zero energy, i.e. $\det Q(k)\neq 0$, the winding number of $\det Q(k)=|\det Q(k)|e^{i\theta(k)}$ around the origin of the complex plane
	\begin{equation}\label{SPR:eq:W}
		\mathcal W = \frac{1}{2\pi} \int_0^{\frac{2\pi}{a}} \dd k \,\dv{\theta(k)}{k}
	\end{equation}	 
	is a well defined topological invariant of the Hamiltonian. Its absolute value then equals the number of topologically protected edge states at zero energy. 
	
	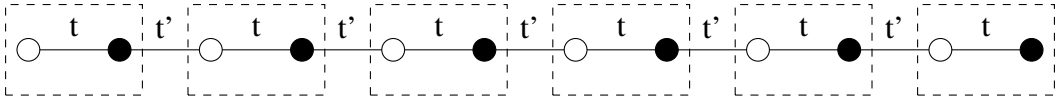
\begin{figure}[htb]\centering
		\begin{tikzpicture}
		\def\s{0.15}
		\def\a{1.2}
		\def\h{4*\s}
		\draw (0,0)--(11*\a,0);
		\foreach \x in {0,2,4,6,8,10}
			\filldraw[fill=white] (\x*\a,0) circle (\s);
		\foreach \x in {0,2,4,6,8,10}
			\filldraw[fill=black] (\x*\a+\a,0) circle (\s);
		\foreach \x in {0,2,4,6,8,10}
			\draw (\x*\a+.5*\a,2*\s) node { t};
		\foreach \x in {0,2,4,6,8}
			\draw (\x*\a+1.5*\a,2*\s) node { t'};
		\foreach \x in {0,2,4,6,8,10}
			\draw[dashed] (\x*\a-2*\s,-\h) rectangle (\x*\a+\a+2*\s,\h);
		\end{tikzpicture}
		\caption[The Su-Schrieffer-Heeger lattice]{The Su-Schrieffer-Heeger lattice with the alternating hopping amplitudes $t$ and $t'$ indicated between the lattice sites. The boxes (dashed lines) enclose the unit cells.}
		\label{SPT:fig:ssh}
	\end{figure}
	
	The SSH model, depicted in \cref{SPT:fig:ssh}, is a simple special case with only one site per sublattice ($n=1$). The tight-binding Hamiltonian is given by
	\begin{equation}\label{SPT:eq:H_SSH}
		\hat H_{SSH} = \sum_x ( t \hat b_x^\dagger \hat a_x + t' \hat a_{x+1}^\dagger \hat b_x + H.c.),
	\end{equation}
	where $\hat a_x$ and $\hat b_x$ are the annihilation operators for the two sublattices at site $x$, and $t$ and $t'$ are, respectively, the intracell and intercell hopping amplitudes\footnote{In a finite system, the cells are defined with respect to the boundary, as in \cref{SPT:fig:ssh}}. The momentum space Hamiltonian is now
	\begin{equation}
	H = |Q(k)|
	\begin{pmatrix}
		0	&	e^{-i\theta(k)} \\
		e^{i\theta(k)}	&	0
	\end{pmatrix},
	\end{equation}
	with eigenvalues $\frac{1}{\sqrt 2} \begin{pmatrix}\pm e^{-i\theta(k)} & 1\end{pmatrix}^T $. By \eqref{eq:berry_connection}, the Berry connection is then $\scalA^\pm(k)=\frac{1}{2}\dv{\theta}{k}$, and we see that the winding number \eqref{SPR:eq:W} is proportional to the Zak phase \eqref{eq:Zak}
	\begin{equation}
		\mathcal W = \frac{1}{\pi} \int_0^{\frac{2\pi}{a}} \dd k \, \scalA^\pm(k) = \frac{1}{\pi} \gamma_{Zak}.
	\end{equation}
	Due to inversion symmetry, this means that the winding number assumes two possible values depending on the relative values of $t$ and $t'$. For $t>t'$, the system is in the $\mathcal W=0$ phase, while for $t<t'$, it is in the $\mathcal W=1$ phase. 
	In an infinite system, it is possible to group the sites into unit cells in another way ($t\rightleftarrows t'$), which switches the winding numbers of the two phases ($0\rightleftarrows 1$).     Thus, there is no essential difference between the two phases. On the other hand, in a bounded system, the boundary imposes a unique choice of the unit cell, i.e. the two sites next to the boundary belong to the same cell (see \cref{SPT:fig:ssh}). Now, the phase $\mathcal W=0$ can be considered topologically trivial, i.e. equivalent to the vacuum, while the $\mathcal W=1$ phase is nontrivial, and possesses a zero-energy edge (gap) state due to a discontinuous transition in $\mathcal W$ at the boundary \cite{ozawa}. Equivalently, a gap state should appear at the topological defect created by joining two semi-infinite chains with different winding numbers, since this also marks a boundary between two phases.

\subsection{Experimental platforms}\label{sec:TQM:platforms} 
	The study of topological phases of matter has its roots in condensed matter physics. Both the integer and the fractional QHE, largely responsible for diverting attention to topological quantum phases of matter, were discovered in semiconductor systems \cite{kdp,TSG}. Condensed matter remains a relevant platform for scientific and technological progress in topological matter, even as other methods arise.
	
	With technological advancement, it is becoming increasingly possible to realize equivalent (as well as novel) physical models in entirely different platforms, such as clouds of ultracold atomic gases and photonic systems. These systems can be designed to obey the same approximate mathematical laws that arise in condensed matter systems. This is akin to numerical simulation, but it is not subject to the same computational limitations. These platforms can allow for measurements that would be difficult or impossible in traditional condensed matter setups, and can have their own technological applications.
	
	Ultracold atomic gases \cite{metcalf99} are systems composed of one or more species of atoms trapped and cooled by interaction with optical, electric and magnetic fields. These interactions can also be used to modulate low-energy atomic interactions, and create synthetic gauge fields for neutral atoms \cite{goldman2014}. This makes them well suited for realization of low-temperature gapped phases that may have topological properties.
	
	Photonics is a field concerned with controlling the behavior of light. While light can be thought of as obeying the laws of classical electrodynamics, its propagation modes may observe band structure and be governed by the same equations as seemingly unrelated quantum systems. We will encounter examples of this in \cref{ch:topofoto}. Since photonic systems may possess band structure, it is possible to create topologically nontrivial phases in this framework. Developments in the field of topological photonics were reviewed in \cite{ozawa}.

%% file: 2_photo/photo.tex
\newcommand{\polarangle}{\varphi}

In this chapter, we will present the results of two recent papers in the field of topological photonics coauthored by the author of this thesis. The work presented in this chapter has been published in:
\begin{itemize}
	\item[\cite{zb}] \bibentry{zb}
	\item[\cite{solitonssh}] \bibentry{solitonssh} 
\end{itemize} 
	
	The first paper presents the results of a theoretical and experimental study of the effect of the valley degree of freedom on propagation of light through an inversion-symmetry-broken photonic honeycomb lattice (a $(2\!+\!1)$D system). Valleys are the local minima (maxima) in the conduction (valence) bands. The honeycomb lattice (HCL) is composed of two triangular lattices which together form an equilateral hexagonal (honeycomb) tiling pattern (see circles in panel a) of \cref{IQHE:fig:HaldaneModel}). The valleys of the HCL are located in the corners of the hexagonal first Brillouin zone. Only two of the six corners are inequivalent, and therefore the HCL is  a two-valley system. The valleys are labelled $K$ and $K'$ and are found in the neighbouring corners. Near the $K$ and $K'$ points, the dispersion and the wave dynamics is described by the 2D Dirac equation (see eq. \eqref{zb:eq:dirac} below), which is why they are called the Dirac points. If the system is inversion symmetric with respect to the midpoint between two nearest-neighbouring sites, the conduction and the valence bands touch in the $K$ and $K'$ points, and the dispersion is conical. This dispersion corresponds to the massless Dirac equation. Breaking the inversion symmetry opens the gap, allowing us to view the two bands as truly separate. The gapped parabolic dispersion reveals the presence of the mass term in the Dirac equation. 
	
	The experiments%
	\footnote{\label{note:exp-sec}The details of the experiment are given in the Experimental section of the published article \cite{zb}:\\ \url{https://onlinelibrary.wiley.com/doi/full/10.1002/lpor.202000563\#lpor202000563-sec-0060-title}.}
	 were performed by shining a light beam with a Gaussian profile and a triangular lattice structure on a 2D hexagonal array of straight waveguides "written" into a photonic crystal. The profile of the beam at a certain point in time is described by a quantum state, since the propagation of light is governed by a Schrödinger-like equation (see eq. \eqref{zb:eq:schro} below).
	 In the following, we will see how an initially Gaussian wavepacket with zero angular momentum can obtain a vortex component with a nonzero angular momentum when propagating through a HCL. This is the effect of a topological singularity, i.e. nontrivial winding of the Berry curvature, at the Dirac points. The full band is topologically trivial, as the winding is opposite for $K$ and $K'$ points, but when the light is restricted to propagate close to one of the valleys, the topological singularity is imprinted onto the propagating beam. As shown in the images below, this manifests as a rotating spiral pattern of the light intensity (for inversion-symmetry-broken HCLs), caused by the interference of the vortex and nonvortex components of the beam. Further analysis also reveals the Zitterbewegung phenomenon in the evolution of the pattern.
	 
	 The second paper is a numerical investigation of a $(1\!+\!1)$D array of soliton beams propagating in a nonlinear medium in the following configuration: an SSH lattice is formed from two arrays of beams propagating at an angle, so that they periodically cross (see \cref{ssh:fig:2} a) below).
	 Recall that the SSH system assumes one of the two SPT phases, depending on the relative strengths of the intracell and the intercell tunnelling coefficients. In our case, these coefficients arise due to the soliton beam interaction, mediated by nonlinearity, and are controlled by the spatial separation of the beams at a certain point during the propagation (see \cref{ssh:fig:1} below). We study the spectrum of the system which reveals the topological phase through the presence or absence of the topological gap states. The role of energy in the spectrum is played by the propagation constant ($\beta$) which prescribes the evolution of the complex electric field envelope, analogous to how the energy prescribes the evolution of the energy eigenstate. As the separation between the neighbouring beams changes during the propagation, the system repeatedly gains the edge states in dynamical topological phase transitions, and then loses them in the meantime as they decouple from the lattice, thus leaving the remaining lattice in the trivial phase until the next phase transition. The existence of different topological phases is emergent from nonlinearity of the medium, since the beams would not couple in a linear medium.

\markedsection{Self-Rotation and Zitterbewegung in HCLs}{Wavepacket Self-Rotation and Helical Zitterbewegung in Symmetry-Broken Honeycomb Lattices}
\label{sec:zb}
\subsection{Introduction}\label{sec:zb:intro}
	Electric charge is the key quantity for controlling signals in conventional electronics and semiconductor devices. However, advances in manipulating spin and valley degrees of freedom have reshaped the traditional perspective, leading to the development of spintronics \cite{zutic2004} and valleytronics \cite{schaibley2016}. Amplitude, phase, and polarization are the key quantities of usual recipes for controlling the flow of light. However, the understanding and development of optical spin-orbit interactions \cite{bliokh2015}, photonic pseudospins \cite{song2015}, and valley degrees of freedom\cite{ma2016,dong2017,gao2018,wu2017,noh2018,chen2017,ni2018,shalaev2019} have offered us new knobs that can be used for manipulation of light in photonic structures, in analogy with parallel advances in electric systems.
	
	The pioneering achievements exploiting valley degrees of freedom in photonics include, e.g., the prediction \cite{ma2016,dong2017} and experimental demonstration \cite{wu2017} of photonic valley-Hall topological insulators, topologically protected refraction of robust kink states in valley photonic crystals \cite{gao2018}, topological valley-Hall edge states \cite{noh2018}, and spin and valley-polarized one-way Klein tunnelling \cite{ni2018}. Photonic valley systems can be implemented at telecommunication and terahertz wavelengths on a silicon platform \cite{shalaev2019,yang2020}, at subwavelength scales on plasmonic platforms \cite{wu2017,jung2018,proctor2020}, and they can be used for the development of topological lasers \cite{zeng2020,zhong2020,gong2020,smirnova2020a}, which opens the possibilities for many applications. Besides electromagnetic waves, valley topological materials have been used for manipulation of other waves such as sound waves \cite{lu2017} and elastic waves \cite{yan2018}. All these exemplary successes unequivocally  point at the need and importance of discovering valley-dependent wave phenomena, for both fundamental understanding and advanced applications.
	
	To this end, it is important to understand the behaviour of physical quantities that distinguish different valleys. Among the most studied examples in photonics are the two inequivalent valleys of the honeycomb lattices, located at the lattice high-symmetry $K$ and $K'$ points in the Brillouin zone, which are furnished with nontrivial Berry phase winding \cite{lu2014,ozawa}. Since the Berry curvatures are in opposite directions at $K$- and $K'$-valleys in a (symmetry-broken) HCL (e.g., see \cite{ozawa}), they can be used to distinguish the two valleys. Besides the Berry curvature, in electronic systems, the electron magnetic moment can also be used to distinguish the two valleys \cite{xiao2007,xu2014}. The magnetic moment occurs from the self-rotating electric wavepacket \cite{xiao2007,xu2014,xiao2010}, which is virtually impossible to be directly observed with electrons.
	
	Here, we study valley-dependent propagation of light in an inversion-symmetry-broken photonic HCL. We establish the lattice by employing a direct laser-writing technique \cite{xia2018}, and we demonstrate experimentally and numerically the valley-dependent helicity in spiraling intensity patterns related to wavepacket self-rotation. Specifically, we show that, by selective excitation of the valleys in a gapped HCL, a probe beam undergoes distinct spiralling during propagation through the lattice, characterized by its helical intensity pattern and "center-of-mass" oscillation, even though no initial orbital angular momentum is involved. We theoretically demonstrate that the observed phenomenon dwells upon the existence of the Berry phase \cite{berry1984}, leading to the fundamental phenomenon of Zitterbewegung, first introduced by Schrödinger \cite{schrodinger1930} in the context of relativistic electrons. We find that the helicity of Zitterbewegung in our system is a valley-dependent quantity.
	
	Zitterbewegung refers to a prediction that elementary particles such as electrons described by the relativistic Dirac equation would exhibit rapid oscillatory motion in vacuum, with angular
frequency on the order of $2mc^2/\hbar$ \cite{schrodinger1930}. It was studied in attempts to provide a deeper understanding of the electron spin \cite{huang1952,chuu2010} and even to interpret some aspects of quantum mechanics \cite{hestenes1990}, but the Zitterbewegung of electrons in vacuum has never been observed owing to its inherent ultra-small amplitude and ultra-high frequency. However, electrons in Bloch bands of some materials are driven by equations analogous to the relativistic Dirac equation, e.g., Zitterbewegung of electrons was predicted to occur in semiconductor quantum wells \cite{schliemann2005}. In a full analogy, Zitterbewegung was also predicted with ultracold atoms in optical lattices \cite{vaishnav2008} and with photons in 2D photonic crystals \cite{zhang2008}. Experimental observation of Zitterbewegung-like phenomena was, however, mostly limited to 1D domain in systems including trapped ions \cite{gerritsma2010}, photonic lattices \cite{dreisow2010}, and Bose–Einstein condensates \cite{leblanc2013,qu2013}, or to surface acoustic waves in an integrated phononic graphene \cite{yu2016}. The Zitterbewegung term in the quantum expectation value of the position operator vanishes if the wave-packets are made up with solely positive (or negative) energy states, thus leading to its interpretation in terms of interference of positive and negative energy states. In periodic lattices, this amounts to the interference of Bloch waves from different bands. In this work, we show theoretically that the Zitterbewegung can be interpreted via interference between the incident nonvortex beam component and the vortex component arising from the universal momentum-to-real space mapping mechanism, which inherently has a topological
origin \cite{liu2020}. Thus, we provide a different perspective about the Zitterbewegung phenomenon, which gives rise to a simpler visualization than the original interpretation.

\subsection{Results}\label{sec:zb:results}
\subsubsection{Experimental Results and Numerical Simulations}
	We study light propagation in (2+1)D photonic lattices, which in the paraxial approximation is governed by the Schrödinger-like equation (e.g., see ref. \cite{ozawa} and references therein)
	\begin{equation}\label{zb:eq:schro}
		i\dv{\Psi}{z} = -\frac{1}{2k_0}\nabla^2\Psi - \frac{k_0\delta n(x,y)}{n_0}\Psi(x,y,z).
	\end{equation}
	Here $\Psi(x,y,z)$ is the complex amplitude of the electric field, $k_0$ is the wave number in the medium, $n_0$ is the background refractive index, and $\delta n(x,y)$ is the induced refractive-index change forming the HCL with broken inversion symmetry, as illustrated in \cref{zb:fig:1} a). Equation \eqref{zb:eq:schro} is mathematically equivalent to the Schrödinger equation describing electrons in 2D quantum systems, with $z$ playing the role of time. The HCL is comprised of two sublattices (A and B), and the inversion-symmetry breaking is achieved with a refraction index offset between the sublattices, see \cref{zb:fig:1} a). In $k$-space, the HCL has two distinct valleys located at the $K$ and $K'$ Dirac points in the Brillouin zone, as illustrated in \cref{zb:fig:1} b). In the vicinity of Dirac points, the band structure is described by $\pm\sqrt{k^2+m^2}$
	, and the wave dynamics is approximately described by the 2D massive Dirac equation (see theoretical analysis below). Here, $k$ is the magnitude of the wave vector with origin at the Dirac point, and $m$ is the "effective mass" determined by band dispersion, namely, the band gap. The band gap is proportional to the refraction index offset between the two sublattices as illustrated in \cref{zb:fig:1} c) for the photonic lattices used in our numerical simulations and experiments, and its size is $2m$. This implies that the effective mass $m$ is expressed in the same units as the Hamiltonian in eq. \eqref{zb:eq:schro}, i.e. the inverse length which indeed corresponds to the units of $k$. 
	For direct comparison, the insets in \cref{zb:fig:1} c) show the band structure close to the $K$-point for $m=0$ (gapless Dirac-cone type band structure), and for $m>0$ (gapped $\pm\sqrt{k^2+m^2}$ type band structure).
	
	Our main finding is sketched in \cref{zb:fig:1} b). The probe beam which is formed by interfering three broad Gaussian beams excites the modes in the vicinity of three equivalent $K$-points (or $K'$-points) in momentum space, i.e., the modes in one valley, with both sublattices equally excited in real space. The output beam exhibits self-rotation during propagation through the HCL, which has a spiraling intensity pattern with the helicity depending on the valley ($K$ or $K'$) that is initially excited. It will be shown below that this spiraling self-rotating motion is attributed to a root of the Zitterbewegung of the wavepacket, identified through the rotation of its "center-of-mass" (COM), which is defined as the space average value $\ev\vr=\int \vr I(\vr)\dd a$ weighted with respect to intensity $I(\vr)$.
	
\afterpage{
\addtocounter{footnote}{-2}
	\begin{figure}[htb]\centering
		\includegraphics[width=.95\textwidth]{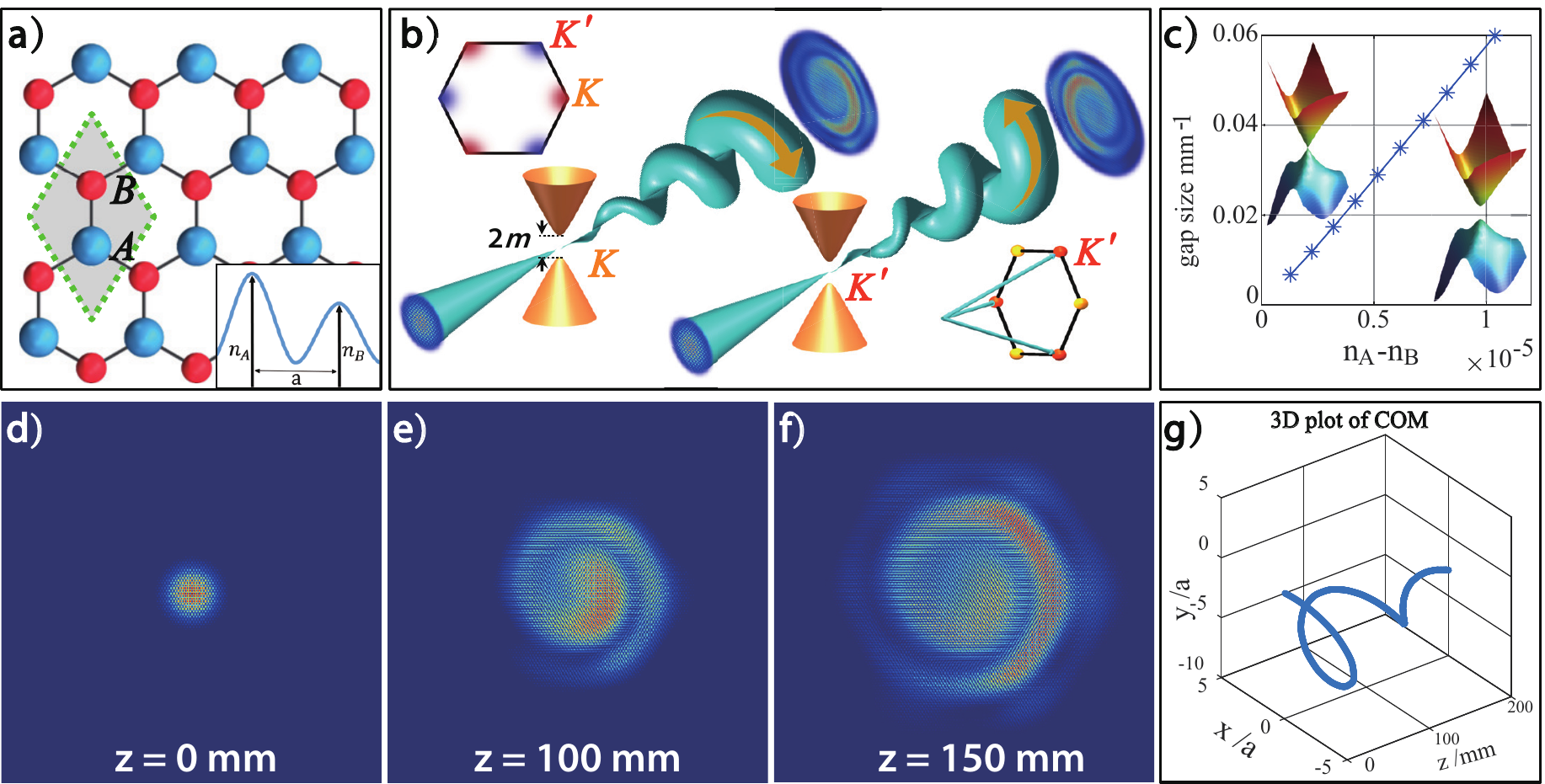}
		\caption[Valley-dependent wavepacket self-rotation in a symmetry-broken HCL]{Valley-dependent wavepacket self-rotation in a symmetry-broken HCL. a) Illustration of an inversion-symmetry-broken HCL consisting of A and B sublattices. The inset sketches the refractive-index offset ($n_A>n_B$ is shown, e.g.). b) Illustration of wavepacket self-rotation when the modes in the vicinity of the $K$-valley (or $K'$-valley) are excited, showing spiralling intensity patterns with valley-dependent helicity. Top inset shows the valley locations at the edges of the first Brillouin zone in $k$-space; the Berry curvature is opposite at two inequivalent valleys (sketched with red and blue colors). Bottom inset shows the scheme when three $K$-valleys are simultaneously excited. c) The size of the gap as a function of the index offset ($n_A-n_B$) for photonic
lattices used in our experiments. The insets show the band-gap structures in the vicinity of the $K$-point for $n_A-n_B=0$ (upper inset), and $n_A-n_B>0$ (lower inset). d)-f)  Spiralling intensity patterns at different propagation distances indicating self-rotation of the wavepacket. g) Plot of the COM trajectory obtained numerically. The probe at $z = 0$ shown in d) has a Gaussian envelope with no initial orbital angular momentum - see video file in the
supporting information\protect\footnotemark\phantom{ } to \cite{zb}. }
		\label{zb:fig:1}
	\end{figure}
	\footnotetext{\label{note:support}\url{https://onlinelibrary.wiley.com/doi/full/10.1002/lpor.202000563\#support-information-section}}
}
	
\addtocounter{footnote}{1}
	In \cref{zb:fig:1} d)-f), we show numerical results of the output patterns of the probe beam at different propagation distances in the inversion-symmetry-broken HCL, obtained by solving equation \eqref{zb:eq:schro}, with the refractive index offset between the sublattices set by the ratio ${n_A:n_B=1.2:1}$, exciting only the $K$-valley. The parameters used in the simulations correspond to that of the experiment: ${n_0=2.35}$ for the SBN:61 (strontium barium niobate) crystal \cite{sbn}, ${k_0=2\pi n_0/\lambda}$ and ${\lambda=488\,\mathrm{nm}}$, the lattice constant is $16\,\mathrm{\mu m}$ (i.e., the distance between nearest neighbouring lattices is $9\,\mathrm{\mu m}$, and the maximal index change (depth of the lattice) is about $1.3\times10^{-4}$. The overall envelope of the probe beam is Gaussian-like (\cref{zb:fig:1} d)), but with a triangular lattice structure (due to three-beam interface) at $z=0$ that can be positioned to excite one or both sublattices. In simulations displayed in \cref{zb:fig:1} d)-f), the probe beam excites the middle points between the A and B sublattices, i.e., both sublattices are equally excited. We find that the output wavepacket exhibits self-rotation during propagation (see the video in supporting information%
	\footnote{See footnote \ref{note:support}.}
	 to the published article \cite{zb}), and the initially symmetric probe beam evolves into an asymmetric spiralling intensity pattern as displayed in \cref{zb:fig:1} e), f). It expands during propagation because of diffraction, whereas the spiral helicity and the direction of rotation are valley-dependent. In \cref{zb:fig:1} g), the dynamical evolution of the beam’s COM is plotted in 3D, showing spiral-like Zitterbewegung oscillations (in the plot we subtracted the drift which occurs alongside
Zitterbewegung phenomenon for better visualization).

\addtocounter{footnote}{1}

\begin{figure}[htb]\centering
		\includegraphics[width=.95\textwidth]{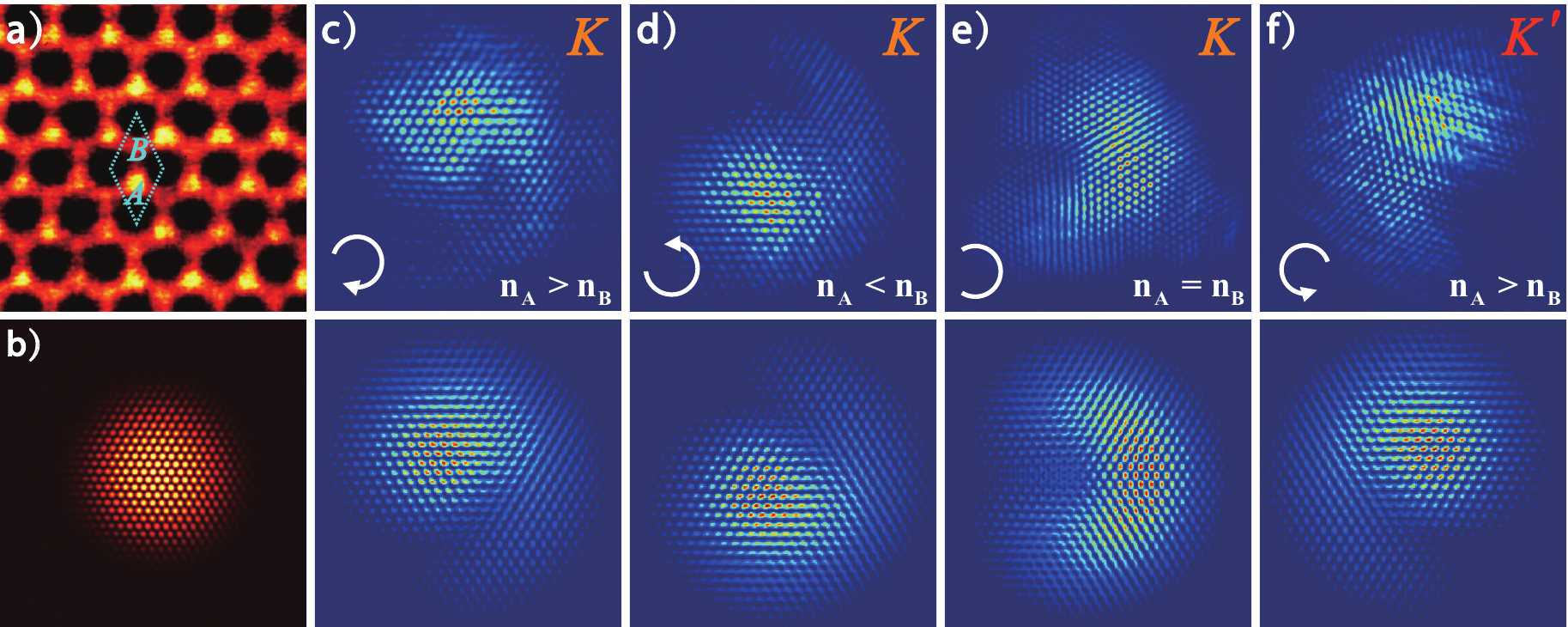}
		\caption[Experimental and numerical results demonstrating valley-dependent wavepacket self-rotation.]{Experimental and numerical results demonstrating valley-dependent wavepacket self-rotation. a) Zoom-in image of a laser-written HCL with broken-inversion-symmetry; in this plot, $n_A>n_B$, corresponding to \cref{zb:fig:1} a). b) Input triangular lattice pattern used in experiment as the probe beam. c–f) Experimental (top row) and numerical (bottom row) results of output intensity patterns for different excitation conditions: c–e) Results obtained under $K$-valley excitation where the index ratio is c) $n_A:n_B=1.2:1$, d) $n_A:n_B=1:1.2$, e) $n_A:n_B=1.:1$. f) Result obtained under $K'$-valley excitation with $n_A:n_B=1.2:1$. Note that the helicities of the spiralling patterns in (c) and (f) (as well as in (c) and (d)) are in opposite directions, as illustrated by curved arrows.}
		\label{zb:fig:2}
	\end{figure}

	Next, we present corresponding experimental results obtained in a HCL established in a $20\,\mathrm{mm}$ long nonlinear crystal by a cw-laser-writing method \cite{xia2018}. Instead of using a single Gaussian beam for writing, here the two sublattices are separately written and controlled by a triangular lattice pattern. The refractive-index difference of the two sublattices $n_A:n_B$ is readily tuned by the writing time for each sublattice (see the experimental section%
	\footnote{See footnote \ref{note:exp-sec}.} in \cite{zb}). 
	 A typical example of experimentally generated symmetry-broken HCL with $n_A>n_B$ is shown in \cref{zb:fig:2} a). As in simulation, the probe beam is a truncated triangular lattice pattern
formed by interfering three broad Gaussian beams (see \cref{zb:fig:2} b)) with their wavevectors matched to the three $K$- or $K'$-points. In real space, we excite both sublattices with equal amplitude and phase by positioning the probe beam at middle points between the two sublattices. The observed intensity patterns of the probe beam at the lattice output under different excitation conditions are shown in the top panels of \cref{zb:fig:2} c)-f),  with corresponding numerical simulation results plotted in the bottom panels.
	
	When the input beam excites the $K$-valley with the refractive-index offset between sublattices such that $n_A>n_B$, the beam evolves into a spiralling pattern (\cref{zb:fig:2} c)).  The helicity of the spiralling pattern and therefore the rotation direction of the output beam is reversed if the offset is changed to be $n_A<n_B$ (\cref{zb:fig:2} d)). As we will show theoretically below, such spiralling intensity pattern is related to the circular motion of the COM of the wavepacket and the Berry-phase-mediated Zitterbewegung. We emphasize that the rotation can only be realized when the inversion symmetry of the HCL is broken and the gap opens; for comparison, when $n_A=n_B$, the output pattern exhibits conical diffraction \cite{peleg2007} rather than a spiralling pattern (\cref{zb:fig:2} e)) under the equal excitation condition. Importantly, we experimentally demonstrate that the rotation direction depends on the valley degree of freedom. If we excite the $K'$-valley instead of the $K$-valley, while keeping all other conditions unchanged, we observe that the spiraling direction (i.e., helicity) of the intensity pattern is reversed. This can be seen by comparing the experimental results shown in \cref{zb:fig:2} c) and f). These observations are corroborated by numerical beam propagation simulations using equation \eqref{zb:eq:schro}, which is shown in the bottom panels of \cref{zb:fig:2}. We point out that altering the helicity of the spiralling pattern by reversing the index offset between the two sublattices  (\cref{zb:fig:2} c) vs 2d)) is fully equivalent to altering the helicity via exciting different valleys (\cref{zb:fig:2} c) vs f)). In both cases, the helicity of the spiralling pattern is correlated with the direction of the Berry curvature around the gapped Dirac cone. In other words, the spiralling intensity is a valley contrasting quantity, analogous to the orbital momentum of electrons in
condensed matter systems \cite{xiao2007,xu2014,xiao2010}, manifested when the inversion symmetry is broken.

\subsubsection{Theoretical analysis}
	For excitations in the vicinity of the $K$-valley, equation \eqref{zb:eq:schro} is approximated by $i\dv{\psi}{z}=H\psi$, where the Hamiltonian (in $k$-space) is an effective 2D massive Dirac equation
	\begin{align}\label{zb:eq:dirac}
		H = \kappa(\sigma_x k_x + \sigma_y k_y) + \sigma_z m &=
		\begin{pmatrix}
			m	&	\kappa(k_x-ik_y) \\
			\kappa(k_x+ik_y)	&	-m
		\end{pmatrix} \nonumber
		\\
		&=
		\begin{pmatrix}
			m	&	\kappa k e^{-i\varphi_k} \\
			\kappa k e^{i\varphi_k}	&	-m
		\end{pmatrix},
	\end{align}	 
	where $\sigma_i$ are the Pauli matrices. The numerical coefficient $\kappa$ depends on the coupling strength between adjacent waveguides in the lattice (e.g., see ref. \cite{xu2014}). Without any loss of generality, we set $\kappa=1$ in all analytical expressions, because they can be rescaled to any value of $\kappa$ with the substitution $\q\rightarrow\kappa \q$. The complex amplitude of the electric field $\psi=\begin{pmatrix} \psiup & \psidn \end{pmatrix}^T$ is a two-component spinor, because the HCL has two sublattices. Pseudospin components $\psiup$ and $\psidn$ describe the field amplitudes in the A and B sublattices (e.g., see \cite{liu2020}). The eigenmodes of the Hamiltonian in equation \eqref{zb:eq:dirac} are given by $H\psink=\betank\psink$,
	\begin{equation}\label{zb:eq:eigenmodes}
		\psink=\frac{1}{\sqrt{\Nnk}}\begin{pmatrix}
			\frac{m+\betank}{k} e^{-i\varphi_k} \\ 1		
		\end{pmatrix}
		,\qquad
		\Nnk = 2 + \frac{2m(m+\betank)}{k^2},
	\end{equation}
	where $\betank=n\sqrt{k^2+m^2}$; $n=\pm 1$ is the band number, $\vk$ is the wavevector with origin at the $K$-point, and $\varphi_k$ its polar angle with respect to the origin. It is important to note that the $k$-space topological charges of the two spinor components in equation \eqref{zb:eq:eigenmodes} differ by one, i.e., the vorticity of the two spinor components in $k$-space is different. This difference is independent of the gauge used, and it gives rise to the Berry phase winding around the $K$-point.
	
	Dynamics around the $K'$-valley is described analogously with substitution $k_x\rightarrow-k_x$ in equation \eqref{zb:eq:dirac} \cite{xu2014}. The eigenmodes at the $K'$-valley are given by $\psink^*$, i.e., the winding of the spinor vorticity in $k$-space is in the opposite direction. Thus, the geometry of the eigenmodes gives rise to the Berry curvature which is in opposite directions at the $K$- and $K'$-points \cite{ozawa,zhang2008,xiao2010} (see \cref{zb:fig:1} b)). This is the origin of the opposite helicity of the spiralling patterns observed in \cref{zb:fig:2} c), f).
	
	We are interested in the dynamics from an axially symmetric initial excitation
	\begin{equation}\label{zb:eq:ic}
		\psi (r,\polarangle,z=0) = \psi_0 \sqrt{I_0(r)} = \int \dd^2 k\psi_0 f(k) e^{i\vk\cdot\vr} ,
	\end{equation}
	where we have introduced the radial coordinates ($x=r\cos\polarangle$ and $y=r\sin\polarangle$), $f(k)$ corresponds to the spatial power spectrum of the initial excitation $f(k)=(2\pi)^{-2}\int{\sqrt{I_0(r)}e^{-i\vk\cdot\vr} \dd a}$, with $\dd a=r\dd r\dd\polarangle$ the infinitesimal area element, and it determines the distribution of excitations around the $K$- or $K'$-point; and finally, $\psi_0=\begin{pmatrix} \cos\theta e^{i\alpha} & \sin\theta \end{pmatrix}^T$ is the most general initial spinor, where $\alpha$ is the relative phase between the fields in the sublattices at $z=0$, and $\theta$ determines the amplitude in each sublattice.
	
\begin{figure}[htb]\centering
	\includegraphics[width=.95\textwidth]{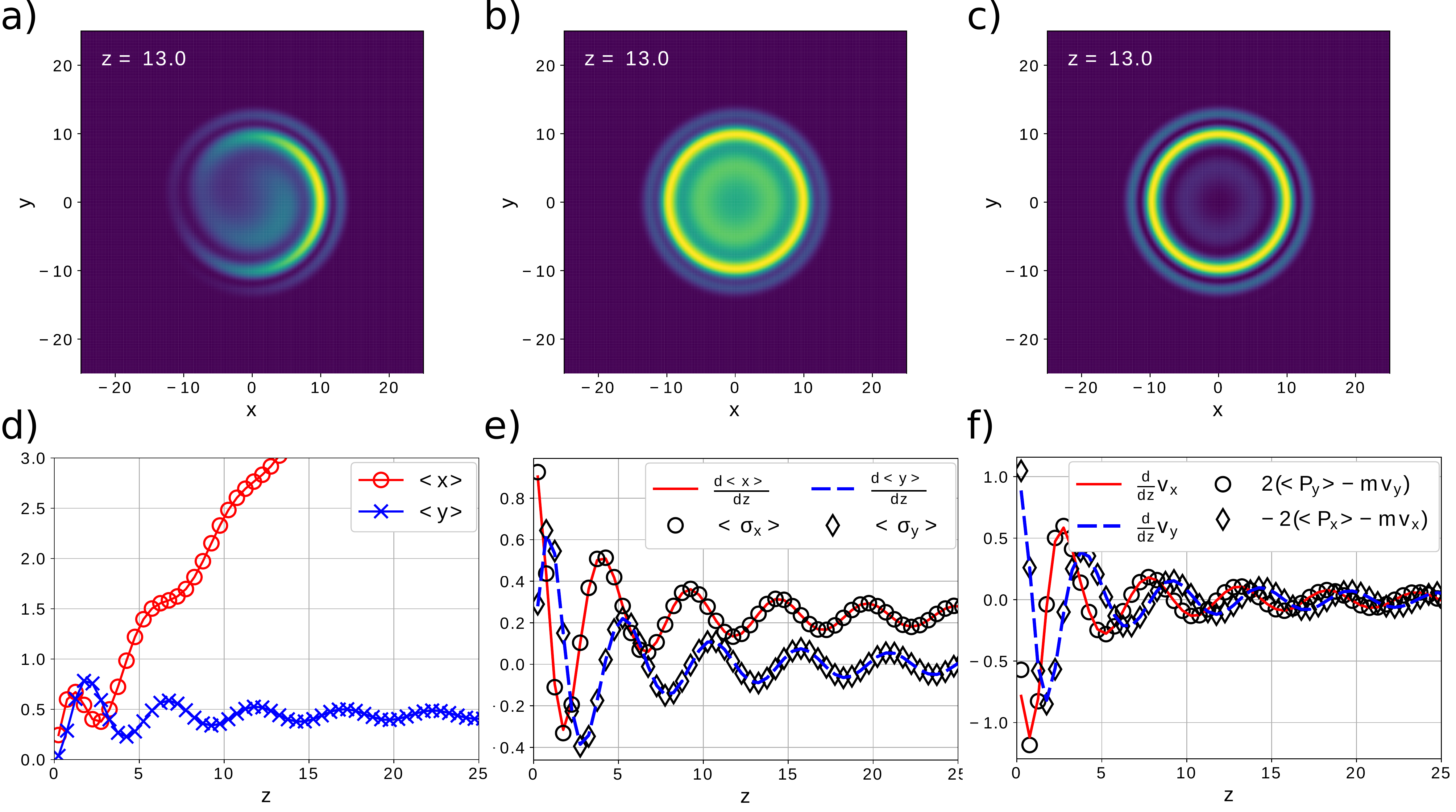}
		\caption[Theoretical analysis of wavepacket self-rotation from Dirac equation]{Theoretical analysis of wavepacket self-rotation from Dirac equation. Top panels are the intensity structure of the $1/2$ spinor component of the spiralling beam and of its subcomponents, and bottom panels show the motion of its "center-of-mass (COM)." In the figure, $z$ is in units $(\kappa k_0)^{-1}$, $x$ and $y$ are in units $k_0^{-1}$. a) Intensity structure of the pseudospin component $|\psiup(r,\polarangle,z)|^2$, b) the nonvortex component $|\gupN(r,z)|^2$, and c) the vortex component $|\gupV(r,z)|^2$. d) The position of the COM of the wavepacket (average values of $x$ and $y$) as functions of $z$. e) Propagation of the velocity components of the COM, and the (identical) expectation values $\ev{\sigma_x}$ and $\ev{\sigma_y}$. f) Propagation of the acceleration components and numerical verification of equation \eqref{zb:eq:10}. See the text for details.}
		\label{zb:fig:3}
	\end{figure}

	Dynamics from the initial condition \eqref{zb:eq:ic} is readily found by expanding into eigenmodes of the system. After a straightforward calculation one finds
	\begin{equation}\label{zb:eq:solution}
		\psi(r,\polarangle,z)=
		\begin{pmatrix}
			\psiup(r,\polarangle,z) \\ \psidn (r,\polarangle,z)
		\end{pmatrix}
		=
		\begin{pmatrix}
			\gupN(r,z)+\gupV(r,z)e^{-i\polarangle} \\
			\gdnV(r,z)e^{i\polarangle} + \gdnN(r,z)
		\end{pmatrix},
	\end{equation}	 
	where $z=13/\kappa _0$, and the $g$-functions can be expressed as integrals in $k$-space (see the supporting information%
	\footnote{See footnote \ref{note:support}.}
	 to \cite{zb} for derivation):
	\begin{align}\label{zb:eq:g_funcs}
		\gupN (r,z) &= 2\pi \cos\theta \, e^{i\alpha} \sum_n \int_0^\infty \frac{k\dd k}{\Nnk} \oble{\frac{m+\betank}{k}}^2 J_0(kr) \, e^{-\betank z}, \nonumber \\
		\gupV (r,z) &= 2\pi i \sin\theta \sum_n \int_0^\infty \frac{k\dd k}{\Nnk} \, \frac{m+\betank}{k} \, J_1(kr) \, e^{-\betank z}, \nonumber \\
		\gdnV (r,z) &= 2\pi i  \cos\theta \, e^{i\alpha} \sum_n \int_0^\infty \frac{k\dd k}{\Nnk} \, \frac{m+\betank}{k} \, J_1(kr) \, e^{-\betank z}, \nonumber \\
		\gdnN (r,z) &= 2\pi \sin\theta \sum_n \int_0^\infty \frac{k\dd k}{\Nnk} J_0(kr) \, e^{-\betank z},
	\end{align}		 
	 where $J_0(x)$ and $J_1(x)$ are the Bessel functions of the first kind. In \cref{zb:fig:3}, we plot the spiralling intensity pattern $|\psiup(r,\polarangle,z)|^2$ obtained with the Hamiltonian in equation \eqref{zb:eq:dirac}; the envelope of the initial excitation is Gaussian, $f(k)=\exp(-k^2/k_0^2)$, and both sublattices are equally excited with same phase, $\psi_0=\begin{pmatrix} 1 & 1 \end{pmatrix}^T$. The mass term is $m=0.6\kappa k_0$, which determines the gap size. For our initial condition, the intensity in the lower spinor component $|\psidn(r,\polarangle,z)|^2$ is equal to that in the upper component $|\psiup(r,\polarangle,z)|^2$, and consequently the whole intensity ${|\psi|^2=|\psiup|^2+|\psidn|^2}$ has the same spatial dependence as $|\psiup|^2$ illustrated in \cref{zb:fig:3} a) (see the supporting information to \cite{zb}). For this reason, in what follows we focus on understanding the intensity in just one spinor component. It is evident that the spiralling intensity pattern obtained with the "low-energy" Hamiltonian equation \eqref{zb:eq:dirac} agrees with those obtained from numerical simulations of the Schrödinger equation \eqref{zb:eq:schro} as well as from experiments
shown in \ref{zb:fig:2}. 

	It is important to note from Equation \eqref{zb:eq:solution} that each spinor component is a superposition of a nonvortex (Gaussian-like) amplitude and a vortex amplitude. At $z=0$, there are no vortex components, because $\gupV(r,0)=\gdnV(r,0)=0$. The underlying mechanism behind the emergence of these vortices was explained previously \cite{liu2020}: the vortex that is present in the $k$-space of each eigenmode of this system (related to inherent topological singularity at the Dirac point) is mapped to the real space during linear propagation dynamics (analogous to the far-field mapping of the power spectra to real space intensity).
	
	To explain the spiraling pattern observed in our experiments, we calculate the intensity in the pseudospin components
	\begin{align}\label{zb:eq:6}
		\abs{\psiup}^2 &= \abs{\gupN(r,z)}^2+\abs{\gupV(r,z)}^2+ \nonumber \\
		& + 2\abs{\gupN}\abs{\gupV}\cos( -\arg\gupN(r,z)+\arg\gupV(r,z)-\polarangle )
	\end{align}
	and equivalently for the other pseudospin component. The last term describes the interference between the vortex and nonvortex field amplitudes, which depends on their relative phase. The intensities of the nonvortex term $|\gupN(r,z)|^2$ and the vortex term $|\gupV(r,z)|^2$ are radially symmetric, as shown in \cref{zb:fig:3} b), c) (see the supporting information to \cite{zb} for plots of all the $g$-functions). Therefore, the spiralling pattern must arise from the interference. The interference term has a maximum when
	\begin{equation}\label{zb:eq:7}
		\polarangle=[-\arg\gupN(r,z)+\arg\gupV(r,z)]\!\!\mod 2\pi.
	\end{equation}

	When $-\arg\gupN(r,z)+\arg\gupV(r,z)$ is monotonically increasing (or decreasing) with $r$, the function implicitly given in equation \eqref{zb:eq:7} is a spiral in the $(r,\polarangle)$-plane; the spiral helicity depends on whether the right-hand side in equation \eqref{zb:eq:7} decreases or increases. Evidently, the spiralling self-rotating pattern arises from the interference of the vortex and the nonvortex components.
	
	We now present the theory for the wavepacket self-rotation and Zitterbewegung phenomenon in our system. Dynamics of the COM of the wavepacket $\vr_C=x_C\uvec x+y_C\uvec y$ is given by
	\begin{equation}\label{zb:eq:8}
		\vr_C(z)=\ev \vr = \int\psi^\dagger(r,\polarangle,z) \vr \psi(r,\polarangle,z) \dd a,
	\end{equation}
	where $\vr=x\uvec x+y\uvec y$, and $\dd a=r\dd r\dd\polarangle$ is the infinitesimal area element. It can be understood by observing the velocity of the COM
	\begin{equation}\label{zb:eq:9}
		\bm v_C = \dv{\vr_C}{z} = \int \psi^\dagger(r,\polarangle,z) i\uglate{H,\vr} \psi (r,\polarangle,z) \dd a = \ev{\sigma_x}\uvec x + \ev{\sigma_y}\uvec y
	\end{equation}
	and its acceleration
	\begin{equation}\label{zb:eq:10}
		\dv{\bm v_C}{z} = -2\uvec z \times \bm P + 2m\, \uvec z\times \bm v_C,
	\end{equation}
	where we have introduced  the vector $\bm P = \ev{k_x\sigma_z}\uvec x + \ev{k_y\sigma_z} \uvec y$ (see the supporting information%
	\footnote{See footnote \ref{note:support}.}
	 to \cite{zb} for the derivation). Calculated results from equations \eqref{zb:eq:9} and \eqref{zb:eq:10} are illustrated in \cref{zb:fig:3} d)-f).

	The second term in Equation (10) is the Zitterbewegung term; it corresponds to the oscillations of the COM with frequency $2m$ (the size of the spectral gap). Oscillations are clearly visible in
all \cref{zb:fig:3} d)–f). Moreover, it is evident from \eqref{zb:eq:10} that the helicity of Zitterbewegung oscillations depends on the sign of $m\propto n_A-n_B$, which corroborates our experimental findings. The first term in equation \eqref{zb:eq:10} yields the drift of the COM of the wavepacket, visible in \cref{zb:fig:3} d), which is an expected feature of the Zitterbewegung effect (e.g., see refs. \cite{schliemann2005,david2010}). The direction of the drift depends on the initial conditions. More specifically, the expectation value of the pseudospin operator $\sigma=\uvec x\sigma_x+\uvec y \sigma_y$ at $z=0$ sets the direction of the initial velocity of the COM (see \cref{zb:fig:3} e)). Such drifting of the COM is also observed in our numerical simulations using equation \eqref{zb:eq:schro}. We note that for better visualization of the spiraling dynamics, we did not include the drift when plotting \cref{zb:fig:1} g).

	The components of the vector $\bm P$ are interpreted as the difference of the expectation value of the momentum between the pseudospin-up and -down components, i.e., the difference of the momentum between the two sublattices. The acceleration of the COM in the $x$-direction is proportional to $P_y$, which can therefore be interpreted as a pseudo-force exerted on the COM. From the example shown in \ref{zb:fig:3} f), we see that this pseudo-force $\bm P$ shows damped oscillations around zero. Thus, it induces some oscillations, which should be distinguished from those of the Zitterbewegung term. Our calculations indicate that the circular Zitterbewegung motion in symmetry-broken HCLs exists only when $m$ is nonzero and thus the gap opens, which is in agreement with the Zitterbewegung of electrons \cite{schrodinger1930}. Yet, our finding is in contradistinction with similar oscillations that were called Zitterbewegung in gapless HCL systems \cite{zhang2008,yu2016,liang2011}. The oscillations reported there could be linked to the oscillations of the pseudo-force P described above, rather than to the Zitterbewegung term which is absent for $m=0$.
	
\subsection{Discussion}
	The theory of the Zitterbewegung has been addressed in numerous papers \cite{huang1952,hestenes1990,schliemann2005,vaishnav2008,zhang2008,cserti2006,cserti2010,ye2019}. The Zitterbewegung effect was originally associated with oscillatory motion of electrons in 3D space \cite{schrodinger1930,huang1952}, but such motion has never been observed. Here, we focus on the novel aspects of this phenomenon using optical wavepackets in 2D photonic lattices. We discuss connection between the experimentally observed valley-dependent spiralling intensity pattern (related to self-rotation of the wavepacket) and the Zitterbewegung phenomenon. This leads to a novel interpretation of the phenomenon, and sheds light on the role played by the Berry
phase.

	First, we mention a seemingly unrelated simple example. Considering two coupled single-mode waveguides, the 
	system has a symmetric and an antisymmetric eigenmode, $\frac{1}{\sqrt 2} (u_L\pm u_R)$, with  two propagation constants (eigenvalues) whose difference depends on the strength of the coupling (here the letter L stands for the left waveguide, and R for the right waveguide). By launching a beam, e.g., into the left waveguide, both modes will be excited and they will undergo beating; the field amplitude will thus jump from the left to the right waveguide and back and forth, with the frequency given by the coupling strength. The COM of the beam will oscillate at this frequency.
	
	The very same mechanism, albeit a bit more complicated, leads to Zitterbewegung in our 2D system. First, we excite both sublattices of the HCL equally and with the same phase, i.e. $\psi_0=\begin{pmatrix} \cos\theta \,e^{i\alpha} & \sin\theta \end{pmatrix}^T = \begin{pmatrix} 1 & 1 \end{pmatrix}^T $. The envelope of the initial excitation is Gaussian-like with azimuthal symmetry. In experiments and numerical simulations, the intensity fine structure under this envelope is a triangular lattice (it allows tuning the excitation of the two sublattices). In "low-energy" theory equation \eqref{zb:eq:dirac}, this means that the continuous field amplitudes $\psiup(r,\polarangle,z=0)$ and $\psidn(r,\polarangle,z=0)$ are independent of the azimuthal angle $\polarangle$. Because of the nontrivial Berry phase winding around the Dirac points, i.e., the topology of the system, a vortex beam component (with $\polarangle$-dependent amplitude) will dynamically emerge. The underlying universal mechanism which maps the topological singularity (vortex) from the $k$-space to the real space was discovered recently \cite{liu2020}. As such, for $z>0$, a single pseudospin component
is furnished with both the nonvortex and the vortex beam components, which naturally interfere. It is demonstrated in the previous section and shown in \cref{zb:fig:3} that without the interference of these two components, the intensity pattern of the beam retains its azimuthal symmetry. The shape of the interference fringes depends on the evolution of the phase fronts of these two components, i.e., on $\arg \gupN(r,z)-\arg\gupV(r,z)$, which yields a spiraling self-rotating pattern (see \cref{zb:fig:3}). This rotation breaks the azimuthal symmetry of the initial beam and leads to oscillation of the COM of the beam, $\vr_C(z)$, in analogy to the two-mode beating discussed above. This alternative interpretation of the Zitterbewegung oscillations is perhaps more easily visualized than the original one invoking interference between positive and negative energy states. Both interpretations are correct; however, ours gives a simple picture for the circular oscillations of the COM associated with wavepacket self-rotation.

	It should be emphasized that without the gap, there is no Zitterbewegung (see equation \eqref{zb:eq:10}). This means that the gap is crucial for the existence of radial dependence of the phase fronts $\arg\gupN(r,z)-\arg\gupV(r,z)$ that yields the spiralling intensity patterns. This can be understood because evolving phase fronts correspond to the dispersion curves. The dispersion curves drastically change when the gap opens, from the linear (conical) dispersion to the "parabolic" one. The helicity of the spiraling self-rotating motion determines the helicity of the Zitterbewegung of the COM. Consequently, it is valley-dependent in the symmetry-broken HCLs.
	
	Let us digress a bit and comment on results of conical diffraction shown in \cref{zb:fig:2} e), i.e., when $m = 0$ and the lattice band structure is a conical intersection. The far field intensity structure, i.e., the outcome of conical diffraction, depends on the initial excitation conditions, i.e., the weights and the relative phases of the Bloch modes excited around Dirac points. For example, if one excites both sublattices in phase, and simultaneously at both $K$- and $K'$-valleys, the output would be circularly symmetric \cite{peleg2007}. If one excites only a single sublattice at either $K$-valleys or $K'$-valleys, the output is also circularly symmetric \cite{song2015}. The crescent-like output in \cref{zb:fig:2} e) is the result when both sublattices are initially excited but at a single set ($K$ or $K'$) of valleys. 
	
	Finally, we discuss the crucial role played by the Berry phase. The existence of the Berry phase at each valley is responsible for the existence of the momentum to real space mapping which produces a vortex component in the field, even though the initial excitation beam is Gaussian-like. The connection between the Berry phase and Zitterbewegung has been analysed previously in literature \cite{vaishnav2008}. These analyses relied on the fact that the COM of the beam can be expressed as
$\ev{\vr} = \int \tilde{\psi}^\dagger (\vk,z) i \Nabla_{\vk} \tilde{\psi}(\vk,z) \dd^2 k$ in the momentum space representation of the field amplitude \cite{vaishnav2008,cserti2010}. When $\tilde{\psi}(\vk,z)$ is expressed in eigenmodes of the system, some terms in the expression for $\ev \vr$ will contain the Berry connection $\A(\vk)=i\psink^\dagger \Nabla_{\vk} \psink$; however, the terms corresponding to Zitterbewegung will be nonzero only if the interband matrix elements $i \psi_{-1\vk}^\dagger\Nabla_{\vk} \psi_{1\vk}$ and $i \psi_{1\vk}^\dagger\Nabla_{\vk} \psi_{-1\vk}$ are nonzero (see the supporting information%
	\footnote{See footnote \ref{note:support}.}
 of \cite{zb} for the derivation). These matrix elements take a very similar form to that of the Berry connection, except that the operator $i\Nabla_{\vk}$ is evaluated between modes of different bands. This is consistent with our experimental setting where both bands are excited. Thus, we conclude that in our observations, the key role of the Berry phase is to generate the vortex term enabling its interference with the nonvortex component, and hence the Zitterbewegung. The direction of the Berry curvature sets the helicity of the spiralling pattern, and therefore the valley-dependence of the spiralling self-rotating wavepacket.
 
\newpage

\markedsection{Topological Phase Transitions in Nonlinear Interacting Soliton Lattices\\}{Dynamically Emerging Topological Phase Transitions in Nonlinear Interacting Soliton Lattices}\label{sec:ssh}

\subsection{Introduction}
	Topological photonics offers a unique path for manufacturing photonic devices immune to scattering losses and disorder \cite{lu2014,ozawa}. Since the pioneering theoretical predictions \cite{raghu2008} and experimental demonstrations \cite{wang2009} of topologically protected electromagnetic edge states, most studies have
focused on linear topological photonic structures \cite{lu2014,ozawa}. However, by combining topology with nonlinearity \cite{lumer2013,ablowitz2014,leykam2016,malkova2009a,hadad2016,zhou2017,maczewsky2020,katan2016,mukherjee2020,solnyshkov2017,dobrykh2018,xia2020,bisianov2019,zhang2019,kruk2019,wang2019}, many opportunities for fundamental discoveries and new device functionalities arise \cite{smirnova2020b}. This is appealing also because nonlinearity inherently exists or is straightforwardly activated in most of the currently used linear topological photonic systems. The studies of nonlinear topological phenomena in photonics include, for example, nonlinear topological edge states and solitons \cite{lumer2013,ablowitz2014,leykam2016,malkova2009a,mukherjee2020,solnyshkov2017,dobrykh2018,xia2020,bisianov2019,zhang2019},
topological phase transitions activated via nonlinearity \cite{hadad2016,zhou2017,maczewsky2020,katan2016}, nonlinear frequency conversion \cite{kruk2019,wang2019}, topological lasing \cite{bandres2018,harari2018,pilozzi2016,st2017,bahari2017,zhao2018,parto2018}, and nonlinear tuning of non-Hermitian topological states \cite{xia2021,pernet2021}.

	In a recent study, we have introduced the concepts of inherited and emergent nonlinear topological phenomena \cite{xia2020}. In this classification, inherited phenomena occur when nonlinearity is a small perturbation on an otherwise linear topological system. For example, in the Su-
Schrieffer-Heeger lattice \cite{ssh1979}, nonlinearity can easily break the chiral symmetry and therefore the underlying topology; this enables coupling into an otherwise topologically protected edge state \cite{xia2020,bisianov2019}. However, many of the system properties, such as the structure of the nonlinear topological edge states and/or solitons \cite{lumer2013,ablowitz2014,leykam2016,mukherjee2020,solnyshkov2017,dobrykh2018,xia2020,bisianov2019,zhang2019}, are inherited from the corresponding linear system \cite{xia2020}. In contrast, emergent nonlinear topological phenomena occur when the underlying linear system is not topological, but the nonlinearity induces nontrivial topology \cite{xia2020}. Nonlinearity induced topological phase transitions \cite{hadad2016,zhou2017,maczewsky2020,katan2016} are examples of emergent nonlinear topological phenomena. In a recent experiment utilizing a nonlinear waveguide lattice structure \cite{maczewsky2020}, such a transition was shown to happen when power (i.e., nonlinearity)
exceeded a certain threshold value. Emergent nonlinear topological phenomena are intriguing but were scarcely explored in nonlinear topological photonics.

\begin{figure}[htb]\centering
	\includegraphics[width=.8\textwidth]{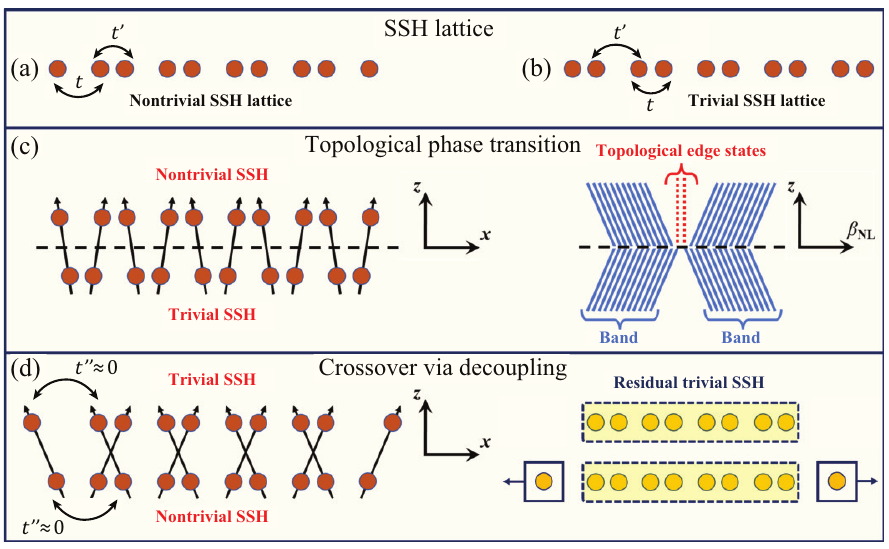}
		\caption[Illustration of the topological phase transition and crossover found in the evolving SSH soliton lattice.]{Illustration of the topological phase transition and crossover found in the evolving SSH soliton lattice. a) The SSH lattice in the topologically nontrivial regime with $t<t'$, characterized by two localized topological edge states. b) The SSH lattice in the topologically trivial regime with $t>t'$. c) Sketch of the topological trivial-to-nontrivial phase transition in real space (left) and in the spectrum (right). At the phase transition, the gap closes, and two extended eigenmodes are pulled from the bands into the gap to become topological edge states. d) Sketch of the crossover from the nontrivial to the trivial
phase via decoupling of the outermost lattice sites. The NNN coupling is negligible in our SSH lattice, $t''\approx0$, which results in decoupling during evolution in our system (left). This is equivalent to pulling off the outermost SSH lattice sites to infinity, leaving the residual lattice in the trivial phase (right).}
		\label{ssh:fig:1}
	\end{figure}

	Here we report on dynamical topological phase transitions entirely driven by nonlinearity, which constitute an example of emergent nonlinear topological phenomena. These phase transitions occur in colliding soliton lattices and are enabled by elastic soliton collisions. In optics, spatial solitons are stable localized optical beams, which occur when diffraction is balanced by nonlinearity \cite{chen2012}. Here we create two 1D soliton sublattices and initially kick them in opposite directions. As the sublattices evolve and collide, they form a paradigmatic model of topological physics: the SSH lattice \cite{barisic1,barisic2,barisic3,ssh1979}. Recall that this lattice can be in both the
topologically nontrivial (\cref{ssh:fig:1} a)) and trivial (\cref{ssh:fig:1} b)) phase, depending on the Zak phase \cite{zak}. We find two kinds of interesting phenomena, which periodically occur: (i) a dynamical topological phase transition from the topologically trivial to nontrivial phase, characterized by a gap closing and reopening at a single point, where two extended states are pulled from the bands into the gap to become localized topological edge states (\cref{ssh:fig:1} c)), and (ii) a crossover from the topologically nontrivial to the trivial regime, which occurs via decoupling of the edge states from the bulk of the lattice (\cref{ssh:fig:1} d)).

	We emphasize up front that there is a distinction between our system and those from refs. \cite{hadad2016,zhou2017,maczewsky2020}, which all exhibit nonlinearity-induced topological phase transitions. In the theoretical models of refs. \cite{hadad2016,zhou2017}, the photonic lattices are fixed in the $x$ space. In \cite{maczewsky2020} they are fixed in the $x$-$z$ space (i.e., "spacetime"); the power of an external excitation can change the coupling via nonlinearity to induce a phase transition. In our system, the whole lattice autonomously and nonlinearly evolves in the $x$-$z$ space, resulting in different topological phases along $z$ (i.e., "time"). The surprising connection between interacting soliton lattices and nontrivial topology is revealed by the phase transitions and crossovers accompanied by the "birth" and "death" of topological edge states. This is reminiscent of the connection between topology and quasicrystals, also revealed by the phase transitions \cite{kraus2012}. In addition, our work is also distinct from a recent endeavor in the topological control of nonlinear extreme waves \cite{marcucci2019}. 
	
	We first outline a few basic facts about the SSH lattice. It is a 1D topological system, which exists due to the underlying chiral symmetry \cite{ozawa,ssh1979}. In its topologically nontrivial phase, the intercell coupling $t'$ is stronger than the intracell coupling $t$ ($t$ < $t'$) (see \cref{ssh:fig:1} a)). The nontrivial SSH lattice has two topological edge modes with propagation constants residing in the band gap and a characteristic phase structure \cite{ssh1979,malkova2009b}. In the trivial phase $t$ > $t'$ (see \cref{ssh:fig:2} b)), there are two bands separated by a gap, and all eigenmodes are extended. This model has been implemented in versatile systems, including photonics
and nanophotonics \cite{malkova2009b,keil2013,xiao2014,blanco2016,weimann2017}, plasmonics \cite{poddubny2014,bleckmann2017}, as well as quantum optics \cite{kitagawa2012,cardano2017,blanco2018,bello2019}. Some of the aforementioned nonlinear topological phenomena have been studied also in the nonlinear SSH model \cite{malkova2009a,hadad2016,zhou2017,solnyshkov2017,dobrykh2018,xia2020,bisianov2019,kruk2019,zhao2018,parto2018}.

\subsection{Results}
	We consider the propagation of a linearly polarized optical beam in a nonlinear medium, which is described by a nonlinear Schrödinger equation (NLSE),
	\begin{equation}\label{ssh:eq:1}
		i\pdv{\psi}{z} + \frac{1}{2k}\pdv[2]{\psi}{x} + \gamma \abs{\psi}^2 \psi(x,z) = 0,
	\end{equation}
	where $\psi(x,z)$ refers to the electric field envelope, $\gamma$ defines the strength of the nonlinearity (we assume a Kerr-type nonlinearity), and $k$ is the wave number in the (isotropic) medium. The NLSE possesses a family of soliton solutions, with the hyperbolic-secant soliton being the most representative \cite{agrawal}:
	\begin{equation}\label{ssh:eq:2}
		\psi_S(x,z;\kappa,\theta) = \sqrt{I_0} \sech(\frac{x}{x_0}-\frac{\kappa z}{k x_0^2})
							\exp[i\oble{ \frac{\kappa}{x_0}x + \frac{1-\kappa^2}{2kx_0^2}z + \theta }].
	\end{equation}
	Here, $x_0$ is a scaling factor, $\kappa/x_0$ is the initial momentum, $I_0$ defines the peak intensity, and $\theta$ is an arbitrary phase. The stationary propagation is achieved when diffraction (quantified by the diffraction length $kx_0^2$ ) is balanced by nonlinearity (quantified by the nonlinear length $1/\gamma I_0$ ), that is, when $\gamma I_0 = (kx_0^2)^{-1}$.

	Nontrivial topology in photonics is usually implemented via specially designed topological photonic structures where light propagates \cite{ozawa}, whereas light propagation in a homogeneous and isotropic nonlinear medium described by equation \eqref{ssh:eq:1} is usually unrelated to nontrivial topology. Unexpected nontrivial topology emerges from the initial condition(s) given by
	\begin{equation}\label{ssh:eq:3}
		\psi(x,0) = \sum_{j=-M}^M \psi_S (x-T-jd,0;-\kappa,0) 
					+ \sum_{j=-M}^M \psi_S (x+T-jd,0;\kappa,\theta),
	\end{equation}
	where the first sum relates to sublattice B, and the second to sublattice A. The parameter $d$ defines the size of the unit cell, and $T$ is the initial offset between the two sublattices.
The next-nearest-neighbour (NNN) tunnelling in our SSH lattice is negligible, $t''\approx0$. Due to the presence of nonlinearity, soliton interaction results in a dynamically evolving optically induced lattice. To understand its properties, we study the eigenvalues $\beta_{\mathrm{NL},n}(z)$ and the eigenmodes $\phi_{\mathrm{NL},n}(x,z)$ of the (nonlinearly) optically induced lattice potential $V(x,z)=-\gamma\abs{\psi(x,z)}^2$, defined by ${H \phi_{\mathrm{NL},n}= \beta_{\mathrm{NL},n}\phi_{\mathrm{NL},n}}$, where $H=-(2k)^{-1}\partial_{xx} + V$. Here we use the continuous potential $V(x,z)$ for better correspondence with experiments; one could in principle use the SSH Hamiltonian in the tight-binding approximation as NNN tunnelling is negligible. An equivalent approach for evolving nonlinear topological lattices was adopted in ref. \cite{xia2020}.

\begin{figure}[htb!]\centering
	\includegraphics[width=.8\textwidth]{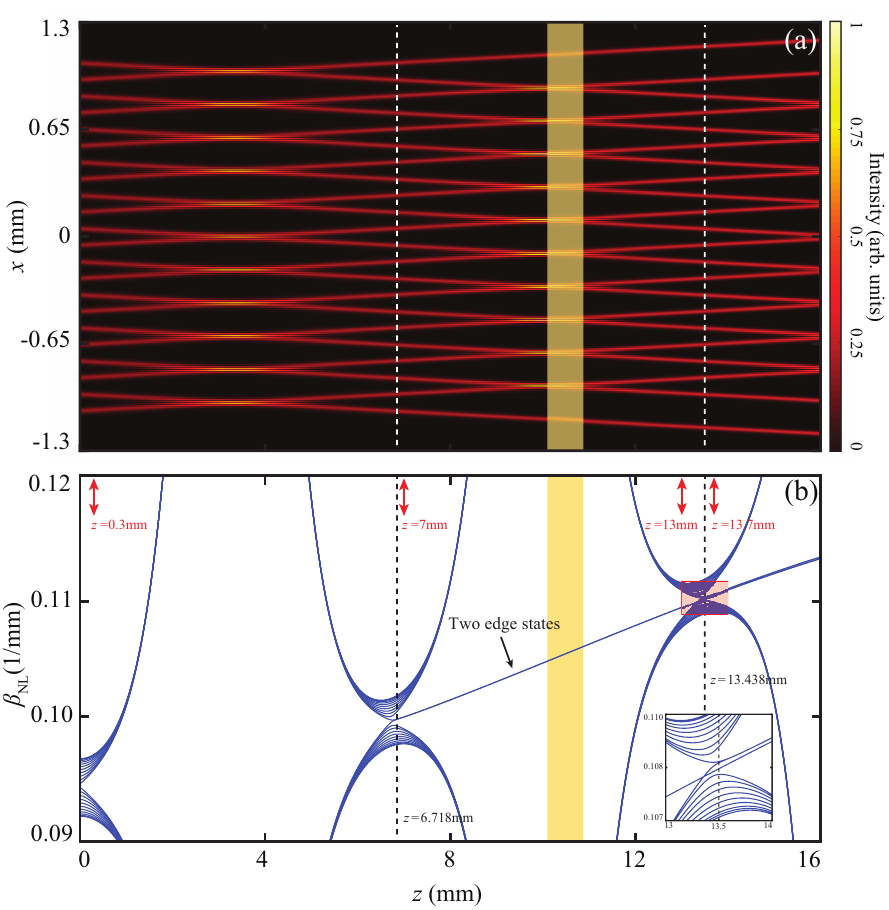}
		\caption[Intensity and spectrum of the SSH soliton lattice
evolving with propagation distance $z$.]{Intensity (a) and spectrum (b) of the SSH soliton lattice
evolving with propagation distance $z$. Locations where topological phase transitions take place are indicated with vertical dashed lines, while the crossover region is highlighted by the yellow
stripe. Topological phase transitions occur at the gap-closing points, after which two extended eigenmodes are pulled from the bands into the gap and become topological edge states (i.e., the
phase transition here is from the trivial to the nontrivial phase). Between these closing points, there is a crossover from the nontrivial to the trivial phase via decoupling of the edge states from
the bulk of the lattice, which can be understood by comparing a) with \cref{ssh:fig:1} d). At the second transition stage, two new edge states emerge, as clearly seen in the enlarged inset in b), while the old ones had by this point point turned into decoupled walk-off solitons. Parameters: $M=5$, $T=d/4=50\,\mathrm{\mu m}$, $\theta=\pi$, $x_0=18.0\,\mathrm{\mu m}$, $\kappa=5$, $k=1.71\times 10^7 \,\mathrm{m}^{-1}$, and $\gamma I_0 =(kx_0^2)^{-1}$.}
		\label{ssh:fig:2}
	\end{figure}

	In \cref{ssh:fig:2} a) we show the numerically calculated intensity of the evolving soliton lattice. The two sublattices propagate in opposite directions and periodically collide, but they keep their sublattice structures and propagation directions intact after every collision, which is ensured by the colliding properties of (Kerr-type) solitons \cite{chen2012}. The intercell and intracell distances are equal at $z=0$, because we have chosen $T=d/4$; $\kappa>0$ implies that the sublattices initially approach each other. Thus, in the $z$ interval from
$z=0$ until the first collision, the soliton lattice has the structure of the trivial SSH lattice (see \cref{ssh:fig:2} a)). After the first collision, the lattice retains its trivial topology until the intercell and intracell distances became equal again for the first time after $z=0$. This point is denoted with a vertical dashed line at ${z=6.718\,\mathrm{mm}}$ in \cref{ssh:fig:2}. At that point, the lattice undergoes a topological phase transition from the trivial to the nontrivial SSH soliton lattice, illustrated in real space in \cref{ssh:fig:2} a) and the left-hand side of \cref{ssh:fig:1} c).

\begin{figure}[htb!]\centering
		\includegraphics[width=.8\textwidth]{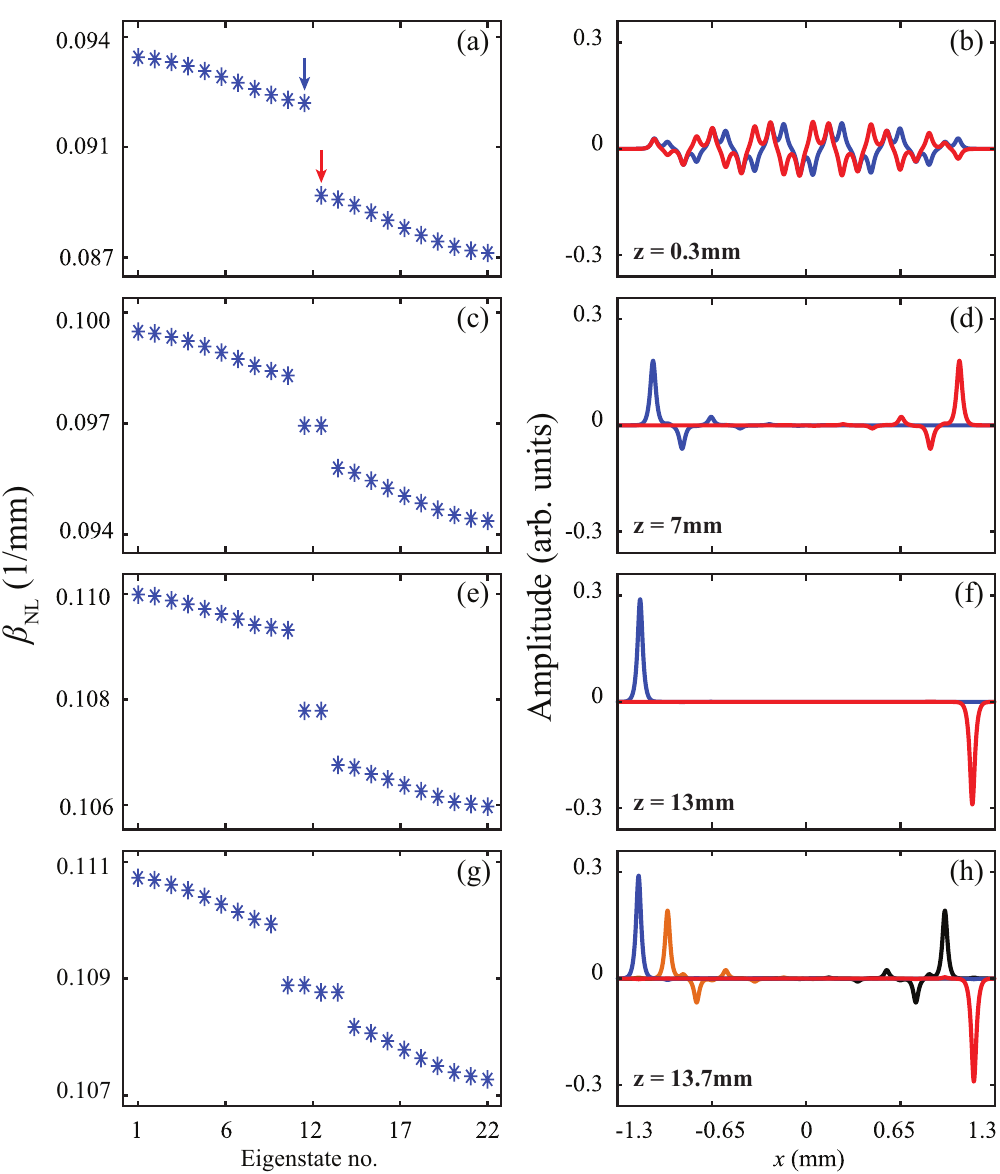}
		\caption[Spectra of the evolving soliton lattice and
selected eigenmodes]{Spectra of the evolving soliton lattice (left column) and
selected eigenmodes $\phi_{\mathrm{NL},n}$ (right column) at propagation distances indicated by red arrows in \cref{ssh:fig:2}. a) Spectrum and b) two eigenmodes in the trivial phase at ${z=0.3\,\mathrm{mm}}$. The two eigenmodes are closest to the gap as indicated with arrows in a).
c) Spectrum and d) topological localized states in the nontrivial phase at ${z=7\,\mathrm{mm}}$, just after the first topological phase transition. e) Spectrum and f) localized states after the nontrivial-to-trivial crossover at ${z=13\,\mathrm{mm}}$. The states are localized solely in the outermost solitons and their amplitude is zero in the bulk of the soliton lattice, which is in contrast to the amplitude-phase structure of topological edge states shown in d). g) Spectrum and h) localized states at ${z=13.7\,\mathrm{mm}}$, after the
second phase transition. Two of the localized states are topological (black and orange lines), whereas the other two are outermost solitons (blue and red lines).}
		\label{ssh:fig:3}
	\end{figure}

	An ultimate signature of the dynamical topological phase transition is illustrated in \cref{ssh:fig:2} b), which shows the band gap structure of the evolving soliton lattice. We see that for $z$ values up to the first topological phase transition point at ${z=6.718\,\mathrm{mm}}$, there are two bands without any states in the gap. At the transition point, the gap closes and
immediately reopens, while two eigenvalues are pulled from the bands to stay within the gap. These isolated eigenvalues correspond to the topologically nontrivial edge states of the SSH soliton lattice, with characteristic phase and amplitude structure, illustrated in \cref{ssh:fig:3} d) \cite{xia2020,ssh1979,malkova2009b}. They dynamically emerge at the transition point. Gap closing is an inevitable and necessary ingredient of the topological phase transition that is clearly illustrated in \cref{ssh:fig:2} b).

	In order to fully unveil the behaviour of our system, we explore the band gap structure and the modes of the SSH soliton lattice before the transition in figs. \ref{ssh:fig:3} a), b) (for concreteness we consider ${z=0.3\,\mathrm{mm}}$), and just after the transition in figs. \ref{ssh:fig:3} c), d) (at ${z=7\,\mathrm{mm}}$). At ${z=0.3\,\mathrm{mm}}$ there are two bands separated by the gap (see
\cref{ssh:fig:3} a)). All eigenmodes of the lattice are extended. In \cref{ssh:fig:3} b) we plot the two extended modes with eigenvalues closest to the gap. At the phase transition, these two
extended modes are pulled from the band into the gap (see \cref{ssh:fig:3} c)); at this point they became localized topological edge modes of the SSH soliton lattice, illustrated in \cref{ssh:fig:3} d). We see that both of them have the characteristic features of the topological edge modes: their amplitude is nonzero only in odd lattice sites (counting from the edge inward), and the neighbouring peaks in the mode amplitude are out of phase (see, e.g. \cite{xia2020,ssh1979,malkova2009b}).

	A glance at \cref{ssh:fig:2} b) reveals an interesting feature of the
evolving spectrum at $z=13.438\,\mathrm{mm}$: another gap closing and reopening occurs, where two eigenstates bifurcate from the bands to become localized in the gap; see the inset in \cref{ssh:fig:2} b) and figs. \ref{ssh:fig:3} g), h). This appears to be another topological phase transition from the trivial to the nontrivial SSH lattice. However, if this interpretation is correct (as we show below), it means that the system is converted from the nontrivial to the trivial regime in between the two gap-closing points depicted in \cref{ssh:fig:2} b). This conversion is not a topological phase transition because the gap remains open at all propagation distances between ${z=6.718\,\mathrm{mm}}$ and ${z=13.438\,\mathrm{mm}}$.

	To explain this intriguing phenomenon, we need to resort to the real space dynamics in \cref{ssh:fig:2} a), and explore the region shaded in yellow where the soliton collisions take
place. In this region the two outermost solitons become separated from the lattice, because the distance to their nearest neighbours becomes $d$, which is the NNN distance in the SSH lattice, and thus the probability of tunnelling from these outermost solitons to the bulk of the SSH lattice is practically zero. The eigenvalues corresponding to the outermost solitons are in the gap (see \cref{ssh:fig:3} e)), so the eigenmodes are obviously localized (see \cref{ssh:fig:3} f)), but their amplitude-phase structure does not possess the feature of the topological edge states illustrated in \cref{ssh:fig:3} d). Thus, in the yellow region, two outermost solitons are actually
decoupled from the SSH lattice, which leads to the crossover from the topologically nontrivial to the trivial phase. Instead of a discontinuous change of the system's phase, the lattice sheds the outermost sites, thus forcing us to redefine the boundary of the system, as well as the unit cell of the lattice. The nontrivial phase of the old lattice (pre-crossover) corresponds to the trivial phase of the new latice (post-crossover). This crossover is fully equivalent to a gradual process of pulling two outermost lattice sites of the nontrivial SSH lattice into infinity, as illustrated in \cref{ssh:fig:1} d).

	The existence of the crossover is in full agreement with the observation and interpretation of the gap-closing point at ${z=13.438\,\mathrm{mm}}$ in \cref{ssh:fig:2} b) described above. This
pattern of an alternating sequence of events—dynamical topological phase transitions (trivial to nontrivial phase) → crossover via decoupling of the outermost solitons (nontrivial to trivial phase)—repeats itself during propagation until the two sublattices become fully separated. The
sublattice constant $d$ is chosen sufficiently large so that the NNN tunneling probability is negligible; therefore, when sublattices become separated, we can regard this system as a set of independent solitons.

	The evolving nonlinear lattice has chiral symmetry in those $z$ intervals where $V(x,z)$ corresponds to either the trivial or the nontrivial SSH lattice. When the two sublattices collide/overlap, there is no more chiral symmetry, but each on-site potential has two bound modes leading
to two bands. The overall structure of the dynamically evolving lattice in real space is stable with respect to perturbations of the initial state. For perturbations preserving the lattice and the chiral symmetry, the evolving nonlinear spectrum is robust, as the gap-closing points
indicating topological phase transitions are present, and the edge states have characteristic topological features. Perturbations which break the chiral symmetry will, strictly speaking, destroy the topological phase. However, if they are sufficiently small, the characteristic features of the topological states will be inherited and present in the perturbed system; see the supplemental material\footnote{\label{note:ssh_suppl}\url{http://link.aps.org/supplemental/10.1103/PhysRevLett.127.184101}} to \cite{solitonssh}.

	In conclusion, we have found dynamically emerging topological phase transitions in SSH soliton lattices, which are classified as emergent nonlinear topological phenomena because they cease to exist if nonlinearity is turned off. These phase transitions convert the SSH soliton lattices from the topologically trivial to the nontrivial phase and are characterized by the gap closing and reopening, accompanied by the emergence of two localized topological edge states. In addition, we have found crossovers from the topologically nontrivial to the trivial regime, which occur via decoupling of the edge states from the bulk of the lattice. These two events occur one after the other in succession. Our results are presented in a spatial optical system; however, they are
accessible also in nonlinear fiber optics with realistic parameter values \cite{bohac2010} (see supplemental material \cite{solitonssh}). In nonlinear saturable media such as photorefractive crystals, soliton collisions are typically not elastic; thus, an observation of the proposed phenomena should be more
challenging (in the low saturation regime however, Kerr nonlinearity that we used here is typically a good approximation). We envisage that this work will lead to exciting fundamental research in nonlinear topological photonic systems, including the recently demonstrated nonlinear higher-order topological insulators \cite{hu2021,kirsch2021}.

%% file: 3_anyons/synth_anyons.tex
The work presented in this chapter has been published in:
\begin{itemize}
	\item[\cite{prb}] \bibentry{prb}
\end{itemize}

	In this chapter, we explore an unconventional way to realize anyonic statistics by perturbing a noninteracting integer quantum Hall effect system with specially tailored localized probes. Rather than exciting quasiparticles (which would obey integral statistics), the probes serve to induce synthetic anyons by modifying the ground state in such manner that braiding of the probes leads to Abelian fractional phases. Since the probes can be used as handles to induce and manipulate the synthetic anyons, an experimental realization of our model could potentially be a step towards topological quantum computers.

	In the introduction to the next section, we present the motivation for our model, give a brief overview of the related theoretical and experimental developments, and summarize the results to be presented. In \ref{sec:PRB:gswf} we present the system in detail, and discuss the solutions to the model. We consider many-body ground state wave function, as well as the single-particle states in the lowest Landau level, and the gap between the lowest and the first excited Landau levels. In \cref{sec:PRB:BPH}, we obtain the statistical phase analytically and numerically. We discuss the distinction between synthetic anyons and emergent quasiparticles in \cref{sec:PRB:notQPs}, and in \cref{sec:PRB:fusion} we explain how this impacts the anyon fusion rules. Finally, we discuss the results and conclude the chapter in \cref{sec:PRB:discussion}.

\markedsection{Synthetic anyons in a noninteracting system}{Exact solutions of a model for synthetic anyons in a \\noninteracting system}\label{sec:PRB}

\subsection{Introduction}
	Since the possibility of fractional statistics was discovered \cite{LeinaasMyrheim,wilczek1982}, a considerable amount of theoretical and experimental work was done to explore the topologically ordered 2D systems that may host anyonic excitations. Recall from \cref{sec:TQM:anyons} that the various fractional Hall effect systems are believed to support anyons obeying Abelian and non-Abelian fractional statistics \cite{arovas,camino2005,nayak}. While the FQHE is the paradigmatic example, anyons are also believed to arise in other systems. For example, some interacting spin systems may be topologically ordered and have anyonic excitations \cite{kitaev2003,kitaev2006,dai2017,jansa2018,savary2016}. On the other hand, non-Abelian Majorana zero modes may be associated with excitations of topologically ordered systems, but also with defects in topological superconductors \cite{sarma2015,mourik2012}.
	While the fractional charge of the Laughlin quasiparticles was observed in the quantum shot noise of the electrical current through a point contact in a 2D electron gas \cite{picciotto1998}, it has proven difficult to make anambiguous measurements of their anyonic phases (though detections were claimed \cite{camino2005}). Convincing signatures of anyons were obtained only recently \cite{bartolomei2020, nakamura2020}.
	
	As explained in \cref{sec:TQM:anyons}, the chief source of motivation for studying anyons is their potential application in topologically protected quantum computation \cite{kitaev2003,nayak}. However, efficient methods for creation, detection and manipulation of non-Abelian anyons are necessary. 
	Many experiments have addressed anyonic systems, of which we only mention some. In condensed matter, this includes the already mentioned experiments on the FQHE \cite{camino2005}, Majorana zero modes \cite{mourik2012}, and the Kitaev paramagnetic state of the honeycomb magnet $\mathrm{RuCl_3}$ \cite{jansa2018} (see refs. \cite{sarma2015,nayak} for reviews). Experiments on other platforms were also performed.
	 For example, a minimal variant of the Kitaev toric model \cite{kitaev2003}, concieved as a platform for topological quantum computing, was experimentally realized in ultracold atomic gases \cite{dai2017}, and with trapped ions using dissipative pumping processes \cite{barreiro2011}. Anyonic statistics was simulated in photonic quantum simulators \cite{lu2009,pachos2009}, in superconducting quantum circuits \cite{zhong2016}, and by using nuclear magnetic resonance \cite{li2017}. The body of theoretical proposals is larger (we will not attempt to provide a review) and, besides condensed matter systems \cite{sarma2015,nayak}, includes proposals in ultracold atomic gases based on emulating the FQHE \cite{paredes2001,zhang2014} or the Kitaev model \cite{duan2003,jiang2008}, or by employing synthetic gauge potentials \cite{burrello2010}. Different mechanisms to achieve FQH states of light have also been proposed \cite{kapit2014,umucalilar2017}. Furthermore, it was recently proposed that anyonic charge-flux composites can be achieved by sandwiching a charged magnetic dipole between two semi-infinite blocks of a high-permeability material \cite{marija2018}.
	 
	 Despite considerable progress, there is still a long way to go before fault tolerant topological quantum computation is experimentally feasible \cite{sarma2015,nayak}. Thus, there is a motivation to explore less traditional schemes for realizing and manipulating anyons. For example, it was proposed that anyons could be synthesized by coupling weakly interacting (or noninteracting) electrons to a topologically nontrivial background (or topologically nontrivial external perturbations) \cite{weeks2007,rosenberg2009,rahmani2013}. In refs. \cite{weeks2007,rosenberg2009}, anyons are proposed in a system of an artificially structured type-II superconducting film, adjacent to a two-dimensional electron gas in the integer QHE with unit filling fraction \cite{weeks2007,rosenberg2009}. A periodic array of pinning sites imprinted on the superconductor will structure an Abrikosov lattice of vortices \cite{weeks2007}. Anyons are bound by
vacancies (interstitials) in the vortex lattice, which carry a deficit (surplus) of one-half of a magnetic flux quantum \cite{weeks2007}. In ref. \cite{rahmani2013} anyons were proposed in integer QHE magnets. Magnetic vortices in this system are topologically stable and have fractional electronic quantum numbers yielding anyonic statistics. Anyons were also proposed by using topological defects in graphene \cite{seradjeh2008}.

	Here we present exact solutions of a model for synthetic anyons, which was considered in refs. \cite{weeks2007,rosenberg2009} (it was referred to as the continuum model therein). The model is represented by the Hamiltonian for noninteracting 2DEG, in a uniform magnetic field, with $N$ external solenoids (probes), which introduce localized fluxes at positions $\veta_k$, for $k=1,\dots,N$. 
	We find analytically and numerically the ground state of this Hamiltonian when the Fermi energy is such that only the lowest Landau-level (LLL) states are populated. When the flux through each solenoid is a fraction of the flux quantum, i.e. $\Phi=\alpha\flqnt$, the ground-state wave function is anyonic in the coordinates of the external probes $\veta_k$.  
	In other words, by braiding the probes one imprints the Berry (statistical) phase \cite{berry1984} on the ground state. We calculate this Berry phase analytically and numerically. A potential experimental realization of this model must have a mechanism that fixes the flux in external solenoids to an identical value in order for synthetic anyons to be identical entities.
	 From the solutions we find that around every solenoid probe there is a cusp-like dip of missing electron charge $\Delta q$. We demonstrate that the missing charge should not be identified with the concept of an emergent quasiparticle by showing that $\frac{\Delta q}{\hbar} \oint\VP\cdot\dd \bm l$ does not correspond to the Aharonov-Bohm phase \cite{AB} acquired as the probe traverses a loop in space. One could arrive at the same conclusion by using gauge invariance arguments \cite{rahmani2013}. 
	 This has consequences on the fusion rules of the synthetic anyons: the fusion rules depend on the microscopic details of the fusion process as discussed below. Even though we consider Abelian anyons, if an analogous scheme for synthetic non-Abelian anyons is developed, it will be a potential path towards a platform for quantum computers.
	 
\subsection{Ground-state wave function}\label{sec:PRB:gswf}
\begin{figure}[htb]\centering
		\includegraphics[width=0.75\textwidth]{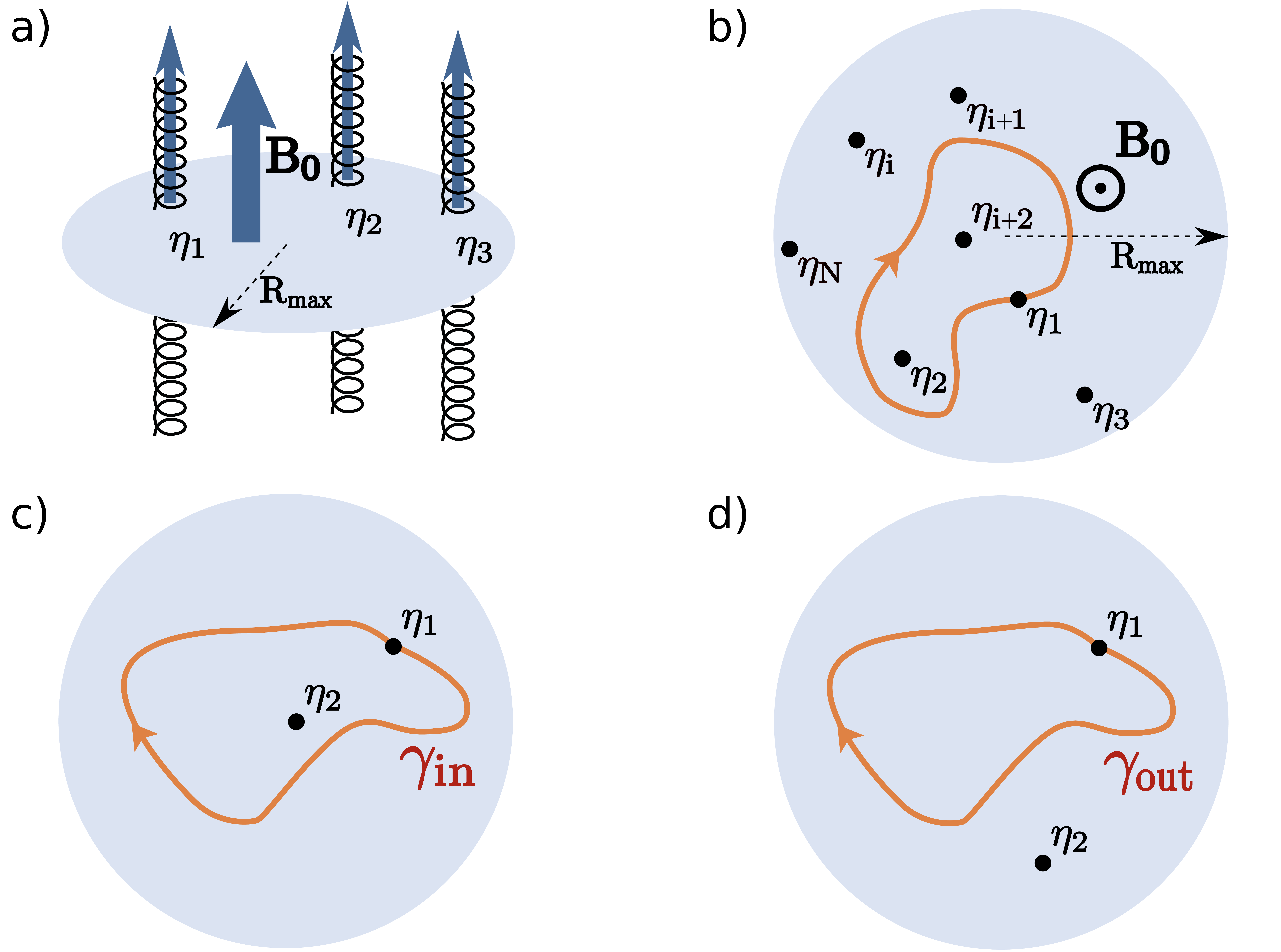}
		\caption[Sketch of the model for synthetic anyons]{Sketch of the model. a) We explore a 2DEG in a magnetic field $\B_0$ on a disk of radius $R_\mathrm{max}$. The solenoid probes with flux $\Phi$ pierce the 2DEG at positions $\eta_j$  (coordinates are in complex notation). b) The contour path of one probe, which adiabatically traverses a closed loop in space; we are interested in the Berry phase accumulated along such paths. Contours are illustrated corresponding to c) $\gamma_\mathrm{in}$ and d) $\gamma_\mathrm{out}$. See text for details.}
		\label{PRB:fig:model}
	\end{figure}

	In our theoretical model we consider $N_e$ noninteracting spin-polarized electrons in 2D configuration space (in the xy-plane), in a uniform magnetic field $\B_0 = \Nabla\times\VP_0=B_0\uvec{z}$ where $\VP_0(\vr)=\frac{1}{2}\B_0\times\vr$ is the vector potential in the symmetric gauge ($B_0>0$). The system is perturbed with $N$ very thin solenoids at locations $\veta_k=\eta_{x,k}\uvec x + \eta_{y,k}\uvec y$. The vector potential of each solenoid is
	\begin{equation}\label{PRB:eq:solVP}
		\VP_k(\vr)=\frac{\Phi}{2\pi} \frac{\uvec z \times (\vr-\veta_k)}{|\vr-\veta_k|^2},
	\end{equation}
	where $\Phi$ is the magnetic flux through each solenoid. The Hamiltonian representing the model is then
	\begin{equation}\label{PRB:eq:H}
		H	= \sum\limits_{j=1}^{N_e} \uglate{ \frac{1}{2m} \left( 
						\bm{p_j} - q\VP_0(\bm{r_j}) - q\sum_{k=1}^N\VP_k(\vr_j) \right)^2 + V(\vr_j) },
	\end{equation}
	where $V(\vr)$ is zero for $r<R_\mathrm{max}$, and infinite otherwise; $q<0$ and $m$ are the electron charge and mass, respectively.  The model is illustrated in \cref{PRB:fig:model}
	Compare this Hamiltonian to that of the continuum model for the integer quantum Hall effect \eqref{IQHE:eq:cont_H}. The only difference is in the vector potential $\VP_k(\vr_j)$ due to the thin solenoid probes, which causes spectral flow of the IQHE single-particle states, but preserves the Landau level structure far away from the probes. We assume that the Fermi level is such that only the states from the LLL of energy $\hbar\omega_B/2$ are populated, and we assume they are all populated. The ground-state wave function with energy $N_e\hbar\omega_B/2$ is given by
	\begin{equation}\label{PRB:eq:MBGS}
		\psi = \frac{1}{\sqrt{Z}}
			\left[ \prod\limits_{j=1}^{N_e} \prod\limits_{k=1}^{N} \left| z_j-\eta_k \right|^{-\alpha} \conj{z_j-\eta_k} \right] \left[ \prod\limits_{i<j}^{N_e}\conj{z_i-z_j} \right]
			\exp(-\sum\limits_{i=1}^{N_e}\frac{|z_i|^2}{4l_B^2}),
	\end{equation}
	where $\alpha=\Phi/\flqnt$, $Z=Z(\{\eta_k\},\{\conj{\eta_k}\})$ is the normalization factor, $z_j$ and $\eta_k$ are (respectively) the electron and probe coordinates in complex notation, and the functional dependence $\psi=\psi(\{z_j\}, \{\conj{z_j}\}; \{\eta_k\}, \{\conj{\eta_k}\})$ is implied. We consider $\alpha\in\ointerval{0}{1}$; results for fractional values outside the $\ointerval{0}{1}$ interval are easily deduced. The "electron-electron" factors (in second brackets) are the first powers of $\conj{z_i-z_j}$, which is a form characteristic of the IQHE states. We may compare this state to the $1/m$ FQHE Laughlin state with $N$ quasiholes \eqref{FQHE:eq:eta_state} in which the same factors are raised to the $m$-th power. As we will show, the system is nevertheless anyonic in coordinates $\eta_k$, since the "electron-probe" factors (in first brackets) are now modified with the $|z_j-\eta_k|^{-\alpha}$ factor. 
	
\begin{figure}[htb]\centering
		\includegraphics[width=0.9\textwidth]{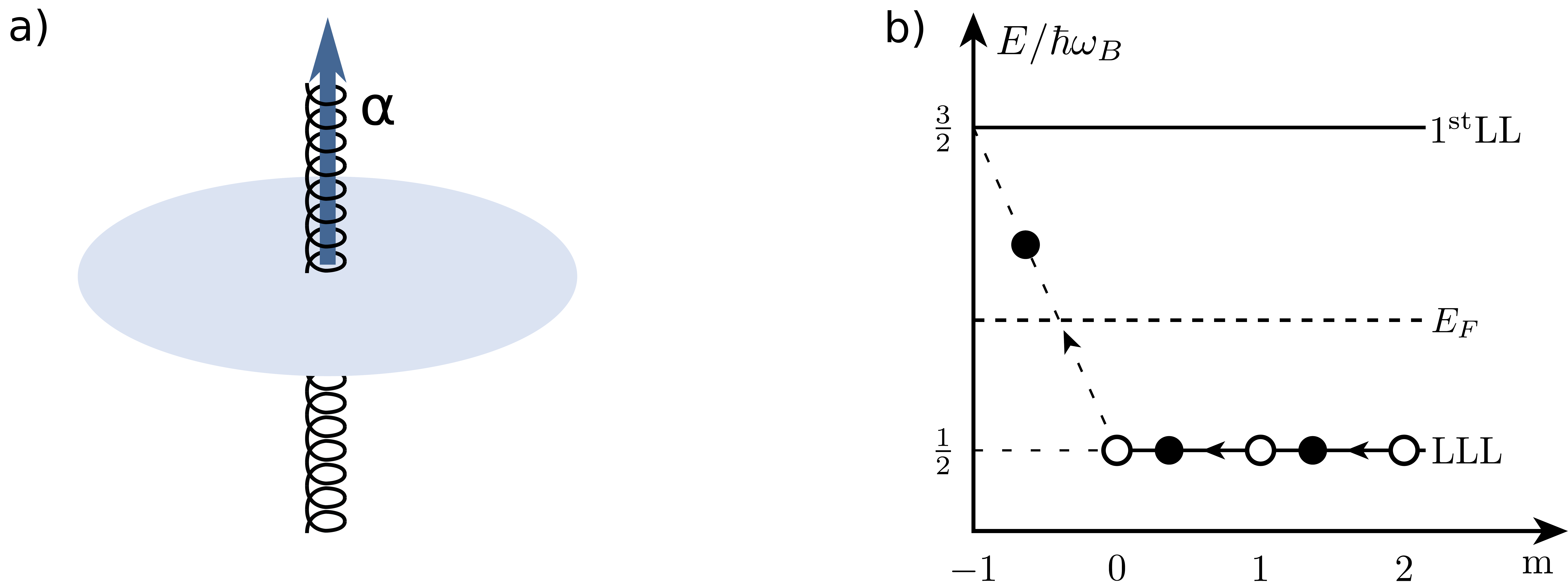}
		\caption[Sketch of the energy scales and the spectral flow for just one probe.]{Sketch of the energy scales and the spectral flow for just one probe. a) A probe is centered in the system; its flux is such that $0\leq\alpha=\Phi/\flqnt\leq 1$. b) As $\alpha$ is increased, there is a spectral flow as illustrated. The Fermi energy $E_F$ is always set such that only the LLL states are populated. See text for details.}
		\label{PRB:fig:spflow}
	\end{figure}

	For the clarity of the presentation, we first present what happens when only one probe is placed in the system, and subsequently what happens when two probes are inserted. For a single probe, the single-particle states of the system at the LLL energy are given by (see Appendix A in \cite{prb} for details of the calculation)
	\begin{equation}\label{PEB:eq:wf_1sol}
		\psi_m = |z-\eta|^{-\alpha} \conj{z-\eta} \,\conj{z}^m \exp(-\frac{|z|^2}{4l_B^2}), 
		\quad m=0,1,2,\dots \,.
	\end{equation}
	Note that as $\eta\rightarrow 0$, the state becomes \eqref{IQHE:eq:LLL_corbino_alpha}, except $m$ is shifted by $1$, thus avoiding a spurious divergent state\footnote{We note that
in ref. \cite{weeks2007} this spurious state was used to construct the
many-body ground state, and as a result the ground state from
ref. \cite{weeks2007} is in fact not anyonic (see below our discussion on
gauge invariance in calculating the Berry phase).
} $|z|^{-\alpha}\exp(-\frac{|z|^2}{4l_B^2})$ (see \cite{prb}). 
	There is one state localized at the position of the probe, with energy $\hbar\omega_B(1+2\alpha)/2$ in between the LLL and the first excited LL
	\begin{equation}\label{PRB:eq:LS}
		\psi_{LS} = |z-\eta|^\alpha \exp( -\frac{|z-\eta|^2+\conj\eta z - \eta\conj{z}}{4l_B^2} ).
	\end{equation}
	Suppose that one introduces the solenoid probe at some point in time, and adiabatically increases the flux through it. As $\alpha$ increases from zero to one, spectral flow occurs as illustrated in \cref{PRB:fig:spflow}. As we can see, one state from the LLL rises in energy and flows
towards the first Landau level. Here we assume that the flux is fixed at some value $\alpha$, and the Fermi energy is between the LLL energy and $\hbar\omega_B(1+2\alpha)/2$, and thus, this localized state is not populated.  The many-body ground state is constructed by inserting all LLL states in a Slater determinant and it is given by \eqref{PRB:eq:MBGS} for $N=1$. 

	For the case of two probes, the single-particle states of the system at the LLL energy are
	\begin{equation}
		\psi_m = |z-\eta_1|^{-\alpha}|z-\eta_2|^{-\alpha} \conj{z-\eta_1}\,\conj{z-\eta_2} \, \conj{z}^m \exp(-\frac{|z|^2}{4l_B^2}), \quad m=0,1,2,\dots \, .
	\end{equation}
	Now there are two localized states in between the LLL and the first excited Landau level. We did not find analytical expressions for these states, but they are visible in numerical calculations. The energies of these localized states are in the gap, between the LLL and the first excited Landau level. They increase with increasing $\alpha$ and join the first excited Landau level when $\alpha=1$ as expected. The many-body ground state is given by \eqref{PRB:eq:MBGS} for $N=2$.
	
	Now we generalize our results for any number of probes $N$. To this end, we employ the following singular gauge transformation
	\begin{equation}\label{PRB:eq:singulargauge}
		\psi' = \psi \prod_{i=1}^{N_e}\prod_{j=1}^N \exp(i\alpha\phi_{ij}),
	\end{equation}
	where $\phi_{ij}$ denotes the argument of $z_i-\eta_j=|z_i-\eta_j|\exp(i\phi_{ij})$. In this gauge, the vector potential of the probes is $\VP_k'=\bm 0$ everywhere except at the positions of the probes, and the Hamiltonian $H'$ is given by \eqref{PRB:eq:H} with $\VP_k$ replaced by $\VP_k'=\bm 0$. It is straightforward to verify that $\psi'$ is an eigenstate of $H'$ with energy $N_e\hbar\omega_B/2$, and hence the ground state.
	
	It should be pointed out that in the limit $\alpha\rightarrow 0$ the wave function \eqref{PRB:eq:MBGS} does not approach the IQHE ground state with all LLL states filled, but rather it becomes an IQHE state with $N$ of the LLL states left empty. Namely, the localized states which appear at the position of the probes for $\alpha>0$ are not included in the Slater determinant used to construct the ground state \eqref{PRB:eq:MBGS}, as discussed above. For $\alpha=0$ they enter the LLL, but since they were not used in constructing \eqref{PRB:eq:MBGS}, the wave function does not approach the IQHE ground state (with all LLL states filled) in the limit $\alpha\rightarrow 0$. Strictly speaking, \eqref{PRB:eq:MBGS} is the ground state for $\alpha\in\ointerval{0}{1}$, provided that only the LLL states are filled. It is not the ground state for $\alpha=0$ and all LLL states filled. 
	
	In a potential experimental implementation of the proposed system, one should not populate the localized states such as $\psi_{LS}$. With this state populated, the ground state is no longer
anyonic in the coordinates of the probes. For this state to remain empty, the temperature $T$ must be sufficiently low in order that $k_BT\ll\hbar\omega_B\alpha$, which is difficult to obtain for small $\alpha$. However, an additional localized repulsive scalar potential at the location of the probes (e.g., the delta function potential), which may be present naturally depending on the realization,
would lift the energies of the localized states to remedy this issue.

	\begin{figure}[htb]\centering
		\includegraphics[width=.55\textwidth]{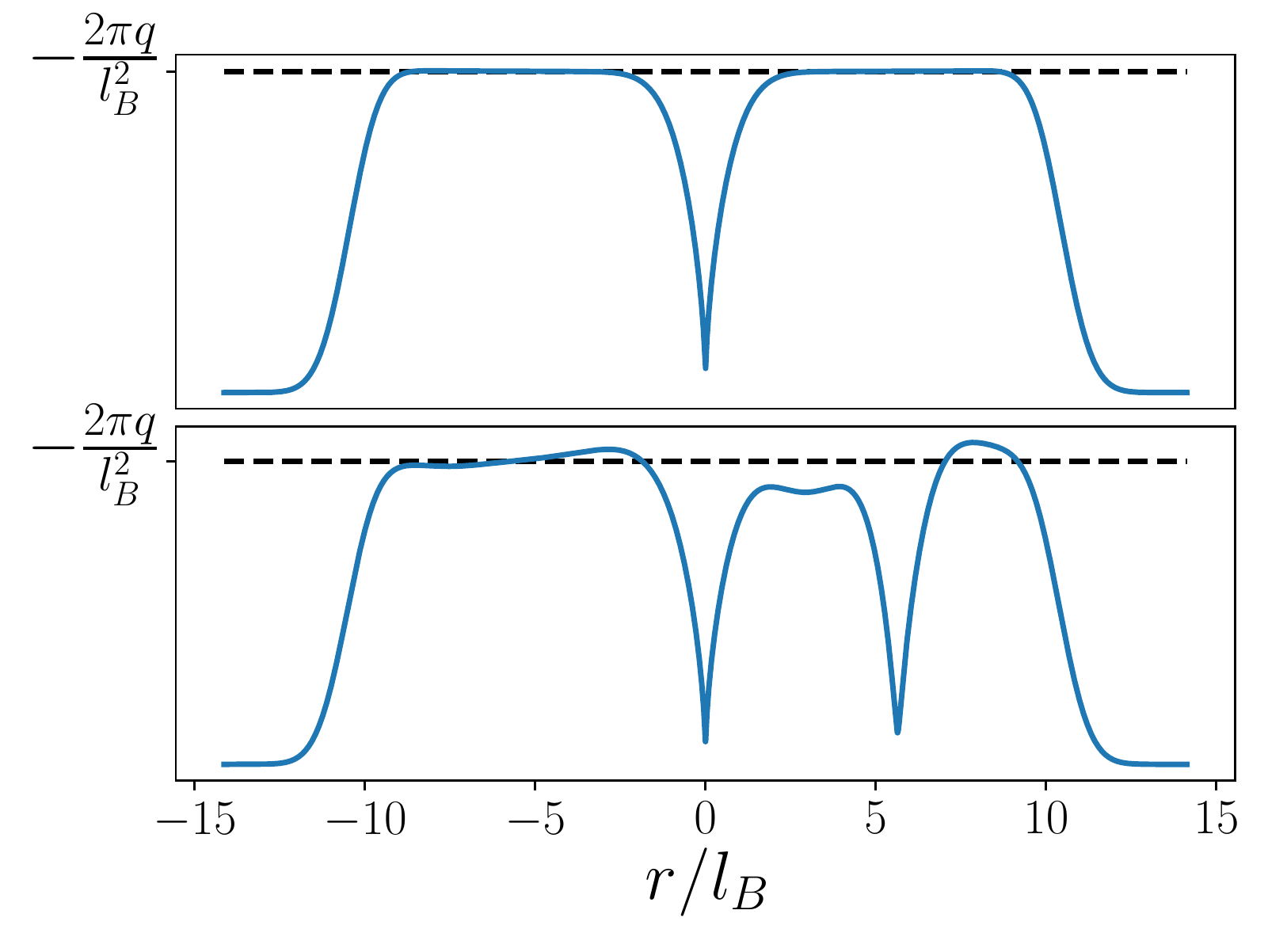}
		\caption[Ground state single-particle density for one and two probes]{The single-particle densities (cross sections) of the ground states with one probe (at $r=0$) and two probes (at $r=0$ and $5.264l_B$). The flux is given by $\alpha=0.7$. The horizontal dashed line depicts the density of an infinite system (see text for details).}
		\label{PRB:fig:sp_dens}
	\end{figure}

	In \cref{PRB:fig:sp_dens} we illustrate the single-particle density (cross section) for the system with one and two probes. Clearly, the single-particle density has a cusplike dip at the position of a probe, i.e., a missing electron charge $\Delta q>0$. It is tempting to identify the composite of the missing electron charge $\Delta q$ and the probe with flux  $\Phi$ with Wilczek's charge-flux-composite anyons \cite{wilczek1982}. However, a careful analysis of the Berry phase
below shows that this identification would be erroneous (see \cref{sec:PRB:notQPs}).

\subsection{Anyonic properties of the wave function; calculation of the Berry phase}\label{sec:PRB:BPH}
	In this section we calculate the Berry phase as one of the probes undergoes adiabatically a closed loop in space as illustrated in \ref{PRB:fig:model} b). More specifically , we calculate the Berry phases $\gamma_\mathrm{in}$ when a single probe is within the loop as in \cref{PRB:fig:model} c), and $\gamma_\mathrm{out}$ when all of the other probes are outside of the loop as in \cref{PRB:fig:model} d). The calculation proceeds in the same manner as the calculation of the statistics of Laughlin quasiholes (pages \pageref{QPSTATS1}-\pageref{QPSTATS2}). Once again, the difference between the two phases is the statistical phase, which we find to be ${\gamma_S = \gamma_\mathrm{in}-\gamma_\mathrm{out}=2\pi(\alpha-1)}$, where $\alpha=\Phi/\flqnt$. This result means that in the coordinates of the external probes the wave function $\psi$ is anyonic when $\alpha$ is fractional. 
	
	We assume that the probes remain sufficiently far apart from each other at any time. Without loss of generality, we assume that the probe $\eta_1$ traverses the path. Recall that the Berry phase accumulated along the path $\C$ is given by
	\begin{equation}\label{PRB:eq:bph_cplx}
		\gamma = i \oint_\C (\scalA_{\eta_1} \dd\eta_1 + \scalA_{\conj{\eta_1}} \dd\conj{\eta_1} ),
	\end{equation}
	where $\scalA_{\eta_1} = i \braket{\Psi}{\pdd_{\eta_1}\Psi}$ and $\scalA_{\conj{\eta_1}} = i \braket{\Psi}{\pdd_{\conj{\eta_1}}\Psi}$. After employing the plasma analogy to take normalization into account (see Appendix B in \cite{prb}), once again, we find 
	\begin{equation}\label{PRB:eq:bph_n}
		\gamma = 2\pi\ev{n}_\C,
	\end{equation}
	where $\ev{n}_\C$ is the mean number of electrons in the area encircled by the path $\C$. Since the inner probe expels some charge as illustrated in \cref{PRB:fig:sp_dens}, and thus the mean number of electrons inside the contour differs in the two cases: $\ev{n}_{C,\mathrm{in}}\neq\ev{n}_{C,\mathrm{out}}$, there is a statistical phase
	\begin{equation}\label{PRB:eq:statph_n}
		\gamma_S = 2\pi(\ev{n}_{\C,\mathrm{in}}-\ev{n}_{\C,\mathrm{out}}).
	\end{equation}
	We calculate the expelled charge from the single-particle density $\rho$ of the many-body wave function $\psi$. It can be found by employing the plasma analogy:
	\begin{equation}\label{PRB:eq:SPdensity}
		\rho(z) = \frac{1}{2\pi l_B^2} - (1-\alpha)\sum_{k=1}^N \delta^2(z-\eta_k).
	\end{equation}
	The missing charge at the probe is thus $\Delta q=-q(1-\alpha)$, and the statistical phase (in the thermodynamic limit) is
	\begin{equation}\label{PRB:eq:statph_alpha}
		\gamma_S=2\pi(\alpha-1).
	\end{equation}
	Thus, $\gamma_S\!\mod 2\pi$ is equal to $2\pi\alpha$.
	
	Let us briefly comment on the fact that $\Delta q\rightarrow-q$ as $\alpha\rightarrow0$, and $\Delta q\rightarrow0$ as $\alpha\rightarrow1$, which may seem awkward at first sight. This is related to our discussion in the previous section on the behavior of the wave function \eqref{PRB:eq:MBGS} as $\alpha\rightarrow0$. In constructing the wave function, we do not populate  the localized states which appear at the position of the probes for $\alpha>0$. Therefore, as $\alpha\rightarrow0$, they are not in the Slater determinant, leaving a hole of charge $\Delta q=-q$   at the position of the probe. When $\alpha\rightarrow1$,  the localized states at the position of the probe enter the first LL (which is empty by assumption); however, the corresponding state in the LLL below is now filled, yielding $\Delta q=0$, as the spectrum has flown back on itself when $\alpha$ flows from zero to one.
	
	In order to further underpin \eqref{PRB:eq:statph_alpha}, and explore the dependence of the statistical phase on the separation between the probes (we assumed above that they are sufficiently far apart along the path $\C$) and the details of the path, we perform numerical calculations. We numerically consider the cases with one and two probes. In all our calculations presented here, the magnetic field is given by $B_0R_\mathrm{max}^2 \pi/\flqnt = 100$ ($R_\mathrm{max}$ is the radius of the finite numerical system, see \cref{PRB:fig:model} a) and b)), and we construct the numerical ground state by filling the first $N_e=55$ states to minimize the boundary (finite-size) and numerical grid effects while still successfully mimicking an infinite system. The method for the numerical calculation of the Berry phase is as follows \cite{mukunda1993}: instead of performing the integral in \cref{PRB:eq:bph_cplx} we discretize the evolution parameter, here called time for simplicity, and evaluate it at $N_t$ equidistant points. Let $\psi_i(t)$ be the $i$-th numerical single-particle eigenstate lying in the LLL at time $t$, and let $M_{ij}(t_k,t_l)=\braket{\psi_i(t_k)}{\psi_j(t_l)}$ be the elements of the overlap matrix $M(t_k,t_l)$ at two different times. Then the Berry matrix
	\begin{equation}\label{PRB:eq:Berry_mtrx}
		U=M(t_0,t_1)M(t_1,t_2)\dots M(t_T,t_0)
	\end{equation}
	leads directly to the Berry phase 
	\begin{equation}\label{PRB:eq:}
		\gamma \approx -\arg(\det U ).
	\end{equation}
	This relation is exact in the limit $N_t\rightarrow\infty$. A possible concern about numerical error arises if the value of $\alpha$ is close to a whole number (or two values of $\alpha$ are close). However, this is not an issue for the values we choose in the following.

\begin{figure}[htb]\centering
		\includegraphics[width=.873\textwidth]{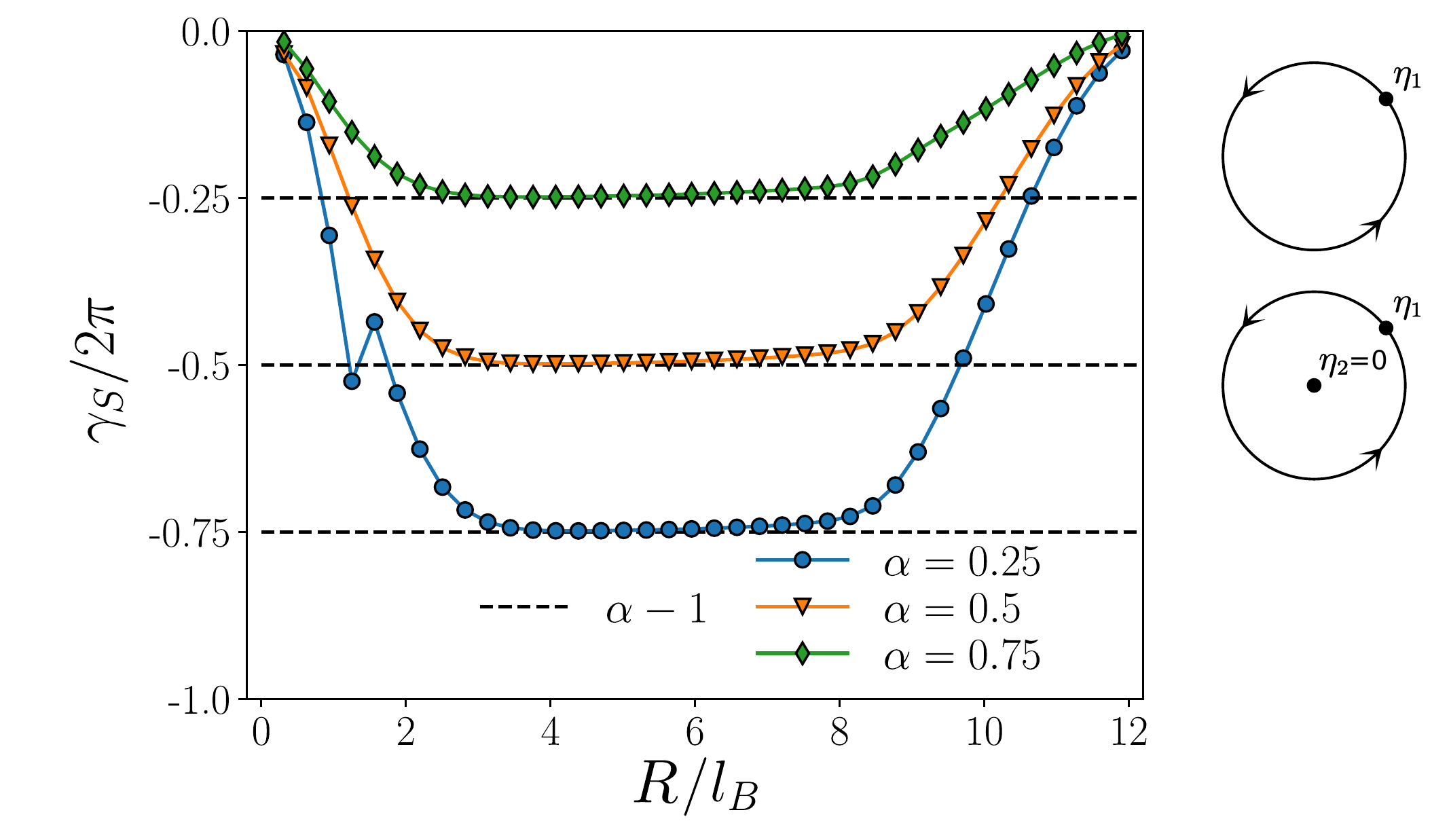}
		\caption[The statistical phase $\gamma_S$ as a function of the separation between the probes]{The statistical phase $\gamma_S$ as a function of the separation between the probes at three different flux values $\alpha$.  In every calculation, one of the probes is at $z=0$, and the other one adiabatically traverses a circle of radius $R$ in the anticlockwise direction (shown on the right). The dashed lines denote the $2\pi(\alpha-1)$ values corresponding to the analytical prediction.}
		\label{PRB:fig:stph_num}
	\end{figure}

	In \cref{PRB:fig:stph_num} we illustrate $\gamma_S$. as a function of the separation between the probes R. The dashed lines denote the analytical prediction in \eqref{PRB:eq:statph_alpha}. We see that if the probes are too close they will influence each other’s cusp dip in the density, and consequently the statistical phase will not be given by $2\pi(\alpha-1)$. However, after they are sufficiently apart, $\gamma_S$ exhibits a plateau at the value $2\pi(\alpha-1)$. As the outer probe gets close to the edge of our (numerical) finite-size system, the phase departs from the analytical solution. We conclude that the numerical calculations agree with the analytical prediction when the path of the moving probe does not come too close to other probes, and if they are not too close to the edges of the sample. The system exploited in numerical calculations is very small (practically mesoscopic). In reality, the system would be much larger providing a much broader region in space where a constant plateau would be observed.

\begin{figure}[htb]\centering
	\includegraphics[width=.9\textwidth]{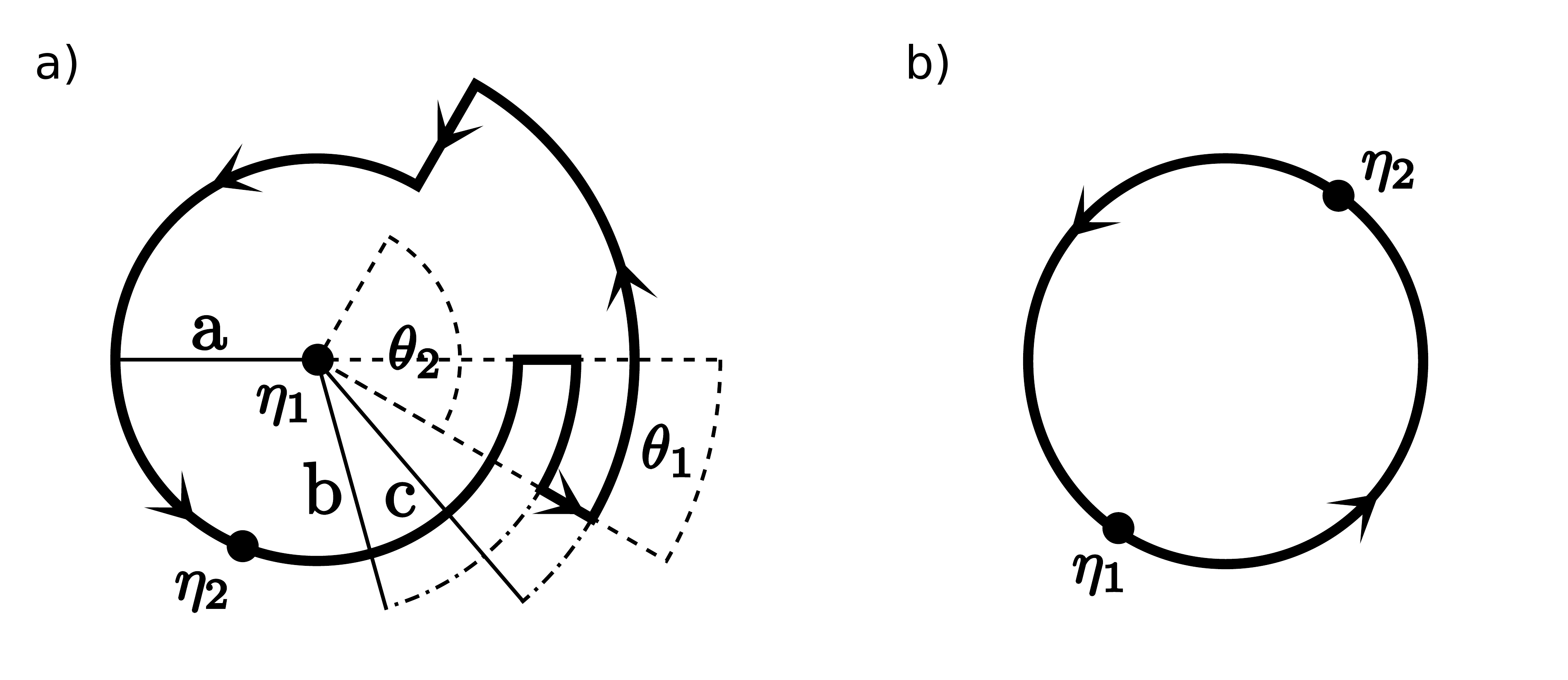}
	\caption[Additional contours (numerics)]{Two different contours. a) One of the probes undergoes a closed loop, visiting three different radii $a$, $b$, and $c$, such that each is sufficiently far from the probe at zero and from the edge of the system. b) Two probes at opposite radii ($|\eta_1|=|\eta_2|=3.13 l_B$) are exchanged leading to an exchange phase $\pi(\alpha-1)$ (see text for details). The parameters used in the calculation are $\alpha=0.37$, $a=4.76 l_B$, $b=6.14 l_B$, $c=7.52 l_B$, $\theta_1=\pi/6$, and $\theta_2=\pi/2$.}
	\label{PRB:fig:defform}
	\end{figure}

	Next we perform the same calculation, but deform the contour $\C$ as illustrated in \cref{PRB:fig:defform} a). The contour is such that the probes are sufficiently separated at all times, and away from the sample edges. We obtain the statistical phase $\gamma_S=-0.631\times 2\pi$, which is in agreement with the analytical result $\gamma_S=2\pi(\alpha-1)$ for $\alpha=0.37$, with relative error of about $0.2\%$. We conclude that the statistical phase does not depend on the
details of the contour.
	
	Next we consider the exchange of two probes. We numerically calculate the exchange phase obtained when two of the probes are exchanged along the path illustrated in \cref{PRB:fig:defform} b). We then subtract the two Berry phases obtained when each of the probes $\eta_1$ and $\eta_2$ traverses its respective path (semicircles), without the other probe present. We find the result to be $-0.636\times\pi$ for $\alpha=0.37$, once again in agreement with the analytical calculations. The relative error is about $1\%$. From the viewpoint of the relative coordinate, when one of the probes encircles the other probe, this corresponds to a double exchange of the two probes illustrated in \cref{PRB:fig:defform} b). Thus, we conclude that if we exchange two of the probes adiabatically along a path illustrated in \cref{PRB:fig:defform} b) (with no other probes within the closed contour), the exchange phase accumulated by the wave function will be $\pi(\alpha-1)$. This means that the wave function $\psi$ is anyonic in the coordinates of the probes, with the statistical parameter given by $\theta=\alpha-1$.
	
	We end this section by a note on the gauge invariance of the Berry phase calculated along the closed path $\C$. The wave function $\psi$ in \eqref{PRB:eq:MBGS} is a single-valued function of the positions of the external probes $\eta_k$, provided that the normalization $Z(\{\eta_k\},\{\conj{\eta_k}\})$ is also chosen to be a single-valued function of $\eta_k$. In contrast, the singular gauge wave function $\psi'$ in \eqref{PRB:eq:singulargauge} is a multivalued function of $\eta_k$. Equation \eqref{PRB:eq:bph_cplx} for calculating the Berry phase yields different results when naively used for $\psi$ and $\psi'$. However, the Berry phase calculated along a closed path must be independent of the gauge used. This issue is resolved by noting that \eqref{PRB:eq:bph_cplx} should be used only for single-valued wave functions ($\psi$ in our case) . If one wishes to calculate the Berry phase in the singular gauge by using the multivalued wave function $\psi'$, there is an additional term that should be included in the Berry phase formula (see eq. (5.12) in ref \cite{mukunda1993}) which ensures gauge invariance. We note that our results differ from refs. \cite{weeks2007,rosenberg2009}, which have used multivalued wave functions and \eqref{PRB:eq:bph_cplx} to calculate the Berry phase.

\subsection{Synthetic anyons are not emergent quasiparticles}\label{sec:PRB:notQPs}
	From the illustration of the single-particle density in \cref{PRB:fig:sp_dens} we see that at the position of every solenoid probe there is a cusp-like dip, i.e., a missing electron charge, which is found to be $\Delta q=-q(1-\alpha)$ from the single-particle density. We have already noted that it is tempting to identify the composite of a missing electron charge $\Delta q$, and a solenoid with flux $\Phi$ with Wilczek’s charge-flux-composite anyon \cite{wilczek1982}. Now we show that such an interpretation is erroneous. When a probe traverses a closed path $\C$, the system acquires the Berry phase $\gamma=2\pi\ev{n}_\C$. Let us try to calculate the missing charge by a different route using the Aharonov-Bohm phase, and by assuming that we are dealing with a charge-flux composite. To this end, let us denote the missing charge $q^*$, and check whether we obtain the same result as with the single-particle density. When the charge $q^*$ traverses the path $\C$, it will acquire the Aharonov-Bohm phase $q^*\Phi_\C/\hbar$, where $\Phi_\C=\ev n_\C \flqnt$ is the total magnetic flux within the path $\C$  (we have assumed unity filling of the LLL). To obtain the Berry phase, we should include the Aharonov-Bohm phase acquired by the solenoid with flux $\alpha\flqnt$ that circulates around the charge $q\ev n_\C$, which is equal to $q\ev n_\C \alpha \Phi_0/\hbar$. By identifying
	\begin{equation}
		\gamma = 2\pi\ev n_\C = \frac{q^*\Phi_\C}{\hbar}+\frac{q\ev n_\C \alpha \flqnt}{\hbar},
	\end{equation}
	we find
	\begin{equation}
		q^* = -q(1+\alpha) \neq \Delta q = -q(1-\alpha).
	\end{equation}
	This difference may come as a surprise, because an equivalent calculation for anyons in the FQHE yields identical expressions for the missing charge from the single-particle density and from the Aharonov-Bohm calculation of $q^*$.

	To understand the obtained result, first we note that the external solenoid probe acts as a ladle that stirs the electron sea around, and the Aharonov-Bohm phase depends on the movements of the electrons in the sea, and not of the missing charge. When the missing charge corresponds to the quasiparticle, as in the FQHE, then $q^*=\Delta q$ because the motion of (quasi)holes  uniquely corresponds to the motion of the electron sea. However, the missing charge here is not a quasihole, since a quasihole would produce the same Aharonov-Bohm phase as a particle of the same charge, and we cannot interpret the missing charge attached to the solenoid probe as Wilczek’s charge-flux-tube composite. One way to understand this difference is to assume that the electron sea is a superfluid. Then the Aharonov-Bohm phase acquired by stirring the ladle would be zero. To further corroborate this idea, we mention that a similar point was raised for a charge expelled by a delta function potential barrier on a 1D ring enclosing a solenoid \cite{marija2020}. The Berry phase accumulated as the barrier traverses the ring is due to the particles being reflected from the barrier, rather than the motion of the missing charge around the solenoid.

\subsection{Fusion rules of synthetic anyons}\label{sec:PRB:fusion}
	The conclusion of the previous section has impact on the
fusion rules of synthetic anyons. The fusion rules depend on
the physical microscopic process which corresponds to 
fusion. For example, suppose that we have $N=4$ solenoid
probes in the system with flux $\alpha\Phi_0$,  i.e., we have two pairs
of probes. Next, we slowly bring together (merge) two of
the solenoids from each pair, thereby forming a system
with $N=2$ solenoid probes with flux $2\alpha\Phi_0$. This system is identical to the one we have explored with $\alpha$ replaced by $2\alpha\!\!\mod 1$.  Thus, the exchange phase changes from $\pi(\alpha-1)$ to $\pi[(2\alpha\!\!\mod 1)-1]$. This is not the exchange phase $2^2\pi(\alpha-1)$ expected from fusing two anyons (see discussion in \cref{sec:TQM:anyons}, on page \pageref{page:fusion}).  This is related
to the fact that we cannot interpret the missing charge attached
to a solenoid probe as Wilczek’s charge-flux-tube composite,
because in that case the standard fusion rules would be applicable.

	If we, however, consider the fusion process as pairing the
solenoids two by two in the sense $\eta_2=\eta_1+c$ and $\eta_4=\eta_3+c$, where $c$ is a complex number with magnitude greater than $l_B$, then the standard fusion rules apply. For example, if
we move one of the pairs in a circle of sufficiently large radius
around the other pair, we analytically obtain the expected statistical phase of $2^2\times 2\pi(\alpha-1)$ (see Appendix B of \cite{prb} for details of calculation). 

\subsection{Discussion}\label{sec:PRB:discussion}
	It might be interesting to discuss a potential experimental
realization, and pertinent challenges, of the Hamiltonian in eq. \eqref{PRB:eq:H} in
ultracold atomic gases. Ultracold atomic gases have been experimentally realized in two dimensions \cite{bloch2008,hadzibabic2006}, and a viable path (although not a simple one) for implementing
IQHE states with ultracold atoms is to employ synthetic magnetic fields \cite{lin2016,dalibard2011,bloch2012,goldman2014}. The missing ingredients are the solenoid-like
probes. The synthetic vector potential of a solenoid can in
principle be achieved with vortex laser beams nonresonantly
interacting with two-level atoms \cite{jajtic2018}. Namely, by exploring
eq. (7) in sec. II of ref. \cite{dalibard2011}, one finds that a vortex beam interacting with a two-level atom can yield the Berry connection
which corresponds to the vector potential of a solenoid. The
vortex phase ensures proper direction of the vector potential;
however, to obtain the proper $\sim\!1/r$ dependence one must in addition  properly adjust the detuning and the intensity of the
laser. An additional challenge along this path would be to
ensure that the synthetic magnetic flux through every solenoid
is identical, so that an exchange of any of the two lasers
would depend on the unique statistical parameter (otherwise the
localized perturbations at the probes could not be referred to
as synthetic anyons). The advantages of ultracold atomic systems are long coherence times and the possibility to relatively
easily braid the laser probes.

	In conclusion, we have presented exact solutions of a model for synthetic anyons in noninteracting many-body systems. The key ingredients in the model are the specially tailored
external potentials (that could correspond to some external localized probes), which supply the demanded nontrivial topology in the system. The Hamiltonian representing the model is
that of noninteracting electrons in a uniform magnetic field (in
the IQHE state for LL filling factor $1$), and the probes are solenoids with a magnetic
flux that is a fraction of the flux quantum. The Fermi energy is
such that only the lowest Landau-level states are occupied;
the localized states which appear at the position of every
probe, with energy in the gap, are assumed to be empty. We
have found the ground state of this system, and demonstrated
that it is anyonic in the coordinates of the probes when
the flux through solenoids is a fraction $\alpha$ of the flux quantum $\flqnt$. The statistical parameter of the synthetic anyons is $\pi(\alpha-1)$. We have shown that these synthetic anyons cannot be considered as emergent quasiparticles, and that they cannot
be interpreted as Wilczek’s charge-flux-tube composites. This
observation has consequences on the fusion rules of these
synthetic anyons, which depend on the microscopic details of
the fusion process. In a future study, it would be interesting to consider the forces
that act upon the probes. Geometric forces on point fluxes
carrying integer quanta of fluxes in quantum Hall fluids were
studied in Ref. \cite{avron1998}. Next, it would be interesting to explore
the potential for anyonic physics in a system of solenoids
that does not necessarily rely on the quantum Hall effect. For
example, one such system might be the Aharonov-Bohm billiards \cite{berry2010}.
	 Finally, it would be interesting to extend the ideas
presented here to explore non-Abelian synthetic anyons, and
investigate their capacity for topological quantum computing. A step in this direction was recently taken by solving a related model with spin-unpolarized electrons coupled to non-Abelian probes \cite{bruno}. However, the statistics obtained for the studied system, although of higher dimension, was still commutative. Therefore, a different setup is needed to achieve truly non-Abelian statistics.

%% file: 4_concl/concl.tex
	In this thesis, we have explored several different topics whose unifying theme is the relevance of topology to the physics in question. The main goal of the thesis was to present the papers \cite{zb,solitonssh,prb} published during the course of the author's PhD study, but we have also used the opportunity to set up the stage in \cref{ch:intro} by exploring multiple relevant topics in topological quantum matter, as well as the indispensable concept of geometric phase. Among the topics explored were the integer and the fractional quantum Hall effects, anyons, and the symmetry-protected topological states. 
	
	Besides being one of the first discovered topological states of matter, the conceptual significance of the IQHE lies in its relation to the FQHE, as well as to other related systems such as topological insulators. One could almost consider understanding the IQHE an essential step towards building an understanding of the FQHE, as it enables one to first grasp the Hall conductance quantization in terms of individual electrons following the spectral flow of the quantum states, and then to generalize this understanding to emergent fractional quasiparticles of the FQHE, thus explaining the fractional-value plateaus of the Hall conductance, as well as setting up the stage for grasping the quasiparticles' fractional statistics. On the other hand, the IQHE (despite not being a SPT phase) may also be considered a prototype for topological insulators, as well as other topological phases characterized by simple numerical topological invariants, akin to the first Chern number. Similarly, it may motivate the understanding of the relation between the topology of the bulk and the protected states at the edge. The FQHE is interesting as the first known realization of topological order (in the sense of supporting anyonic excitations), and it is still among the most promising platforms for topological quantum computation, notably the non-Abelian Moore-Read and the $Z_3$ parafermion Read-Rezayi state which are candidates for the $5/2$ and $12/5$ FQHE states, respectively \cite{nayak}. Of course, in working towards fault tolerant quantum computation with the FQHE or other systems, one has the freedom of choosing the most suitable experimental platform, such as solid state materials, ultracold atomic gases (or trapped ions), and topological photonics. Both the fractional and the integer QHE are relevant to our proposal for synthetic anyons in weakly-interacting systems. The IQHE is the foundation on which our model is built, and the Laughlin FQHE states provided the inspiration for the idea of perturbing the IQHE with the probes carrying a fractional magnetic flux (recall that the Laughlin quasiparticles can be created by threading an integer flux). Mathematically, the fractional braiding phases of the Laughlin quasiholes stem from the fact that the factors containing the differences between two electron coordinates are raised to the $m$-th power, while the factors containing the differences between electron and quasihole coordinates are raised to the first power. As we have seen through plasma analogy, this means that the quasiparticles can be interpreted as a fraction of an electron. While $m=1$ in the wave function of the IQHE, it turns out that by using fractional instead of integral probes, anyonic probe-braiding phases can be achieved in this system. 

	Needless to say, we have only scratched the surface in our presentation of topological phases of matter. We have largely focused on the IQHE, the Laughlin and the hierarchical FQHE states, and have briefly discussed some SPT phases, but for example, we have merely mentioned the non-Abelian FQHE states, and topological order in quantum spin liquids \cite{savary2016}. However, the full spectrum of possibilities is even broader; for example, topological materials such as Weyl semimetals \cite{armitage2018} can also be gapless.

	In \cref{sec:zb}, we have presented the results of the photonics experiments carried out by our collaborators, and our theoretical analysis. The idea was to explore the effect of the valley degree of freedom of the honeycomb lattice on the propagation of light through photonic crystals. The valley degree of freedom may play a role in future photonics, as well as electronics applications. An obvious potential application of a two-valley system is in quantum computation, as a two-state system can be a qubit. As we have seen, the propagation in modes close to one of the valleys of the inversion-symmetry broken HCL has an interesting effect on an initially Gaussian beam of light that excites both sublattices. Namely, a vortex component emerges, which interferes with the (initially present) non-vortex component and produces a rotating spiralling pattern, different from the conical diffraction one would see if modes near both valleys were excited. The reason lies in the winding of the Berry curvature around the $K$ and $K'$ points in opposite directions. Essentially, the two points are topological singularities in the momentum space that imprint themselves on the real space, thus producing the vortex components with opposite vorticity. The conical diffraction results when those vortices cancel out, but when only a single valley is excited, a vortex survives, and interferes with the non-vortex component. Plotting the center-of-mass of the spiralling beam against propagation distance (i.e. time) reveals the Zitterbewegung oscillations. This allows us to reinterpret the Zitterbewegung effect, usually interpreted in terms of interference of positive and negative energy states, as  a consequence of the vortex-nonvortex beam interference. This interpretation is equally correct, but easily visualized in our case, and it makes obvious the connection to topology.
	
	In \cref{sec:ssh}, we have presented the results of a numerical simulation of soliton beams arranged so that an SSH lattice is formed from two sublattices propagating at small and opposite angles. The two different tunnelling coefficients needed for an SSH lattice result from the nonlinearity-mediated interaction between solitons. Thus, the topological SSH lattice owes its existence to the nonlinearity. During the propagation, the relative values of the coefficients change, along with the distance of the neighbouring beams. This causes topological phase transitions from the initial trivial phase to the topologically nontrivial phase, manifested by the appearance of the topological edge states. Eventually, as the solitons at the edges separate enough from their nearest neighbour ($\approx$ the next-nearest-neighbour distance), their coupling to the lattice becomes negligible, and they are no longer a part of the lattice. The remaining lattice loses the edge modes, and finds itself in the trivial phase, until the next phase transition. These cycles repeat, until the last of the beams cross, and the two sublattices are fully separate. These nonlinearity-induced topological phase transitions are an emergent nonlinear topological phenomenon. By drawing attention to nonlinearity-induced nontrivial topology, we hope to contribute to unlocking the possibilities at the intersection of nonlinear and topological photonics.
	
	In \cref{ch:prb}, we have proposed to realize synthetic anyons by perturbing a weakly-interacting (or noninteracting) system with specially tailored localized probes. In the model that we have studied, thin solenoid probes carrying a fractional magnetic flux are used  to raise one of the single-particle states (per probe) from the lowest Landau level of an IQHE system, and into the gap. We set the Fermi level below the energy of these localized gap states, so that they are left empty, and some charge is displaced from each probe (to infinity). We have found the ground state of this system and shown that it is anyonic in the coordinates of the probes. As we have stated above, there is an analogy to be drawn between our perturbed system, and the Laughlin state containing anyonic quasiholes. However, we once again caution against interpreting these regions of missing charge as emergent quasiparticles. While this is a correct interpretation of the holes created by threading integer fluxes through the Laughlin state, it is not warranted in case of the synthetic anyons in our proposal, because the fractional flux carried by the probes is not pure gauge, and does not merely excite the underlying system. For this reason, the movement of the electrons as the  probes are braided cannot be captured in terms of the movement of the displaced charges. To show this, we have attempted to calculate the missing charge by assuming that the Berry phase accumulated as the probe traverses a path enclosing an area devoid of other probes corresponds to the Aharonov-Bohm phase (in part) due to the missing charge. Predictably, this calculation does not yield the correct value for the missing charge, and thus it shows that it would be wrong to think of our synthetic anyons as emergent quasiparticles, or as Wilczek's anyons composed of the missing charge and the flux tube. A further consequence of this subtlety was seen when fusing the anyons. Simply bringing the solenoids to the same place does not result in the expected fusion product. However, the expected fusion rule for Abelian anyons still applies if groups of solenoids are braided with other distant groups by moving them in unison, but without bringing them close together.
	
	As our system allows for creation and manipulation of synthetic Abelian anyons, it is a natural step towards topological quantum computation. However, the final product requires a non-Abelian equivalent, as well as a method for pinning all the fractional fluxes to the same constant value. Finally, a practicable realization of the model is needed. Ultracold atomic systems with synthetic magnetic vector potentials hold some promise in this regard.

%% file: Y_additional_materials/papers.tex
\section*{List of publications}

\begin{enumerate}[itemsep=0pt]

\item Šiljić, A., Lunić, F., Teklić, J., Vinković, D., 2018. \textit{Proton-induced halo formation in charged meteors}, Mon. Not. R. Astron. Soc., \textbf{481}, 2858-2870. doi: 10.1093/mnras/sty2357.

\item Lunić, F., Todorić, M., Klajn, B., Dubček, T., Jukić, D., Buljan, H., 2020. \textit{Exact solutions of a model for synthetic anyons in a noninteracting system}. Phys. Rev. B, \textbf{101}, 115139. doi: 10.1103/PhysRevB.101.115139.

\item Liu, X., Lunić, F., Song, D., Dai, Z., Xia, S., Tang, L., Xu, J., Chen, Z., Buljan, H., 2021. \textit{Wavepacket Self-Rotation and Helical Zitterbewegung in Symmetry-Broken Honeycomb Lattices}. Laser Photonics Rev., \textbf{15}, 2000563. doi: https://doi.org/10.1002/lpor.202000563.

\item Bongiovanni, D., Jukić, D., Hu, Z., Lunić, F., Hu, Y., Song, D., Morandotti, R., Chen, Z., Buljan, H., 2021. \textit{Dynamically Emerging Topological Phase Transitions in Nonlinear Interacting Soliton Lattices}. Phys. Rev. Lett., \textbf{127}, 184101. doi: 10.1103/PhysRevLett.127.184101.

\end{enumerate}

%% file: sazetak.tex
\selectlanguage{croatian}

Glavni je cilj ovog rada predstavljanje originalnog istraživanja objavljenog za vrijeme autorovog doktorskog studija. Prva dva rada koja predstavljamo pripadaju području fotonike koje se bavi manipulacijom svjetlosti. U jednom istražujemo propagaciju svjetlosti u dolinskim modovima saćaste fotoničke rešetke s narušenom inverzijskom simetrijom, dok u drugom demonstriramo topološke fazne prijelaze solitonske rešetke koji su posljedica nelinearnosti medija. U trećem radu predstavljamo model za sintetičke anyone u neinteragirajućem sustavu. Poveznica svih triju tema je važnost topologije za fiziku koju proučavamo. Stoga u prvom dijelu dajemo uvod u neke povezane teme i koncepte iz topološke fizike. 



\section*{Teorijska pozadina}
\subsection*{Geometrijske faze}
	Geometrijske se faze javljaju kad se hamiltonijan sustava mijenja u vremenu, a ovisne su o geometriji putanje $\C$ u prostoru parametara hamiltonijana.
	\subsubsection{Berryjeva faza}
	Neka hamiltonijan ovisi o skupu parametara $\R=(R_1,R_2,\dots)$. Parametre adijabatski variramo između vremena $0$ i $T$ uz uvjet cikličnosti $\R(T)=\R(0)$. Za svaki $\R$ odabiremo ortonormalnu bazu $\mR$. Pretpostavimo da je sustav pripremljen u nekom od stanja iz baze, $\psi_n(0)=\ket{n(\R_0)}$, te da ni u jednom trenutku evolucije ne prolazi kroz degenerirano stanje. Adijabatski teorem nam garantira da je $\psint = e^{i\gnt} \exp[-\frac{i}{\hbar} \int_0^t \dd t' E_n(\R(t')) ] \nRt$ rješenje vremenski neovisne Schrödingerove jednadžbe. Drugi faktor je dinamička faza, dok prvi faktor na kraju evolucije postaje Berryjeva faza \cite{berry1984}
	$$\gamma_n\equiv\gamma_n(T)=\oint_\C \dd\R \cdot \An(\R),$$
	gdje smo uveli Berryjevu konekciju
	$\An(\R) \equiv i \braket{n(\R)}{\gradR n(\R)}$. 
	Berryjeva konekcija se pri baždarnim transformacijama ponaša kao vektorski potencijal, dok je Berryjeva faza invarijantna na baždarne transformacije.
	
	Aharonov-Bohmov efekt \citep{AB} je fenomen u kojem nabijene čestice osjećaju utjecaj elektro\-ma\-gne\-tskih potencijala čak i u područjima gdje nema magnetskog polja. Efekt se može interpretirati kao specijalni slučaj Berryjeve faze \cite{berry1984}. Pretpostavimo da je čestica naboja $q$ zarobljena u kutiji bez magnetskog polja, ali osjeća vektorski potencijal beskonačnog solenoida $\VP$. Transpo\-rti\-ramo li kutiju polako oko solenoida duž krivulje $\C$, rezultat je faza $\gab = \frac{q}{\hbar} \oint_\C \VP \cdot \dd \R = \frac{q\Phi_\B}{\hbar}$, gdje je $\hbar$ reducirana Planckova konstanta, a $\Phi_\B$ tok magnetskog polja kroz solenoid. Promatramo li položaj kutije $\R$ kao parametar hamiltonijana, dobivamo da je Berryjeva konekcija $\AnR = \frac{q}{\hbar} \VP(\R)$, a faza $\gamma_n = \gab$. 

	\subsubsection{Neabelovske geometrijske faze \cite{wz1984}}
	Ako je sustav pripremljen u $g_n$-degeneriranom stanju i degeneracija vrijedi za sve $\R$, ciklička adijabatska evolucija može rotirati početno stanje unutar degeneriranog potprostora: $\ketn T = U_\mathrm{dyn} \UnC \ketn 0$. $U_\mathrm{dyn}$ je dinamička faza, a unitarna matrica $\UnC=\mathcal P  \exp( \oint_\C \dd \R \cdot \AnR )$ je Berryjeva holonomija, gdje je $\mathcal P$ tzv. ,,path-ordering'' operator. Komponente neabelovske Berryjeve konekcije su matrice s elementima ${\scalA^n_\mu}_{(ab)}(\R) = i \braket*{n_a(\R)}{\pdd_{R_\mu} n_b(\R)}$, gdje $a,b=1,\dots,g_n$.
	
	\subsubsection{Uloga u Blochovim vrpcama}
	Prostorno-periodički sustavi mogu se opisati periodičkim funkcijama $\unq(\vr)$, gdje je $n$ inde\-ks vrpce u spektru, a $\q$ valni vektor. Funkcije $\unq(\vr)$ su svojstvena stanja Blochovog hamiltonijana $H(\q)=e^{-i\q\cdot\vr} H e^{i\q\cdot\vr}$ koji parametarski ovisi o $\q$, pa tako prva Brillouinova zona poprima ulogu intrinzičnog parametarskog prostora. Berryjeva konekcija $\An(\q) = i \braket{\unq}{\Nabla_{\q}\unq}$ se u 3D rešeci ponaša kao magnetski vektorski potencijal u $\q$-prostoru \cite{stanescu}.
	
	U 1D rešetkama može se javiti konačna geometrijska (topološka) Zakova faza \cite{zak} koja na\-sta\-je kad se $k$ varira preko cijele Brillouinove zone $\gamma_\mathrm{Zak} = \int_{\sq}^{\sq+\frac{2\pi}{a}} \dd \sq \, \mathcal{A}^n(\sq)$. U rešetkama s inverzijskom simetrijom moguće su samo dvije vrijednosti, $0$ ili $\pi$.

\subsection*{Topološka kvantna materija}
	Prije kraja dvadesetog stoljeća postalo je jasno da klasifikacija faza materije temeljena na simetriji nije dovoljna za potpuni opis. Pri apsolutnoj nuli postoje odvojene faze koje karakteriziraju diskretne topološke invarijante. Obično je riječ o sustavima s procijepom izme\-đu vrpca, gdje se fazni prijelazi događaju pri specijalnim vrijednostima parametara gdje je procijep zatvoren, a topološke invarijante pojedine vrpce nedefinirane \cite{stanescu}.
	
	Topološke faze mogu se podijeliti na dalekodosežno (LRE) i kratkodosežno (SRE) spre\-gnute faze. LRE faze nije moguće adijabatski transformirati u nespregnuta produkt-stanja (ili jednu u drugu) bez zatvaranja procijepa \cite{chen2010}, što ih čini topološki netrivijalnim (posjeduju topološki red). Najvažniji primjer su stanja kvantnog Hallovog efekta. LRE faze obično imaju egzotična pobuđenja s frakcijskom statistikom \cite{stanescu}. Za određena se SRE stanja može smatrati da pripadaju topološki netrivijalnim fazama ako razmatramo isključivo transformacije koje ne narušavaju određene simetrije hamiltonijana. Riječ je o tzv. simetrijom zaštićenim topološkim (SPT) fazama. Topološki zaštićena granična stanja su prisutna i kod faza s topološkim redom i kod SPT faza.
	
	\subsubsection{Cjelobrojni kvantni Hallov efekt}
	Cjelobrojni kvantni Hallov efekt (IQHE), otkriven 1980. \cite{kdp}, pojava je  kvantizacije Hallove vodljivosti u 2D sustavima s narušenom simetrijom obrata vremena:
	$$\sigma_{xy} =\frac{q^2}{2\pi\hbar} \nu,$$
	gdje je $q$ naboj čestica nosioca naboja, $\nu=C=\sum_n^N C^n=N$, pri čemu je je $N$ broj popunjenih vrpci prostorno protežnih stanja, a  $C$ je topološka invarijanta sustava zvana Chernov broj (TKNN invarijanta) \cite{stanescu,tknn}. Chernov broj $n$-te vrpce je $C^n=-\frac{1}{2\pi} \int \dd^2\theta \,\hat{\bm e}_3\cdot\curv \in \mathbb Z$, gdje je $\curv=\Nabla_\theta\times\An$ Berryjeva zakrivljenost, $\hat{\bm e}_3$ jedinični vektor u smjeru okomitom na 2D ravninu, a integracija se vrši po parametarskom prostoru koji mora imati topologiju torusa. Obično je riječ o Brillouinovoj zoni kod 2D rešetki. Popunjenost lokaliziranih stanja ne utječe na Hallovu vodljivost, što daje robusnost kvantizaciji (pojava platoa) kod sustava s neredom kod kojih lokalizirana stanja obično zauzimaju širi spektar energija unutar vrpce od protežnih stanja koja se nalaze po sredini vrpce \cite{tong,halperin}.
	
	Simetriju obrata vremena najčešće narušava magnetsko polje. Razmotrimo li neinteragirajući 2D elektronski plin (2DEG) u homogenom magnetskom polju iznosa $B$, pronaći ćemo spektar ravnih Landauovih nivoa s energijama $E_n = \hbar\omega_B (n+1/2)$, gdje je $\omega_B=|q|B/m$ ci\-klo\-tro\-nska frekvencija za čestice naboja $q$ i mase $m$, a $n=0,1,\dots$. U simetričnom baždarenju jednočestična stanja su lokalizirana na radijusu od ishodišta koji određuje kvantni broj $m$. Kad sustav ograničimo na prsten između dva radijusa te adijabatski povećavamo magnetski tok kroz šupljinu prstena, uočavamo pojavu spektralnog toka \cite{tong,halperin}. Spektralni tok podrazumijeva postupni radijalni pomak stanja stopom od $\Delta m=1$ za svaki dodatni kvant magnetskog toka $\flqnt$. Ovakvo ponašanje  posljedica je baždarne invarijantnosti te specifične topologije stanja protegnutih oko prstena. Ova pojava je usko povezana s kvantizacijom Hallove (radijalne) vodljivosti jer vremenski promjenjiv ma\-gne\-tski tok inducira azimutnu elektromotornu silu koja transportira jedan elektron radijalno s vanjskog na unutrašnji rub prstena.
	
	\subsubsection{Frakcijski kvantni Hallov efekt}
	Frakcijski kvantni Hallov efekt (FQHE), otkriven 1982. \cite{TSG}, fenomen je kvantizacije Ha\-llo\-ve vodljivosti sličan IQHE-u, ali s razlikom da se javljaju necjelobrojne vrijednosti $\nu$ kvanta vo\-dlji\-vo\-sti ($q^2/h$), koje odgovaraju djelomično popunjenim Landauovim nivoima gdje se popunjenost ponovo odnosi na protežna stanja. Ključ objašnjenja ove pojave su nezanemarive me\-đu\-ele\-ktro\-nske interakcije koje mogu otvoriti procijep unutar Landauovih nivoa \cite{tong}. Prva skupina FQHE stanja koja su objašnjena su Laughlinova stanja koja se javljaju za faktore popunjenosti oblika $1/m$, gdje je $m$ neparan u fermionskim sustavima: $\psi_m = \prod\limits_{i<j}^{N_e} \conj{z_i-z_j}^\m \GaussMB$ \cite{laughlin1983}, gdje $N_e$ predstavlja broj čestica u sustavu, $z_i=x_i+iy_i$ položaj $i$-te čestice u komple\-ksnom zapisu, $\conj{z_i}$ predstavlja komplesnu konjugaciju, a $l_B=\sqrt{\hbar/B|q|}$ je magnetska duljina. Ova stanja nisu egzaktna rješenja za realistične hamiltonijane, no odlikuje ih odgovarajući topološki red. 
	
	Ako kao i ranije adijabatski provučemo kvant magnetskog toka kroz sustav na prstenu, te pu\-stimo da unutrašnji radijus prstena teži u nulu, baždarna invarijantnost garantira da spe\-ktra\-lni tok uzrokuje pobuđenje jedne kvazičestice u sustavu. Budući da je faktor popunjenosti $1/m$, ovaj put se transportira $1/m$ elektrona \cite{laughlin1983,tong}. Posljedično nastaje lokalizirani manjak (višak) naboja $e/m$ oko solenoida koji odgovara kvazišupljini (čestici). Osim necjelobrojnog naboja Laughlinove kvazičestice imaju i (abelovsku) frakcijsku statistiku \cite{arovas} što smo pokazali u sklopu disertacije pomoću Berryjeve faze koja se javlja pri kruženju jedne oko druge kvazišupljine.
	
	Laughlinova stanja mogu objasniti FQHE i na drugim racionalnim faktorima popunjenosti s neparnim nazivnikom pomoću hijerarhijske konstrukcije \cite{haldane1983,halperin1984}. No postoje i stanja koja se moraju objasniti na druge načine te koja vjerojatno uključuju i sustave s neabelovskim pobuđenjima \cite{nayak}.
	
	\subsubsection{Anyoni}
	\textit{Anyoni} su čestice koje se pri izmjenama ponašaju u skladu s frakcijskom statistikom \cite{LeinaasMyrheim,wilczek1982}. Frakcijska statistika se razlikuje of bozonske i fermionske, a moguća je u 2D zahvaljujući topološkim razlikama u odnosu na 3D prostor. Naime, transformacije valne funkcije pri izmje\-nama čestica su zadane unitarnim reprezentacijama fundamentalne grupe odgovarajućeg ko\-nfi\-gu\-ra\-ci\-jskog prostora. Za skup $N$ identičnih ,,hardcore'' čestica\footnote{"Hardcore" odbijanje podrazumijeva odbojni potencijal koji divergira u konfiguracijama gdje se dvije ili više čestica nalazi u istoj točki čime su te konfiguracije isključene iz konfiguracijskog prostora.} fundamentalna grupa je permutacijska grupa $S_N$ za 3D ili višedimenzionalni prostor, a ,,braid'' grupa $B_N$ za 2D \cite{lerda}. Elementi $B_N$, za razliku od $S_N$, nisu isključivo mapiranja između različitih indeksiranja čestica, već se mogu prikazati kao $N$ niti koje povezuju fiksne početne i konačne konfiguracije točaka. Topološki različite ,,pletenice'' sačinjene od niti predstavljaju različite elemente. Zamjena $i$-te i $(i\!+\!1)$-te čestice je operacija $\sigma_i$, a sve takve operacije čine skup generatora grupe.

	Anyoni mogu biti abelovski i neabelovski. Abelovski se anyoni javljaju kod skalarnih reprezentacija fundamentalne grupe koje opisuje statistički parametar $\theta\in\cointerval{0}{2}$: $\rho_\theta(\sigma_i)=e^{i\theta\pi}$ \cite{wu1984}. Specijalne vrijednosti $\theta=0,1$ odgovaraju bozonskoj i fermionskoj statistici. Višedimenzionalne reprezentacije, koje su relevantne  kad čestice opisuju degenerirana stanja, mogu biti neabelovske. Valne funkcije se tada rotiraju unutar degeneriranog potprostora: $\psi_{a}\rightarrow [\rho(\alpha)]_{ab} \psi_b$. 
	
	,,Fuzija'' anyona rezultira novom vrstom anyona. Kod abelovskih anyona postoji samo jedan ishod, $\theta\times\theta=4\theta$, dok neabelovski imaju više fuzijskih kanala koji razapinju Hilbertov prostor. Ako imamo par anyona u određenom fuzijskom kanalu, dvostruka zamjena jednog anyona iz para s trećim anyonom može promijeniti fuzijski kanal. Budući da kanal fuzije nije osjetljiv na lokalne interakcije s okolinom, neabelovski se anyoni potencijalno mogu iskoristiti za kvantno računanje otporno na greške \cite{nayak}.
	
	\subsubsection{Simetrijom zaštićena topološka stanja}
	SPT stanja ne posjeduju intrinzični topološki red pa ne podržavaju anyonska pobuđenja. No s druge strane, podržavaju netrivijalna topološka granična stanja \cite{stanescu}. Kod neinteragirajućih faza ova stanja uvijek premošćuju procijep između spektralnih vrpca dok kod višedimenzionalnih interagirajućih faza mogu i ne moraju biti u procijepu. SPT faze uključuju (neinteragirajuće) topološke supravodiče i izolatore (npr. kvantni spinski Hallov efekt koji štiti simetrija obrata vremena), te postoje u prostorima svih dimenzija \cite{stanescu}.
	
	Nama je posebno zanimljiv Su-Schrieffer-Heegerov (SSH) model \cite{ssh1979}, koji je 1D rešetka sačinjena od dvije podrešetke s tuneliranjem između prvih susjeda te može biti u dvije SPT faze zaštićene kiralnom simetrijom, ovisno o Zakovoj fazi koja ovisi o relativnim vrijednostima koeficijenta tuneliranja unutar ćelije ($t$) i među ćelijama ($t'$). Netrivijalna faza SSH rešetke posjeduje lokalizirano stanje energije nula na svakom od rubova \cite{ozawa}.
	
\section*{Topološka fotonika}
	Rad predstavljen u ovom dijelu objavljen je u člancima \cite{zb} i \cite{solitonssh}.
	
\subsection*{Rotacija valnih paketa i Zitterbewegung u saćastim rešetkama s narušenom simetrijom } 
	Amplituda, faza i polarizacija su veličine koje se obično koriste za kontrolu toka svjetlosti u fotonici. Napredak u manipulaciji spinskog i dolinskog stupnja slobode otvara nove mogućnosti za  primjene u elektroničkim i poluvodičkim uređajima \cite{zutic2004,schaibley2016} te u fotonici \cite{bliokh2015,song2015,ma2016,dong2017,gao2018,wu2017,noh2018,chen2017,ni2018,shalaev2019}. Doline su lokalni minimumi u vodljivoj ili maksimumi u valentnoj vrpci. 
	
	Motivirani potrebom za istraživanjem pojava ovisnih o dolini proučavali smo propagaciju svjetla u fotoničkim saćastim rešetkama s narušenom inverzijskom simetrijom (IS). Saćaste rešetke imaju dvije neekvivalentne doline u $K$ i $K'$ točkama visoke simetrije u Brillouinovoj zoni. Selektivnim pobuđenjem jedne od dolina postižemo da početno gaussijanska zraka bez kutne količine gibanja tijekom propagacije dobije vrtložnu (,,vorteks'') komponentu. Inte\-rfe\-re\-ncija ,,vorteks'' i ,,nevorteks'' komponenti uzrokuje pojavu rotirajućeg spiralnog uzoraka u profilu intenziteta zrake, gdje smjer rotacije ovisi o pobuđenoj dolini. Posljedica ove rotacije je i \textit{Zitterbewegung} efekt, koji se očituje u oscilacijama ,,centra mase'' valnog paketa.  
	
	\textit{Zitterbewegung} je naziv za predviđene brze oscilacije masivnih relativističkih če\-sti\-ca u vakuumu, opisanih Diracovom jednadžbm, frekvencijom $2mc^2/\hbar$ \cite{schrodinger1930}, gdje je $m$ masa če\-sti\-ce, $c$ brzina svjetlosti u vakuumu, a $\hbar$ reducirana Planckova konstanta. Iako nije opažen u vakuumu, očekuje se i u materijalima u kojim se ponašanje elektrona aproksimativno opisuje istom jednadžbom te u analognim sustavima poput fotoničkih \cite{zhang2008}, uključujući saćastu rešetku čije doline, odnosno niskoenergijsko ponašanje, opisuje 2D Diracova jednadžba $H = \kappa(\sigma_x k_x + \sigma_y k_y) + \sigma_z m$, gdje su $\sigma_x$, $\sigma_y$, $\sigma_z$ Paulijeve matrice, $k_i$ je $i$-ta komponenta valnog ve\-kto\-ra mjerenog od točke $K$, $\kappa$ je koeficijent ovisan o udaljenosti i koeficijentu tuneliranja između susjednih valovoda, a masa $m$ ovisi o širini procijepa koji je otvoren u slučaju narušene inve\-rzi\-jske simetrije. Ovu pojavu uzrokuje interferencija stanja pozitivne i negativne energije, no naš rad upućuje i na interpretaciju iz drugog kuta, preko interferencije ,,vorteks'' i ,,nevorteks'' ko\-mpo\-ne\-nte, što ukazuje na vezu s topologijom. Naime, u disertaciji argumentiramo da ,,vorteks'' komponenta na\-sta\-je kao posljedica topološki netrivijalnog namatanja Berryjeve zakrivljenosti oko točaka $K$ i $K'$.
	
	U eksperimentu je korišten fotonički kristal pozadinskog indeksa loma $n_0$ s $(2+1)$D sa\-ća\-stom rešetkom valovoda indeksa loma $n_A$ i $n_B$. Razlika indeksa loma između susjednih valovoda narušava IS. Probna zraka je formirana interferencijom triju širokih gaussijanskih zraka u blizini triju ekvivalentnih $K$ ili $K'$ točaka u $k$-prostoru, čime se formira trokutasti uzorak koji je prostorno pozicioniran tako da podjednako pobudi obje podrešetke. Propagaciju svjetla u paraksijalnoj aproksimaciji opisuje jednadžba Schrödingerovog tipa
	$$ i\dv{\Psi}{z} = -\frac{1}{2k_0}\nabla^2\Psi - \frac{k_0\delta n(x,y)}{n_0}\Psi(x,y,z),$$
	gdje je $\Psi$ kompleksna amplituda električnog polja, $k_0$ valni broj u mediju, $\Delta n$ odstupanje inde\-ksa loma od $n_0$, a  os $z$ igra ulogu vremena.
	 Rezultati eksperimenta su konzistentni s numeričkim simulacijama propagacije u skladu sa Schrödingerovom jednadžbom te s analitičkim proračunima u okviru niskoenergijske teorije.
	
\subsection*{Dinamički emergirajući topološki fazni prijelazi u nelinearnim \\interagirajućim solitonskim rešetkama}
	Većina dosadašnjih studija se fokusirala na linearne topološke fotoničke strukture, no kombinacija netrivijalne topologije i nelinearnosti u fotonici otvara vrata brojnim novim fundamentalnim otkrićima i funkcionalnostima uređaja \cite{smirnova2020a}. Nelinearne topološke pojave možemo podijeliti na naslijeđene i emergentne \cite{xia2020}. U osnovi, prva skupina se javlja usprkos nelinearnosti, odnosno naslijeđena je iz odgovarajućeg linearnog sustava, dok se druga skupina javlja zahvaljujući nelinearnosti. Primjer emergentnih pojava su nelinearnošću inducirani topološki fazni prijelazi \cite{hadad2016,zhou2017,maczewsky2020,katan2016}.
	
	U ovom dijelu predstavljamo rezultate numeričke studije propagacije linearno polariziranih optičkih zraka u nelinearnom mediju. Pri specijalnim početnim uvjetima uočavaju se dinamički topološki fazni prijelazi između trivijalne i netrivijalne faze optičke SSH rešetke. Propagaciju opisuje nelinarna Schrödingerova jednadžba (NLSE)
	$$ i\pdv{\psi}{z} + \frac{1}{2k}\pdv[2]{\psi}{x} + \gamma \abs{\psi}^2 \psi(x,z) = 0, $$
	gdje je $\psi$ amplituda električnog polja, $k$ valni broj u mediju, a $\gamma$ parametar nelinearnosti koja je Kerrovog tipa i predstavlja optički inducirani potencijal. Kad je difrakcija u ravnoteži s ne\-li\-ne\-a\-rno\-sti, takav medij dozvoljava stacionarno propagirajuća rješenja koja zadržavaju oblik tijekom propagacije, tzv. solitone. 
	Kao početni uvjet postavljamo dvije $(1+1)$D podrešetke zraka koje odgovaraju solitonskim rješenjima, a propagiraju se u suprotnom smjeru u ,,prostornoj'' dime\-nziji, odnosno pod kutom u $(1+1)$D ravnini. Kako svaka solitonska zraka sama ,,usjeca'' vlastiti valovod u nelinearnom mediju, kad dvije zrake imaju nezanemariv preklop, jedna utječe na optički inducirani potencijal druge. Ovim mehanizmom nelinearnost posreduje interakciji među solitonima. Druge najbliže susjede, odnosno najbliže solitone koji pripadaju istoj podrešeci, postavljamo na udaljenost na kojoj je interakcija zanemariva. 
	
	$(1+1)$D periodička konfiguracija s dvije podrešetke i nezanemarivom interakcijom između prvih susjeda odgovara SSH modelu. Rešetka je postavljena tako da je sustav nakon poče\-tnog trenutka u topološki trivijalnoj fazi s procijepom između vrpca u spektru propagacijske ko\-nsta\-nte i bez lokaliziranih (rubnih) stanja u procijepu. Kako se propagiraju pod kutom, podrešetke se presijecaju te potom dolaze do točke gdje interakcija među ćelijama nadjačava interakciju unutar ćelije. Procijep se na trenutak zatvara te dolazi do faznog prijelaza iz netrivijalne u trivijalnu fazu. Fazni prijelaz se odražava pojavom stanja s konstantom propagacije u procijepu. Profil amplitude tih stanja pokazuje lokalizaciju na rub te čvorne točke na parnim mjestima u rešeci što odgovara topološkim rubnim stanjima. S vremenom se rubni modovi dovoljno udalje od ostatka rešetke da više nemaju nezanemarivu interakciju s prvim susjedima. Pritom se odvajaju od rešetke, što se odražava kroz potpunu lokalizaciju uz rub na profilu amplitude, te postaju samostalni solitoni s konstantom propagacije u procijepu. Kroz taj proces redukcije, rešetka postaje topološki trivijalna bez faznog prijelaza. Pri sljedećem faznom prijelazu javlja se novi par topoloških rubnih modova u procijepu. Fazni prijelazi u netrivijalnu fazu i redukcije u trivijalnu fazu ponavljaju se periodički dok se rešetke potpuno ne raziđu.
	
	Zaključno, pronašli smo nelinearnošću inducirane dinamičke fazne prijelaze u solitonskim SSH rešetkama. Ovaj rezultat služi kao demonstracija koja ukazuje na potrebu za daljnjim istraživanjem emergentnih nelinearnih topoloških pojava u fotonici. 
	
\section*{Prijedlog za realizaciju abelovskih anyona}
	Rad predstavljen u ovom dijelu objavljen je u referenci \cite{prb}.
\subsection*{Egzaktna rješenja modela za sintetičke anyone u neinteragirajućem\\ sustavu}
	Sustavi koji podržavaju frakcijsku statistiku privlačno su područje zahvaljujući pote\-nci\-ja\-lnoj primjeni neabelovskih anyona u topološkom kvantnom računanju. Budući da se anyoni obično javljaju u topološki uređenim sustavima s dalekodosežnim korelacijama, jako interagirajući su\-sta\-vi predstavljaju uobičajenu platformu za anyonske sustave. Paradigmatski primjer takvog sustava je FQHE. Anyonski sustavi predmet su brojnih objavljenih eksperimentalnih radova na različitim eksperimentalnim platformama kao što su kondenzirana tvar \cite{camino2005,mourik2012,jansa2018}, ultra\-hladni plinovi \cite{dai2017} i zarobljeni ioni \cite{barreiro2011}, fotonički \cite{lu2009,pachos2009} te drugi \cite{zhong2016,li2017} kvantni simulatori. Usto, literatura sadrži i brojne teoretske prijedloge (vidi npr. rev. članke \cite{nayak,sarma2015} te reference \cite{paredes2001,zhang2014,duan2003,jiang2008,burrello2010,kapit2014,umucalilar2017,marija2018}). 
	
	Topološko kvantno računanje zahtijeva efikasne metode za kreaciju, detekciju i manipulaciju neabelovskih anyona. Usprkos svim istraživanjima, eksperimentalno izvediva metoda još nije nadohvat ruke \cite{nayak,sarma2015} što daje motivaciju da se pokušaju neki manje konvencionalni pri\-stu\-pi realizaciji anyona. Primjerice, frakcijska statistika je moguća u slabo interagirajućim sustavima  s topološki netrivijalnom pozadinom ili smetnjama \cite{weeks2007,rosenberg2009,rahmani2013}. U ref. \cite{weeks2007,rosenberg2009} predložen je kompozitni sustav od 2DEG položenog uz supravodljivi (tip-II) film s periodičkom rešetkom vrtloga koji daju pozadinsko magnetsko polje. Šupljine i međupoložajne primjese u rešeci tada nose frakcijsku statistiku.

	Ovdje proučavamo model za sintetičke anyone čiji specijalni slučaj opisuje sustav iz ref. \cite{weeks2007,rosenberg2009}. Model opisuje hamiltonijan za neinteragirajući 2DEG u homogenom magnetskom polju s $N$ tankih vanjskih solenoida koji nose necjelobrojni magnetski tok $\Phi=\alpha\flqnt$, gdje je $\alpha\in\ointerval{0}{1}$. Daleko od solenoida sustav zadržava spektralnu strukturu IQHE-a, te efekt solenoida možemo objasniti spektralnim tokom. Jednočestična stanja u najnižem Landauovom nivou (LLL) energije $E_0=\hbar\omega_B/2$ su oblika
	$$ \psi_m = |z-\eta|^{-\alpha} \conj{z-\eta} \,\conj{z}^m \exp(-\frac{|z|^2}{4l_B^2}), 
		\quad m=0,1,2,\dots \, , $$
	za $N=1$ solenoid na položaju $\eta$, gdje su $\eta$ i $z$ koordinate u kompleksnom zapisu, a $l_B$ ma\-gne\-tska duljina. Stanja za više solenoida lako se konstruiraju dodavanjem faktora $|z-\eta_i|^{-\alpha} \conj{z-\eta_i}$. Osim stanja u Landauovim nivoima blizu svakog solenoida javlja se jedno lokalizirano stanje po nivou s energijom u procijepu ($\hbar\omega_B(1+2\alpha)/2$ za LLL).
	Za potrebe konstrukcije any\-o\-nskog sustava biramo $\alpha\in\ointerval{0}{1}$ te Fermijevu energiju postavljamo poviše energije LLL-a, no ispod energije lokaliziranih stanja. Osnovno stanje $\Psi$ konstruiramo kao Slaterovu determinantu popunjenih jednočestičnih stanja.
	
	Slično kao za Laughlinove kvazičestice, statistiku izmjene solenoida pokazujemo pomoću Berryjeve faze nakupljene tijekom kruženja jednog solenoida oko drugog \cite{arovas}. Ova faza sadrži doprinos od Aharonov-Bohmove faze koji treba oduzeti da bi se pronašla statistička faza. Za udaljene solenoide statistička faza je necjelobrojna te iznosi
	$$ \gamma_S=2\pi(\alpha-1),$$
	što implicira statistički parametar $\theta=\alpha-1$. Ovaj smo analitički rezultat potvrdili i numerički za više vrijednosti $\alpha$ te za različite radijuse kružne putanje. Provjerili smo i da numerički rezultat nije osjetljiv na deformacije putanje, te da jednostruka izmjena daje fazu $\pi\theta$.
	
	Iako svaki solenoid istiskuje dio naboja iz svoje okoline, kao i kod Laughlinovih kvazi\-čestica, sintetičke anyone u ovom sustavu ne možemo poistovjetiti s istisnutim nabojem jer stanje perturbiranog sustava ne odgovara emergentnoj kvazičestici, kao ni s Wilczekovim kompo\-zi\-tnim česticama od istisnutog naboja te toka u solenoidu. Posljedica ovoga je da  Aharonov-Bohmova faza koja potječe od kruženja istisnutog naboja i solenoida ne odgovara dobivenoj Berryjevoj fazi, odnosno učinak pomicanja solenoida na valnu funkciju nije u potpunosti opisan gibanjem ististnutog naboja. Osim toga, spajanje dvaju solenoida ne stvara anyon očekivane $4\theta$ statistike. Ipak, očekivani rezultat fuzije dobije se u procesu gdje par udaljenih solenoida obilazi identičan par bez promjene relativih pozicija solenoida unutar parova.
	
\section*{Zaključak}
	U ovoj disertaciji proučavali smo više tema koje objedinjuje utjecaj topologije na fiziku sustava. Glavni cilj bio je prezentacija originalnih članaka \cite{zb,solitonssh,prb}, no pritom smo predstavili i neke bitnije teme vezane za topološku kvantnu materiju uključujući IQHE, FQHE, anyone i SPT faze. IQHE i FQHE, iako različiti, su povezani, a obje pojave su bitna pozadina za naše sintetičke anyone. Dok IQHE služi kao temelj za naš model, Laughlinove kvazičestice u FQHE-u daju inspiraciju za ideju perturbacije sustava lokaliziranim frakcijskim solenoidnim probama.
	
	U sklopu prezentacije prvog članka prikazali smo utjecaj netrivijalne topologije dolinskog stupnja slobode saćaste rešetke na propagaciju svjetla kroz fotonički kristal. Ovaj utjecaj se manifestira kroz pojavu ,,vorteks'' komponente koja u interferenciji s "nevorteks" komponentom uzrokuje rotirajući spiralni uzorak u profilu intenziteta te s njim povezanu pojavu Zitterbewegunga. Ove pojave smo vidjeli u rezultatima eksperimenta i u teorijskim proračunima.
	
	U sljedećem dijelu predstavili smo rezultate numeričke studije evoluirajuće solitonske SSH rešetke koja je moguća zahvaljujući nelinearnosti fotoničkog medija. Uočili smo periodičku pojavu rubnih stanja u spektralnom procijepu s faznim profilom topoloških rubnih stanja koja signalizira topološke fazne prijelaze. Ovakvi topološki fazni prijelazi primjer su emergentne nelinearne topološke pojave.
	
	U posljednjem dijelu predlažemo realizaciju sintetičkih anyona perturbacijom neinteragirajućeg sustava posebnim lokaliziranim probama. Naši sintetički anyoni se razlikuju od emergentnih frakcijskih kvazičestica kakve se javljaju u sustavima poput FQHE-a, no realizacija ovog modela je poželjna jer bi njegov neabelovski pandan mogao biti podloga za topološka kvantna računala. Sustavi ultrahladnih atoma sa sintetičkim vektorskim potencijalima su obećavajuća platforma za ovu namjenu.

\selectlanguage{english} 